\documentclass[a4paper,11pt]{article}
\pdfoutput=1 
\usepackage{longtable}
\usepackage[utf8]{inputenc}
\usepackage{array}
\usepackage{jheppub} 
\usepackage{multirow}

\newcommand{\tev}{\, {\rm TeV}}
\newcommand{\gev}{\, {\rm GeV}}
\newcommand{\be}{\begin{equation}}
\newcommand{\ee}{\end{equation}}
\newcommand{\beq}{\begin{equation}}
\newcommand{\eeq}{\end{equation}}
\newcommand{\bea}{\begin{eqnarray}}
\newcommand{\eea}{\end{eqnarray}}
\newcommand{\nn}{\nonumber}

\newcommand{\cO}{{\cal O}}

\newcommand{\cL}{{\cal L}}

\newcommand{\AFB}{A_{\rm FB}}
\newcommand{\met}{\not{\!\!{\rm E}}_{T}}

\title{\boldmath Collider Phenomenology of the 3-3-1 Model}

\author[a,b,c]{Qing-Hong Cao}
\author[a]{Dong-Ming Zhang }

\affiliation[a]{Department of Physics and State Key Laboratory of Nuclear Physics and Technology,\\
Peking University, Beijing 100871, China}
\affiliation[b]{Collaborative Innovation Center of Qantum Matter, Beijing, China}
\affiliation[c]{Center for High Energy Physics, Peking University, Beijing 100871, China}

\emailAdd{qinghongcao@pku.edu.cn}
\emailAdd{zhangdongming@pku.edu.cn}

\abstract{
We study collider phenomenology of the so-called 331 model with $SU(3)_C\otimes SU(3)_L\otimes U(1)_X$ gauge structure at the large hadron collider, including single and double Higgs boson productions, Higgs boson rare decay, $Z^\prime$ boson production, new charged gauge boson pair production, and heavy quark pair production. We discuss all the possible collider signatures of new particle productions. Four benchmark 331 models, $\beta=\pm \sqrt{3}$ and $\beta=\pm 1/\sqrt{3}$, are studied in this work.}

\begin{document}
\maketitle
\flushbottom

\newpage
\section{Introduction}

The 331 model is a simple extension of the SM based on the gauge group $SU(3)_C\times SU(3)_L\times U(1)_X$~\cite{Frampton:1992wt,Pisano:1991ee}. 
There are different versions~\cite{Dias:2003zt,Dias:2003iq,Diaz:2004fs,Dias:2004dc,Dias:2004wk,Dias:2005jm,Dias:2005yh,Dias:2011sq,Ochoa:2005ch} of this model which can be characterized by a
parameter called $\beta$. Models with different $\beta$ have new particles with different electric charges.
But in general, they all have the same features. i) Unlike the SM that anomaly cancellation is fulfilled within each generation, the gauger anomaly is cancelled in the 331 model when considering all the generations. In particular, the number of triplets must be equal to the number of anti-triplets in fermion sector, 
due to the nontrivial $SU(3)$ gauge structure. The number of
generations $N$ must be a multiple of three. On the other hand, in order to ensure QCD an asymptotic free theory, $N$ has to be smaller than six. So the number of generations $N$ is equal to three. That explains why the SM has three generations.  
ii) One of the three quark generations is different from the
other two, making sure that the anomaly is free, which leads to tree-level  Flavour Changing Neutral Current (FCNC) through a new neutral gauge boson $Z^\prime$ or the mixing of $Z$ and $Z^\prime$. And if we choose the third generation of quark as a different one, the heavy top quark mass may be explained.
iii) Peccei-Quinn (PQ) symmetry~\cite{Peccei:1977hh} which can solve the strong CP problem is a natural result of
gauge invariance in the 331 model~\cite{Dias:2003zt,Dias:2003iq}. Thus PQ symmetry does not suffer from quantum corrections, which means it is not a classical symmetry but a quantum-level one. 
iiii) With the extension of gauge group, particles in the 331 model are richer than the SM. For instance, there will be three more gauge bosons, three more heavy quarks or leptons and six more Higgs scalars, which gives rise to very rich phenomenology at the LHC.

The 331 model has already been studied in many aspects. 
For the collider physics, the research of the SM Higgs boson \cite{Ninh:2005su}, charged Higgs boson \cite{Okada:2016whh,Martinez:2012ni,Montalvo:2012qg,Alves:2011mz,
Montalvo:2008cx,CiezaMontalvo:2008sa,CiezaMontalvo:2008ew,Soa:2007zz,VanSoa:2006ea,CiezaMontalvo:2006zt}, $Z^\prime$ \cite{Salazar:2015gxa,Coutinho:2013lta,Martinez:2009ik,RamirezBarreto:2007cie,RamirezBarreto:2006tn,Salazar:2015gxa}, exotic quark \cite{Coutinho:1999hf,Cabarcas:2008ys} have already been 
done by many people. For the neutrino physics, people have studied how to 
generate the small neutrino mass in different versions of the 331 model \cite{Okada:2015bxa,Boucenna:2014ela,Benavides:2015afa,Boucenna:2014dia,Ky:2005yq}. 
The usual way to generate the neutrino mass is through seesaw mechanism, loop induced process or high dimensional operator. A recent review of 
neutrino mass mechanisms in the 331 model can be found in \cite{Pires:2014xsa}.
For the fermion mass mixing, different flavor symmetries including $D_4$ \cite{Vien:2014ica,Vien:2014soa,Vien:2013zra}, $S_3$ \cite{Hernandez:2014vta,Vien:2014vka,Hernandez:2013hea,Dong:2011vb}, $A_4$ \cite{Hernandez:2015tna,Vien:2014pta,Dong:2010gk}, $S_4$ \cite{VanVien:2015xha,Dong:2010zu}, $T_7$ \cite{Hernandez:2015yxx,Vien:2014gza} and $\Delta_{27}$~\cite{Vien:2016tmh,Hernandez:2016eod} have been introduced to explain the fermion mass mixing pattern.
In the dark matter aspect, there is a residual $Z_2$ symmetry~\cite{Mizukoshi:2010ky} after spontaneously symmetry breaking (SSB). The lightest particle in the $Z_2$-odd sector will be a dark matter candidate. Researches related to this aspect can be found in Ref.~\cite{Pires:2016vek,Dong:2015rka,Cogollo:2014jia,Dong:2014esa,Dong:2013ioa}.
Since there will be tree-level FCNC processes through $Z^\prime$ or neutral scalars, it is necessary to consider the constraints from $B$, $D$ and $K$ mesons \cite{Buras:2016dxz,Buras:2015kwd,Dong:2015dxw,Correia:2015tra,Machado:2013jca,Buras:2013dea,DeFazio:2013hla,Cogollo:2012ek,Benavides:2009cn,Promberger:2007py,Rodriguez:2004mw}. 
Lepton flavor violation process can be generated because of introducing the third component of  lepton fields~\cite{Fonseca:2016xsy,Machado:2016jzb,Boucenna:2015zwa,Hue:2015fbb,Hua:2014yna,VanSoa:2008bm}.
New contributions to the electron and neutron electric dipole moment (EDM) \cite{DeConto:2016osh,DeConto:2014fza} and muon $g-2$ \cite{DeConto:2016ith,Binh:2015cba,Kelso:2014qka,Kelso:2013zfa} through new charged scalars and charged gauge bosons have also been studied.  

In this work, we consider different versions of the 331 model, namely $\beta=\pm \sqrt{3},\pm 1/\sqrt{3}$. For
simplicity, we do not consider FCNC and the mixing of $Z$ and $Z^\prime$ since they are too small to affect the processes of interest to us. Detailed discussions of the FCNC interactions in the $331$ model can be found in Refs.~\cite{Buras:2014yna,Teles:2013pea,Cabarcas:2011hb,CarcamoHernandez:2005ka}. 

The paper is organized as follows. In Sec.~\ref{sec:sec2} we introduce the $331$ model briefly. In Sec.~\ref{sec:sec3}, we discuss the Higgs boson phenomenology in the 331 model. We derive constraints on the parameter space of the 331 model from the single Higgs boson measurements and then discuss the Higgs rare decay, $\sigma(pp\to h\to Z\gamma)$, and di-Higgs boson production $\sigma(pp\to hh)$ at the LHC.
We study the production and decay of the new particles~($Z^\prime,~V,~Y,~D,~S,~T$) in this model and give the possible collider signature of those particles in Sec.~\ref{sec:sec4}, \ref{sec:sec5} and \ref{sec:sec6}.
In the last section, we will make a conclusion. In Appendix~\ref{app_a}, we list the Feynman rules of the interaction vertices in unitary gauge in the 331 model. Appendix~\ref{app_b} is used to give a detailed expression of the rotation matrix for the neutral scalars. The loop functions in the loop-induced decay width of the Higgs boson is shown in Appendix~\ref{app_c}.

\section{The Model}\label{sec:sec2}
The 331 model has been studied in details in Ref.~\cite{Buras:2012dp}. In this section we briefly review the model and present the masses of new physics resonances. All the Feynman rules of the interaction 
vertices in unitary gauge among the new scalar sector, the new gauge bosons, the new fermions and the SM particles can be found in Appendix~\ref{app_a}. Three-point couplings of one gauge boson 
to two scalars ($VSS$) are listed in Table~\ref{1gauge-2scalars}. Three-point couplings
of two gauge bosons to one scalar ($VVS$) are listed in Table~\ref{2gauges-1scalar}.
Four-point couplings of two gauge bosons to two scalars ($VVSS$) are listed in Table~\ref{2gauges-2scalars}.
Gauge boson self-couplings are listed in Table~\ref{gauges}. Gauge boson-fermion couplings are listed in
Table~\ref{gauge-fermion}. Scalar-fermion couplings are listed in Table~\ref{scalar-fermion}.

\subsection{Higgs Sector}
\begin{figure}
\centering
\includegraphics[width=0.35\textwidth]{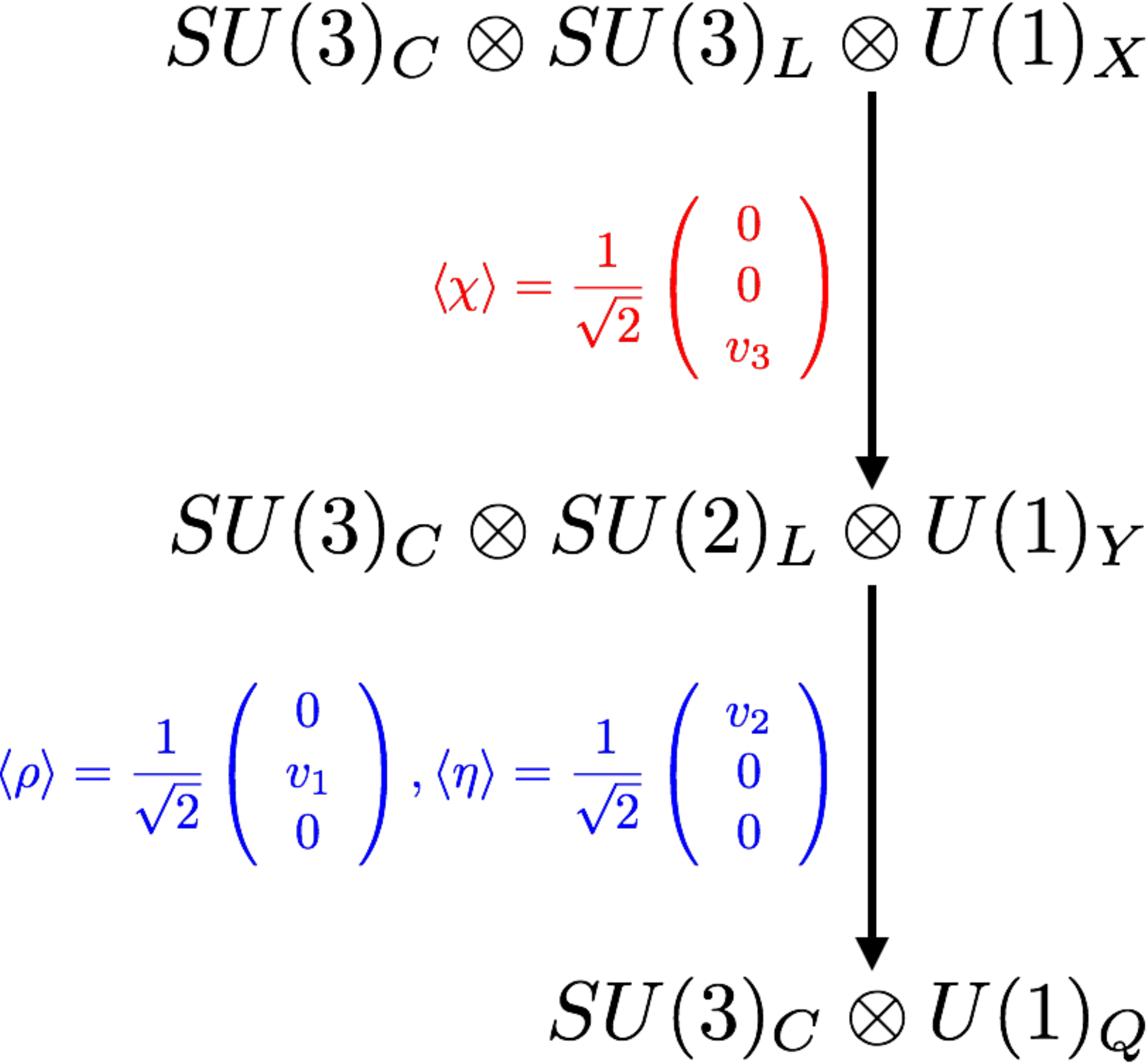}
\caption{The symmetry breaking pattern of the 331 model.}\label{fig:SSB}
\end{figure}

The symmetry breaking pattern of the 331 model is
\be
SU(3)_L\times U(1)_X\rightarrow SU(2)_L\times U(1)_Y\rightarrow U(1)_Q, 
\ee
which is realized by introducing three Higgs triplets $\rho$, $\eta$ and $\chi$
\be
\rho= \left(\begin{array}{c}
\rho^+  \\
 \rho^0   \\
\rho^{-Q_V}  \\
\end{array}\right) \quad
\eta= \left(\begin{array}{c}
\eta^0  \\
 \eta^-   \\
\eta^{-Q_Y}  \\
\end{array}\right) \quad
\chi= \left(\begin{array}{c}
\chi^{Q_Y}  \\
 \chi^{Q_V}   \\
\chi^0  \\
\end{array}\right), \label{1}
\ee
where $Q_V$ and $Q_Y$ are the undetermined electric charges of the corresponding Higgs fields.
The vacuum expectation values (vevs) of $\rho$, $\eta$ and $\chi$ are chosen as follows:
\be
\left<\rho\right>=\frac{1}{\sqrt{2}} \left(\begin{array}{c}
0  \\
v_1   \\
0  \\
\end{array}\right),\quad
\left<\eta\right>=\frac{1}{\sqrt{2}} \left(\begin{array}{c}
v_2  \\
0   \\
0  \\
\end{array}\right),\quad
\left<\chi\right>=\frac{1}{\sqrt{2}} \left(\begin{array}{c}
0  \\
0   \\
v_3  \\
\end{array}\right) .
\ee
At the first step of the symmetry breaking, $\chi$ is introduced to break $SU(3)_L\times U(1)_X$ to $SU(2)_L\times U(1)_Y$ at a very large scale $v_3$, typically at $\tev$ scale. At the second step, we use $\rho$ and $\eta$ to break $SU(2)_L\times U(1)_Y$ down to the residual $U(1)_Q$ electromagnetic symmetry at about the weak scale, i.e. $v_1 \sim v_2 \sim m_W$. Thus, we have $v_3$ $\gg$ $v_{1,2}$, named as the ``decoupling limit". The symmetry breaking pattern is shown in Fig.~\ref{fig:SSB}.

The hypercharge generator is obtained by a linear combination of $X$ and $T^8$, 
\beq
Y=\beta T^8+X ,
\eeq
where $T^8$ is diagonal generator of $SU(3)_L$ and $X$ is the generator of $U(1)_X$. That leads to the electric charge generator as 
\beq
Q=T^3+\beta T^8+X .
\eeq
The $X$ quantum numbers of the three Higgs triplets can be fixed by the electric charge generator as follows:
\beq
X_\rho=\frac{1}{2}-\frac{\beta}{2\sqrt{3}},\quad
X_\eta=-\frac{1}{2}-\frac{\beta}{2\sqrt{3}},\quad
X_\chi=\frac{\beta}{\sqrt{3}},\label{xqn}
\eeq
which give rise to the $Q_Y$ and $Q_V$ of those scalars in Eq.~\eqref{1} as follows:
\beq
Q_Y=\frac{\sqrt{3}}{2}\beta+\frac{1}{2},\quad
Q_V=\frac{\sqrt{3}}{2}\beta-\frac{1}{2}.
\eeq

The Lagrangian of the Higgs sector are
\be
\cL_{Higgs}=(D_\mu \rho)^\dagger (D^\mu \rho)+(D_\mu \eta)^\dagger (D^\mu \eta)+(D_\mu \chi)^\dagger (D^\mu \chi)-V_{Higgs},
\ee
with the Higgs potential
\bea
V_{Higgs}&=&\mu_1^2 (\rho^\dagger \rho)+
 \mu_2^2 (\eta^\dagger \eta)+\mu_3^2 (\chi^\dagger \chi)+ \lambda_1 (\rho^\dagger \rho)^2+
 \lambda_2 (\eta^\dagger \eta)^2+\lambda_3 (\chi^\dagger \chi)^2  \nn \\
 &+& \lambda_{12}(\rho^\dagger \rho)(\eta^\dagger \eta)+
\lambda_{13}(\rho^\dagger \rho)(\chi^\dagger \chi)+
\lambda_{23}(\eta^\dagger \eta)(\chi^\dagger \chi)  \nn \\
 &+& \lambda^\prime_{12}(\rho^\dagger \eta)(\eta^\dagger \rho)+
\lambda^\prime_{13}(\rho^\dagger \chi)(\chi^\dagger \rho)+
\lambda^\prime_{23}(\eta^\dagger \chi)(\chi^\dagger \eta ) \nn  \\
&+& \sqrt{2} f (\epsilon_{ijk}\rho^i \eta^j \chi^k+h.c.)\label{VHiggs},
\eea
where the $\lambda^{(\prime)}_{i(j)}$'s denote the dimensionless parameter while the $\mu_i$ and $f$ are mass-dimensional parameters. For simplicity,  we assume that $f$ is proportional to $v_3$, $f=kv_3$ with $k\sim O(1)$, such that the model has no other scale. 

\subsubsection{Neutral scalars}

The neutral scalar fields can be written in terms of their real and imaginary components explicitly 
\be
\rho^0 ={1 \over \sqrt{2}}(\xi_\rho +i \zeta_\rho), \quad
\eta^0 ={1 \over \sqrt{2}}(\xi_\eta +i \zeta_\eta), \quad
\chi^0 ={1 \over \sqrt{2}}(\xi_\chi +i \zeta_\chi).
\ee
When the real components of those scalars develop the vacuum expectation values, the Higgs potential gives rise to a mass matrix of those read components as following
\be
M_{H}^2=
\left(
\begin{array}{ccc}
 2 \lambda _1 v_1^2+\frac{f v_2 v_3}{v_1} & -f v_3+v_1 v_2 \lambda _{12} & -f v_2+v_1 v_3 \lambda _{13} \\
 -f v_3+v_1 v_2 \lambda _{12} & 2 \lambda _2 v_2^2+\frac{f v_1 v_3}{v_2} & -f v_1+v_2 v_3 \lambda _{23} \\
 -f v_2+v_1 v_3 \lambda _{13} & -f v_1+v_2 v_3 \lambda _{23} & 2 \lambda _3 v_3^2+\frac{f v_1 v_2}{v_3} \\
\end{array}
\right)\label{Neutral even mass mixing}.
\ee

Three CP-even scalars emerge after the symmetry breaking as the eigenstates of the mass matrix in Eq.~\eqref{Neutral even mass mixing} and one scalar is identified as the SM Higgs boson ($h$). The other two scalars are named as $H_1$ and $H_2$. Even though the masses of the three scalars can be solved analytically by diagonalizing the mass matrix, the expressions are too complicated to show here. For simplicity, we expand the CP-even scalar masses in the decoupling limit, $v_3 \gg v_{1,2}$,  which yields three mass eigenvalues at the lowest order as following
\be
m_h^2=\frac{2\left(\lambda_1 v_1^4+\lambda_{12}v_1^2 v_2^2+\lambda_2 v_2^4\right)}{v^2},\quad
M_{H_2}^2=\frac{v^2}{v_1 v_2}fv_3,\quad
M_{H_3}^2=2 \lambda _3 v_3^2,
\ee
where $v^2\equiv v_1^2+v_2^2$. 
As to be shown soon, we require
\beq
v =  \sqrt{v_1^2 + v_2^2}=246~{\rm GeV}.
\eeq
in order to obtain the correct mass of the $W$-boson.
The three neutral scalars $h$, $H_2$ and $H_3$ are related with the weak eigenstates by a rotation matrix $U$
\be
\left(\begin{array}{c}
h\\
H_2\\
H_3\\
\end{array}\right)=
\left(\begin{array}{c c c}
U_{11} & U_{12} & U_{13}\\
U_{21} & U_{22} & U_{23}\\
U_{31} & U_{32} & U_{33}\\
\end{array}\right)^\dag
\left(\begin{array}{c}
\xi_\rho\\
\xi_\eta\\
\xi_\chi\\
\end{array}\right).
\ee
The rotation matrix can also be analytically solved in terms of those three eigenvalues. 
In this work we focus on the three matrix element in the first column and explore the phenomenology of the Higgs boson.  
One can expand the matrix elements in the decoupling limit $v_3 \gg v_{1,2}$.
To the lowest order, the results are
\be
U_{11}=c_{12}+ \frac{g_{11}(v)}{v_3^2},\quad
U_{21}=s_{12}+ \frac{g_{21}(v)}{v_3^2},\quad
U_{31}=\frac{g_{31}(v)}{v_3},\label{U}
\ee
where we have defined
\be
s_{12}=\frac{v_2}{v},\quad
c_{12}=\frac{v_1}{v}
\ee
and $g_{i1}(v)$ are functions independent of $v_3$. The detailed expression of $g_{i1}(v_1)$ is given in Appendix~\ref{app_b}.

For the imaginary part of the neutral scalars, the mass mixing matrix is
\be
M_{H^\prime}^2=
\left(
\begin{array}{ccc}
 v_2 v_3/v_1 & v_3 & v_2 \\
 v_3 & v_1 v_3/v_2 & v_1 \\
 v_2 & v_1 & v_1 v_2/v_3 \\
\end{array}
\right)f.
\ee
The three mass eigenvalues are
\be
M_{H_0}^2=\frac{f \left(v_1^2 v_2^2+v_3^2 v_2^2+v_1^2 v_3^2\right)}{v_1 v_2 v_3},\quad
M_{G_Z}^2=0,\quad
M_{G_{Z^\prime}}^2=0.
\ee
Here, $H_0$ refers to the CP-odd Higgs boson, while $G_Z$ and $G_{Z^\prime}$ represent the goldstone bosons eaten by gauge bosons $Z$ and $Z^\prime$ respectively. The mass eigenstates are linked with the weak eigenstates by the following rotation matrix $O$
\be
\left(\begin{array}{c}
H_0\\
G_Z\\
G_{Z^\prime}\\
\end{array}\right)=
\left(
\begin{array}{ccc}
O_{11} & O_{12} & O_{13} \\
O_{21} & O_{22} & O_{23} \\
O_{31} & O_{32} & O_{33} \\
\end{array}
\right)^\dag
\left(\begin{array}{c}
\zeta_\rho\\
\zeta_\eta\\
\zeta_\chi\\
\end{array}\right).
\ee
See Appendix~\ref{app_b} for detailed expressions of $O_{ij}$'s. In the decoupling limit, we obtain
\be
O_{11}=O_{22}=s_{12},\quad
O_{21}=-O_{12}=c_{12},\quad
O_{i3}=O_{3i}=0,\quad
O_{33}=1.
\ee

\subsubsection{Charged scalars}

In addtion to the neutral scalars mentioned above, there are three kinds of charged scalars whose electric charges are $\pm1$, $\pm Q_Y$ and $\pm Q_V$. The mass mixing matrix of charge $\pm1$ is
\be
M_{\pm1}^2=
\left(
\begin{array}{ccc}
 \dfrac{1}{2} \lambda '_{12} v_2^2+\dfrac{f v_3 v_2}{v_1} &~~& -f v_3-\dfrac{1}{2} v_1 v_2 \lambda '_{12} \\
 -f v_3-\dfrac{1}{2} v_1 v_2 \lambda '_{12} &~~& \dfrac{1}{2} \lambda '_{12} v_1^2+\dfrac{f v_3 v_1}{v_2} \\
\end{array}
\right),
\ee
with two mass eigenvalues
\be
M_{G^\pm}^2=0,\quad
M_{H^\pm}^2=\frac{2fv_3+\lambda^\prime_{12}v_1v_2}{2v_1v_2}v^2.
\ee
The $H^\pm$ scalars are the physical states with electric charge $\pm1$, and $G^\pm$ are the goldstone bosons eaten by $W^\pm$. The rotation matrix from weak eigenstates to mass eigenstates is 
\be
\left(\begin{array}{c}
H^\pm\\
G^\pm\\
\end{array}\right)=
\left(
\begin{array}{cc}
c_{12} & -s_{12}\\
s_{12} & c_{12}\\
\end{array}
\right)
\left(\begin{array}{c}
\eta^\pm\\
\rho^\pm\\
\end{array}\right),
\ee
The mass mixing matrix of charge $\pm Q_Y$ is
\be
M_{\pm Q_Y}^2=
\left(
\begin{array}{ccc}
 \dfrac{1}{2} \lambda '_{23} v_3^2+\dfrac{f v_1 v_3}{v_2} &~~& -f v_1-\dfrac{1}{2} v_2 v_3 \lambda '_{23} \\
 -f v_1-\dfrac{1}{2} v_2 v_3 \lambda '_{23} &~~& \dfrac{1}{2} \lambda '_{23} v_2^2+\dfrac{f v_1 v_2}{v_3} \\
\end{array}
\right),
\ee
with the mass eigenvalues
\be
M_{G^{\pm Q_Y}}^2=0,\quad
M_{H^{\pm Q_Y}}^2=\frac{2fv_1+\lambda^\prime_{23}v_2 v_3}{2v_2 v_3}(v_2^2+v_3^2).
\ee
The $H^{\pm Q_Y}$ scalars stand for the physical states with electric charge $\pm Q_Y$\footnote{Even though $Q_Y$ can be zero when $\beta=-1/\sqrt{3}$, we call $H^{\pm Q_Y}$ as charged scalar for simplicity.} while $G^{\pm Q_Y}$ are the goldstone bosons eaten by new gauge bosons $Y^{\pm Q_Y}$'s. The rotation matrix from weak eigenstates to mass eigenstates is 
\be
\left(\begin{array}{c}
H^{\pm Q_Y}\\
G^{\pm Q_Y}\\
\end{array}\right)=
\left(
\begin{array}{cc}
c_{23} & -s_{23}\\
s_{23} & c_{23}\\
\end{array}
\right)
\left(\begin{array}{c}
\chi^{\pm Q_Y}\\
\eta^{\pm Q_Y}\\
\end{array}\right),
\ee
where we define
\be
s_{23}=\frac{v_3}{\sqrt{v_2^2+v_3^2}},\quad
c_{23}=\frac{v_2}{\sqrt{v_2^2+v_3^2}}.\label{sc23}
\ee
As for the charged scalars with electric charge $\pm Q_V$\footnote{Again, the charge $Q_V$ can be zero when $\beta=+1/\sqrt{3}$, but we call $H^{\pm Q_V}$ as charged scalar for simplicity.}, the mass mixing matrix is
\be
M_{\pm Q_V}^2=
\left(
\begin{array}{ccc}
 \dfrac{1}{2} \lambda '_{13} v_3^2+\dfrac{f v_2 v_3}{v_1} &~~& -f v_2-\dfrac{1}{2} v_1 v_3 \lambda '_{13} \\
 -f v_2-\dfrac{1}{2} v_1 v_3 \lambda '_{13} &~~& \dfrac{1}{2} \lambda '_{13} v_1^2+\dfrac{f v_2 v_1}{v_3} \\
\end{array}
\right),
\ee
with two mass eigenvalues
\be
M_{G^{\pm Q_V}}^2=0,\quad
M_{H^{\pm Q_V}}^2=\frac{2fv_2+\lambda^\prime_{13}v_1 v_3}{2v_1 v_3}(v_1^2+v_3^2).
\ee
Here, $H^{\pm Q_V}$ represent the physical states with electric charge $\pm Q_V$ while $G^{\pm Q_V}$'s are the goldstone bosons eaten by new gauge bosons $V^{\pm Q_V}$'s. The rotation matrix from weak eigenstates to mass eigenstates are
\be
\left(\begin{array}{c}
H^{\pm Q_V}\\
G^{\pm Q_V}\\
\end{array}\right)=
\left(
\begin{array}{cc}
c_{13} & -s_{13}\\
s_{13} & c_{13}\\
\end{array}
\right)
\left(\begin{array}{c}
\chi^{\pm Q_V}\\
\rho^{\pm Q_V}\\
\end{array}\right),
\ee
where we define
\be
s_{13}=\frac{v_3}{\sqrt{v_1^2+v_3^2}},\quad
c_{13}=\frac{v_1}{\sqrt{v_1^2+v_3^2}}.\label{sc13}
\ee

\subsection{Gauge Sector}
The Lagrangian of the gauge sector is
\be
\cL_{Gauge}=-\frac{1}{4}F_{\mu\nu}F^{\mu\nu}-\frac{1}{4}B_{\mu\nu}B^{\mu\nu},
\ee
where $F_{\mu\nu}$ and $B_{\mu\nu}$ are field strength tensors of $SU(3)_L$ and $U(1)_X$
\be
F_{\mu\nu} = \partial_\mu W^i_\nu-\partial_\nu W^i_\mu, \label{2} \quad
B_{\mu\nu} = \partial_\mu X_\nu-\partial_\nu X_\mu.
\ee
Here, $W^i_\mu$ and $X_\mu$ denotes the gauge fields of $SU(3)_L$ and $U(1)_X$, respectively, where the index $i$ runs from 1 to 8.

After the spontaneous symmetry breaking (SSB), gauge bosons obtain their masses from the kinetic terms of the Higgs fields
\be
\cL_{Higgs}^{kinetic}=(D_\mu \rho)^\dagger (D^\mu \rho)+(D_\mu \eta)^\dagger (D^\mu \eta)+(D_\mu \chi)^\dagger (D^\mu \chi). \label{0}
\ee
We explicitly write out the specific terms responsible for the masses of the gauge bosons
\bea
\cL_{Gauge}^{mass}&=&\left<\chi\right>^\dag G_\mu^\dag G^\mu \left<\chi\right>+ \left<\rho\right>^\dag G_\mu^\dag G^\mu \left<\rho\right>+ \left<\eta\right>^\dag G_\mu^\dag G^\mu \left<\eta\right> \notag\\
   &=&(\left<\chi\right>+ \left<\rho\right>+ \left<\eta\right>)^\dag G_\mu^\dag G^\mu(\left<\chi\right>+ \left<\rho\right>+ \left<\eta\right>) \notag\\
   &=&\frac{1}{2}\left(
                  \begin{array}{ccc}
                    v_2 & v_1 & v_3 \\
                  \end{array}
                \right)
     G_\mu^\dag G^\mu
     \left(
       \begin{array}{c}
         v_2 \\
         v_1 \\
         v_3 \\
       \end{array}
     \right)\label{gaugemass},
\eea
where 
\bea
&G_\mu&=gW^i_\mu T^i + g_X X X_\mu T^9  \label{Gmu} \nn \\ 
&=& {1 \over 2} \left(
\begin{array}{ccc}
g(W_\mu^3+{1 \over \sqrt{3}} W_\mu^8)+2g_X X X_\mu & \sqrt{2} gW_\mu^+ & \sqrt{2} gY_\mu^{+Q_Y}  \\ \nn
\sqrt{2} gW_\mu^-  & -g(W_\mu^3-{1 \over \sqrt{3}} W_\mu^8)+2g_X X_\mu & \sqrt{2} gV_\mu^{+Q_V}   \\ \nn
\sqrt{2} gY_\mu^{-Q_Y}  & \sqrt{2} gV_\mu^{-Q_V}  & -{2 \over \sqrt{3}} gW_\mu^8+2g_X X X_\mu
\end{array}\right). \\
\eea
In the above equation, $g$ and $g_X$ are coupling constants of $SU(3)_L$ and $U(1)_X$ respectively. $2T^i$ are eight Gell-Mann Matrices. $T^9$ is just the Identity matrix. $X$ is the $U(1)_X$ quantum number of some specific Higgs fields. We also define the mass eigenstates of the charged gauge bosons as follows 
\be
W^\pm_\mu=\frac{1}{\sqrt{2}}(W^1_\mu \mp iW^2_\mu), \quad
Y^{\pm Q_Y}_\mu=\frac{1}{\sqrt{2}}(W^4_\mu \mp iW^5_\mu), \quad
V^{\pm Q_V}_\mu=\frac{1}{\sqrt{2}}(W^6_\mu \mp iW^7_\mu).
\ee
Inserting Eq.~\eqref{Gmu} into \eqref{gaugemass} and also using the $X$ quantum numbers given in E.q~\eqref{xqn}, one can get the mass terms of gauge bosons as follows
\bea
\cL_{Gauge}^{mass}&=&\frac{1}{4}g^2v^2W_\mu^+W^{\mu,-}+\frac{1}{4}g^2(v_2^2+v_3^2)Y_\mu^{Q_Y}Y^{\mu,-Q_Y}+
\frac{1}{4}g^2(v_1^2+v_3^2)V_\mu^{Q_V}V^{\mu,-Q_V}\label{gauge mass}\nn\\
 \nn
  &+&~\frac{1}{6}v_3^2(-gW_\mu^8+\beta g_XX_\mu)^2+\frac{1}{8}v_1^2\left[-gW_\mu^3+\frac{1}{\sqrt3}gW_\mu^8+\big(1-\frac{\beta}{\sqrt3}\big)g_XX_\mu)\right]^2\\
  &+&~\frac{1}{8}v_2^2\left[gW_\mu^3+\frac{1}{\sqrt3}gW_\mu^8-(1+\frac{\beta}{\sqrt3})g_XX_\mu)\right]^2.
\eea
The masses of the charged gauge bosons are given by 
\be
M^2_{W^\pm}=\frac{g^2}{4}v^2, \quad
M^2_{Y^{\pm Q_Y}}=\frac{g^2}{4}(v_3^2+v_2^2),\quad
M^2_{V^{\pm Q_V}}=\frac{g^2}{4}(v_3^2+v_1^2)\label{MVY},
\ee
where $v$ is the vacuum expectation value of $246~{\rm GeV}$ in the SM.

The $3\times 3$ mass mixing matrix of the neutral gauge bosons is
\be
M_{NG}^2=
\left(
\begin{array}{ccc}
A & \quad B & \quad C \\
B & \quad g^2 v^2/8 & \quad g^2 \left(v_2^2-v_1^2\right)/8 \sqrt{3} \\
C & \quad g^2 \left(v_2^2-v_1^2\right)/8 \sqrt{3} & \quad g^2 \left(v^2+4 v_3^2\right)/24 \\
\end{array}
\right),
\ee
with
\bea
A&=&\frac{1}{24} g_x^2 \left[\left(\beta ^2-2 \sqrt{3} \beta +3\right) v_1^2+\left(\beta ^2+2 \sqrt{3} \beta +3\right) v_2^2+4 \beta ^2 v_3^2\right],\\
B&=&\frac{1}{24} g g_x \left[\left(\sqrt{3} \beta -3\right) v_1^2-\left(\sqrt{3} \beta +3\right) v_2^2\right],\\
C&=&\frac{1}{24} g g_x \left[\left(\sqrt{3}-\beta \right) v_1^2-\left(\beta +\sqrt{3}\right) v_2^2-4 \beta  v_3^2\right].
\eea
The three eigenvalues of the above matrix can be solved analytically. One zero eigenvalue corresponds to the massless photon, the other two eigenvalues correspond to the masses of the $Z$ and $Z^\prime$ bosons. In the decoupling limit $v_3\gg v_{1,2}$, the term $\frac{1}{6}v_3^2(-gW_\mu^8+\beta g_XX_\mu)^2$ can be considered as the mass term of $Z^\prime_\mu$ approximately. Hence, the $Z^\prime_\mu$ field is 
\be
Z^\prime_\mu=-s_{331}W^8_\mu+c_{331}X_\mu,\label{ZP}
\ee
where 
\be
s_{331}=\frac{g}{\sqrt{g^2+\beta^2g_X^2}},\quad
c_{331}=\frac{\beta g_X}{\sqrt{g^2+\beta^2g_X^2}}.
\ee
The state orthogonal to $Z^\prime_\mu$,
\be
B_\mu=c_{331}W^8_\mu+s_{331}X_\mu,\label{B}
\ee
is the gauge field of $U(1)_Y$. From Eq.~\eqref{ZP} and \eqref{B}, one can deduce $W^8_\mu$ and $X_\mu$ in terms of $Z^\prime_\mu$ and $B_\mu$
\be
W^8_\mu=-s_{331}Z^\prime_\mu+c_{331}B_\mu, \quad
X_\mu=c_{331}Z^\prime_\mu+s_{331}B_\mu.
\ee
Substituting the above two equations into Eq.~\eqref{gauge mass} and neglecting the mixing term of $Z^\prime_\mu$ with other fields, one obtains
\beq
\cL_{Gauge}^{mass}\supset\frac{1}{6}(g^2+\beta^2g_X^2)v_3^2Z_\mu^{\prime2}+\frac{1}{8}v^2(gW_\mu^3-g_Y B_\mu)^2
+\frac{\left(g^4+2 \beta ^2 g^2 g_X^2+\beta ^2 \left(\beta ^2+3\right) g_X^4\right)}{12 \left(g^2+\beta ^2 g_X^2\right)}v^2Z_\mu^{\prime2},
\eeq
where 
\be
g_Y=\frac{gg_X}{\sqrt{g^2+\beta^2g_X^2}},
\ee
denotes the coupling constant of $U(1)_Y$. The $Z_\mu$ field is
\be
Z_\mu=\frac{1}{\sqrt{g^2+g^2_Y}} (gW^3_\mu-g_Y B_\mu),
\ee
where we define the Weinberg angle as following
\be
s_W=\frac{g_Y}{\sqrt{g^2+g_Y^2}},\quad
c_W=\frac{g}{\sqrt{g^2+g_Y^2}}.
\ee
The masses of the neutral gauge bosons $Z_\mu$ and $Z^\prime_\mu$ in terms of $g$, $s_W$ and $c_W$ are
\bea
m_Z^2&=&\frac{g^2}{4c_W^2}v^2,\\
M_{Z^\prime}^2&=&\frac{g^2  c_W^2}{3 \left(1-\left(\beta ^2+1\right) s_W^2\right)}v_3^2+\frac{g^2  \left(-2 \sqrt{3} \beta  c_W^2 s_W^2+c_W^4+3 \beta ^2 s_W^4\right)}{12 c_W^2 \left(1-\left(\beta ^2+1\right) s_W^2\right)}v_1^2\nn \\
&+&\frac{g^2  \left(2 \sqrt{3} \beta  c_W^2 s_W^2+c_W^4+3 \beta ^2 s_W^4\right)}{12 c_W^2 \left(1-\left(\beta ^2+1\right) s_W^2\right)}v_2^2.
\eea

\subsection{Fermion Sector}

\begin{figure}
\centering
\includegraphics[width=0.4\textwidth]{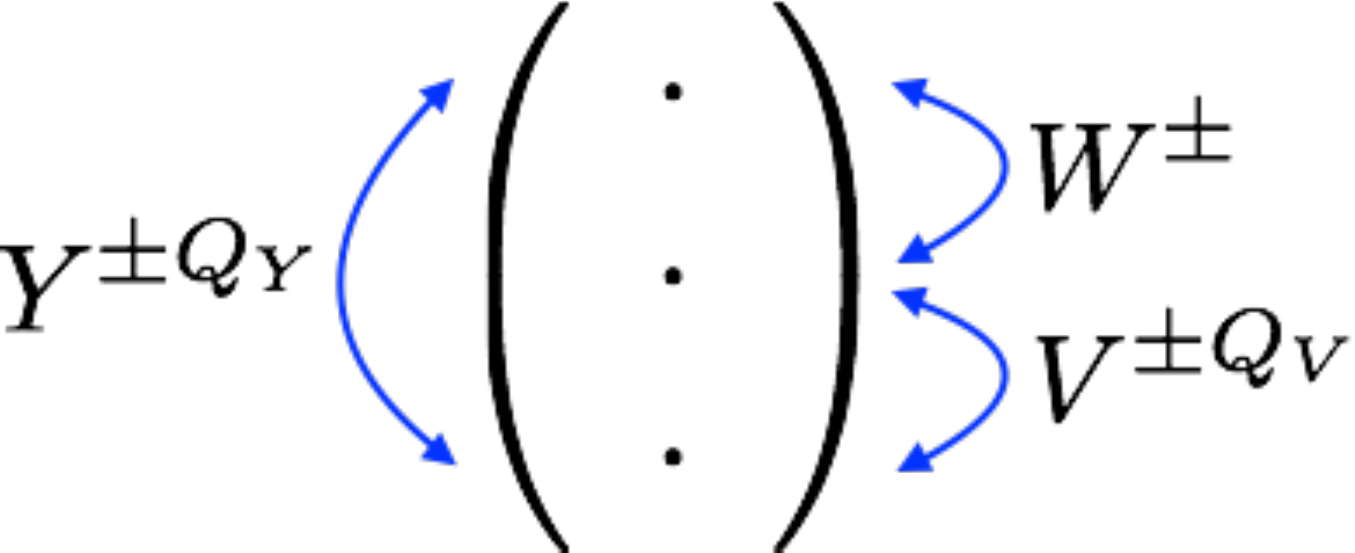}
\caption{The gauge interaction of any triplet or anti-triplet field.}\label{fig:gaugeint}
\end{figure}

The Lagrangian of the fermion section is 
\be
\cL_{\rm Fermion}=\bar\psi^\alpha i D_\mu \gamma^\mu \psi^\alpha,
\ee
where the superscript ``$\alpha$" denotes the fermion flavor. The covariant derivative is defined as follows
\be
D_\mu=\partial_\mu-igW^i_\mu T^i-ig_X X_\mu.
\ee
It is useful to write down the covariant derivatives acting on the fermion fields with different representations:
\begin{itemize}
  \item triplet $\psi_L$ $D_\mu \psi_L =\partial_\mu \psi_L - igW^i_\mu T^i \psi_L - ig_X X X_\mu \psi_L$;
  \item anti-triplet $\bar \psi_L$ $D_\mu \bar \psi_L =\partial_\mu \bar \psi_L + igW^i_\mu (T^i)^T \bar \psi_L - ig_X X X_\mu \bar \psi_L$;
  \item singlet $\psi_R$ $D_\mu \psi_R =\partial_\mu \psi_R - ig_X X X_\mu \psi_R$,
\end{itemize}
where $X$ is the fermions under $U(1)_X$. In this work the quark fields of the first and second genrations are required to be in the triplet representation of the $SU(3)_L$ group while the third generation quarks are in the anti-triplet representation. Therefore we can write the quark fields as follows:
\be
q_{1L}=\left(\begin{array}{c}
u  \\
d  \\
D  \\
\end{array}\right)_L, \quad
q_{2L}=\left(\begin{array}{c}
c  \\
s  \\
S  \\
\end{array}\right)_L, \quad
q_{3L}=\left(\begin{array}{c}
b  \\
-t \\
T  \\
\end{array}\right)_L .
\ee
Note that the $t$ and $b$ assignments are different from the SM as a result of requiring $q_{3L}$ being an anti-triplet. The extra minus sign in front of $t$ is to ensure generating the same Feynman vertices as those in the SM. The new heavy quarks are denoted as $D$, $S$ and $T$ with electric charges
\be
Q_D=Q_S=\frac{1}{6}-\frac {\sqrt{3}}{2}\beta, \quad\quad
Q_T=\frac{1}{6}+\frac {\sqrt{3}}{2}\beta.
\ee
All the lepton fields of the three generations are all treated as anti-triplets to guarantee the gauge anomaly cancellation, which requires equal numbers of triplets and anti-triplets. The lepton fields are given by
\be
l_{1L}=\left(\begin{array}{c}
e  \\
 -\nu_e   \\
E_e  \\
\end{array}\right)_L, \quad
l_{2L}=\left(\begin{array}{c}
\mu  \\
 -\nu_\mu   \\
E_\mu  \\
\end{array}\right)_L, \quad
l_{3L}=\left(\begin{array}{c}
\tau  \\
 -\nu_\tau   \\
E_\tau  \\
\end{array}\right)_L,
\ee
which exhibit the following electron charges
\be
Q_{E_\ell}=-\frac{1}{2}+\frac{\sqrt{3}}{2}\beta,\quad\ell=e,\mu,\tau.
\ee
Figure~\ref{fig:gaugeint} shows the gauge interaction between the fermion triplet or anti-triplet field and the gauge bosons ($W$, $Y$ and $V$), where the ``$\cdot$" symbol denotes some specific matter field. This is also applicable to the scalar triplets.

\subsection{Yukawa Sector}
The Lagrangian of the Yukawa sector is
\be
\cL_{Yuk}=\cL^{q}_{Yuk}+\cL^{l}_{Yuk},
\ee
where $\cL^{q}_{Yuk}$ and $\cL^{l}_{Yuk}$ are the Lagrangians for quarks and leptons as follows:
\bea
-\cL^{q}_{Yuk}&=&y^u_{ij}\bar q_{iL}\eta u_{jR}+y^u_{3j}\bar q_{3L}\rho^* u_{jR} + y^d_{ij}\bar q_{iL}\rho d_{jR}+y^d_{3j}\bar q_{3L}\eta^* d_{jR} \nn \\
             &+&y^J_{ik}\bar q_{iL}\chi J_{kR}+y^J_{33}\bar q_{3L}\chi^* J_{3R}+h.c., \\
-\cL^{l}_{Yuk}&=&y^e_{mn}\bar l_{mL}\eta^* e_{nR}+y^E_{mn}\bar l_{mL}\chi^* E_{nR}+h.c..
\eea
Here, the $i$ ($k$) index runs from 1 to 2 while the $j$ ($m$, $n$) runs from 1 to 3, respectively. The $J_{kR}$ refers to the right-handed heavy quarks $D$ and $S$ while the $J_{3R}$ denotes the right-handed quark $T$. Also, the $E_{nR}$ represents the right-handed heavy leptons $E_e$, $E_\mu$ and $E_\tau$. All the fermions become massive after spontaneously symmetry breaking. 

The mass eigenstates (labelled with the superscript $\prime$) can be related with the weak eigenstates by unitary matrices
\be
\left(\begin{array}{c}
u \\
c \\
t \\
\end{array}\right)^\prime_L=U^{-1}_L
\left(\begin{array}{c}
u \\
c \\
t \\
\end{array}\right)_L, \quad
\left(\begin{array}{c}
d \\
s \\
b \\
\end{array}\right)^\prime_L=V^{-1}_L
\left(\begin{array}{c}
d \\
s \\
b \\
\end{array}\right)_L,
\ee
from which one can define the Cabibbo–Kobayashi–Maskawa (CKM) matrix as
\be
V_{\rm CKM}=U_L^\dag V_L.
\ee
We only introduce the rotation matrix of the left-handed quarks in the SM. The rotation matrices of the right-handed fermions can be absorbed by fermion fields redefinition due to their features of the gauge-singlet and the universality of the fermion-gauge couplings. The lack of the right-handed neutrinos makes it possible for the left-handed neutrinos to have the same rotation matrix as the left-handed electron (muon, tau), which leads to the similar CKM matrix equals to the Identity matrix. Besides, we have made an assumption that the third member of every fermion triplet or anti-triplet is already in its mass eigenstate~\cite{Buras:2012dp} such that we no longer need introduce their rotation matrix.

It is worth mentioning that the unitary matrices $U_L$ and the $V_L$ each individual have the physics meaning in the 331 model. That is owing to the non-universality couplings of the left-handed quarks with the $Z^\prime$ boson, originating from the different representations among the three SM left-handed quark flavors.

\subsection{Electric Charge and Mass Spectrum in the 331 model}
In this section, we summarize the electric charge and mass spectrum of the new particles in the 331 model.
The electric charges of the new gauge bosons $Y$ and $V$ are $\pm (\sqrt{3}\beta+1)/2$ and $\pm (\sqrt{3}\beta-1)/2$ respectively. If we assume that
they are integers, $\beta$ will be restricted to some specific numbers which are
\be
\beta=\pm \frac{n}{\sqrt{3}},~~n=1,3,5,\cdots.
\ee
To ensure $M_{Z^\prime}^2\approx(c_W g v_3)^2/[3(1-(1+\beta^2)s_W^2)]$ being positive definite, $\beta$ must satisfy
\be
\beta^2<\frac{1-s_W^2}{s_W^2},
\ee
which further fix $n=1$ or 3, leading to $\beta=\pm \sqrt{3}$, $\pm 1/\sqrt{3}$. Table~\ref{tbl:charge} shows the electric charges of the new particles in the 331 model for those four different choices of $\beta$.

\renewcommand{\arraystretch}{1.2}
\begin{table}
\centering
\caption{\it Electric charges of the new particles for different choices of $\beta$.}
\label{tbl:charge}
\begin{tabular}{c | c |c |c| c| c }\hline
particles & $Q(\beta)$ & $\beta=-\frac{1}{\sqrt{3}}$ & $\beta=\frac{1}{\sqrt{3}}$ & $\beta=-\sqrt{3}$ & $\beta=\sqrt{3}$ \\ \hline \hline
$D,S$ & $\frac{1}{6}-\frac{\sqrt{3}\beta}{2}$ & $\frac{2}{3}$ & $-\frac{1}{3}$ & $\frac{5}{3}$ & $-\frac{4}{3}$ \\ \hline
$T$ & $\frac{1}{6}+\frac{\sqrt{3}\beta}{2}$ & $-\frac{1}{3}$ & $\frac{2}{3}$ & $-\frac{4}{3}$ & $\frac{5}{3}$ \\ \hline
$E_\ell$ & $-\frac{1}{2}+\frac{\sqrt{3}\beta}{2}$ & $-1$ & $0$ & $-2$ & $1$ \\ \hline
$V$ & $-\frac{1}{2}+\frac{\sqrt{3}\beta}{2}$ & $-1$ & $0$ & $-2$ & $1$ \\ \hline
$Y$ & $\frac{1}{2}+\frac{\sqrt{3}\beta}{2}$ & $0$ & $1$ & $-1$ & $2$ \\ \hline
$H^{\pm Q_V}$ & $-\frac{1}{2}+\frac{\sqrt{3}\beta}{2}$ & $-1$ & $0$ & $-2$ & $1$ \\ \hline
$H^{\pm Q_Y}$ & $\frac{1}{2}+\frac{\sqrt{3}\beta}{2}$ & $0$ & $1$ & $-1$ & $2$ \\ \hline
$H^{\pm}$ & $1$ & $1$ & $1$ & $1$ & $1$ \\ \hline
 \end{tabular}
\end{table}

The mass spectrum of all the new particles introduced in the 331 model in the decoupling limit $v_3 \gg v_{1,2}$. The scalar masses are given by
\begin{itemize}
  \item Neutral Scalars
    \begin{align}
    &m_h^2 = \frac{2\left(\lambda_1 v_1^4+\lambda_{12}v_1^2 v_2^2+\lambda_2 v_2^4\right)}{v^2}, 
    && M_{H_2}^2 = \frac{v^2}{v_1 v_2}fv_3,\label{4}\\
    & M_{H_3}^2 = 2 \lambda _3 v_3^2,\label{5}
    && M^2_{H_0}= (\frac{v_1v_2}{v_3}+\frac{v_1v_3}{v_2}+\frac{v_2 v_3}{v_1})f.
    \end{align}
  \item Charged Scalars
    \bea
    M^2_{H^\pm}&=&\frac{2fv_3+\lambda^\prime_{12}v_1v_2}{2v_1v_2}v^2, \\
    M^2_{H^{\pm Q_Y}}&=&\frac{2fv_1+\lambda^\prime_{23}v_2 v_3}{2v_2 v_3}(v_2^2+v_3^2), \\
    M^2_{H^{\pm Q_V}}&=&\frac{2fv_2+\lambda^\prime_{13}v_1 v_3}{2v_1 v_3}(v_1^2+v_3^2).
    \eea
\end{itemize}
In the above neutral scalars, $h$, $H_2$ and $H_3$ are the CP-even eigenstates, while $H_0$ is the CP-odd eigenstate. Owing to the residue $SU(2)$ symmetry after the first step of spontaneously symmetry breaking at the high energy scale $v_3$, the masses of the $H^\pm$, $H_2$ and $H_0$  scalars are nearly degenerate, i.e.
\beq
M_{H^\pm}^2\approx M_{H_2}^2\approx M_{H_0}^2\approx \frac{fv_3}{v_1v_2}(v_1^2+v_2^2).
\eeq

The gauge boson masses are
\bea
&& M^2_Z= \frac{1}{4c_W^2}g^2v^2, \quad ~~
M^2_{Z^\prime} = \frac{c^2_W g^2 v_3^2}{3\left[1-(1+\beta^2)s^2_W\right]}, \label{3} \\
&& M^2_{W^\pm}=\frac{1}{4}g^2v^2, \qquad
M^2_{Y^{\pm Q_Y}}= \frac{1}{4}g^2(v_3^2+v_2^2),\qquad
M^2_{V^{\pm Q_V}}=\frac{1}{4}g^2(v_3^2+v_1^2).
\eea
where $c_W\equiv \cos\theta_w$ and $s_W\equiv \sin\theta_w$. Note that $M_{Y^{\pm Q_Y}}$ and $M_{V^{\pm Q_V}}$ are very close in the decoupling limit but not equal. The mass splitting should obey the following inequality
\be
\big|M^2_{Y^{\pm Q_Y}}-M^2_{V^{\pm Q_V}}\big|<M_{W^\pm}^2,
\label{deltaM_VY}
\ee
which means that $\big|M_{Y^{\pm Q_Y}}-M_{V^{\pm Q_V}}\big|$ is typically a few\gev.
The relation between the mass of $Z^\prime$ and the mass of $Y^{\pm Q_Y}$ or $V^{\pm Q_V}$ is
\be
M_{Z^\prime}^2=\frac{4c_W^2}{3\left[1-(1+\beta^2)s_W^2\right]}M_{Y(V)^{\pm Q_{Y(V)}}}^2+\cO (v_1^2).\label{zpvy}
\ee

As the heavy fermion masses arise from Yukawa interaction, there's no restriction on them. 

\newpage
\subsection{Residual $Z_2$ symmetry, dark matter and long-lived particles}

As pointed out in Ref.~\cite{Mizukoshi:2010ky}, after the SSB, there is a residual global symmetry $U(1)_G$ which ensures the lightest neutral particle to be dark matter~(DM) candidate. However, the lightest new particle could be charged for some $\beta$ values. Such a lightest charged particle is stable and not allowed by the well measured DM relic abundance. One can assume that  the lightest charged particle decays eventually through high dimensional operators induced by unknown interactions at a much higher energy scale. In many new physics models such a lightest charged particle are long-lived. We name it as a long-lived particle (LLP). Those long-lived charged particles have a very interesting collider signature in high energy collision~\cite{Aaboud:2016uth,ATLAS:2014fka,Chatrchyan:2013oca}. 

Here we introduce a simple ``$Z_2$" symmetry to help understanding the stability of the lightest particle of the charged gauge bosons $(W, V, Y)$. The charged gauge bosons correspond to the raising and lowering operators made by non-Cartan generators. Define the raising and lowering operators as
\beq
I_\pm=T_1\pm i T_2,\qquad V_\pm=T_4\pm i T_5,\qquad U_\pm=T_6\pm i T_7.
\eeq
The communication relations 
\beq
[I_+, V_-]=-U_-,~[I_+, U_+]=V_+,~[U_+, V_-]=T_-,
\eeq
and their hermitian conjugation tell us that the $W$ boson has to couple to a $V$ boson and a $Y$ boson. Hence, in the triple gauge boson interaction, one can assign $Z_2$-even quantum number to one gauge boson while $Z_2$-odd quantum number to the other two gauge bosons. For $W$, $Y$ and $V$ bosons, we take the following $Z_2$ quantum number assignment
\beq
Z_2\left(W\right)=+,\qquad Z_2\left(Y,~V\right)=-.\label{eq:odd1}
\eeq
We further require the gauge bosons associated with the Cartan generators exhibit the $Z_2$-even quantum number, i.e.
\beq
Z_2\left(\gamma,~Z_\mu,~Z^\prime_\mu\right)=+.
\eeq
In terms of the fermions, the upper two components in the fermion triplets are SM fermions ($f_{SM}$) which must be $Z_2$-even. Therefore, the lowest component, i.e. new heavy fermions, is $Z_2$-odd because another $Z_2$-odd gauge boson can connect the lowest component to one of the upper two components. As a result, we have the $Z_2$ assignment of fermions as follows
\be
Z_2\left(f_{SM}\right)=+,\qquad Z_2\left(E_\ell,~D,~S,~T\right)=-,\label{eq:odd2}
\ee
where $f_{SM}$ denote all the SM fermions. For the case of the scalars, the charged scalar whose electric charge is the same as the charged gauge boson will share the same $Z_2$ quantum number due to the Goldstone Equivalence Theorem. Therefore, we have
\be
Z_2\left(h,~H_2,~H_3,~H_0\right)=+,\qquad Z_2\left(H^{\pm Q_Y},~H^{\pm Q_V}\right)=-.\label{eq:odd3}
\ee
There are totally 14 $Z_2$-odd particles from \eqref{eq:odd1}, \eqref{eq:odd2} and \eqref{eq:odd3}.

Since the DM candidate is an electrically neutral particle in the $Z_2$-odd sector. There are only two choices of $\beta$ that can have such a neutral particle:
\begin{itemize}
  \item $\beta=-1/\sqrt{3}$ ($Q_Y=0$): the DM candidate is $Y$ or $H^{Q_Y}$;
  \item $\beta=+1/\sqrt{3}$ ($Q_V=Q_E=0$): the DM candiate is $V$, $H^{Q_V}$ or $E_\ell$.
\end{itemize}
The detailed study on the dark matter is beyond the scope of the current paper, a recent review on this aspect can be found in~\cite{daSilva:2014qba}.

In the case of $\beta=\pm\sqrt{3}$, there will a charged LLP. In general there are many possibilities for LLP. But 
among them, the so-called $R$-hadron~\cite{Aad:2013gva} is more interesting. In this case, one of the three heavy quarks will be the lightest particle in the $Z_2$-odd sector. This heavy quark, say $D$ without any loss of generality, can pick up  another light quark, say $u$ quark, in the vacuum to form a hadronic bound state $D\bar{u}$. In Table~\ref{tbl:Rhadron}, we display all the possible $R$-hadron state with integer electric charge and only considering the first generation light quark component. The lightest $R$-hadron will be stable because there's no other thing it can decay into. For the sake of comparison, we also list the $R$-hadron state for $\beta=\pm\frac{1}{\sqrt{3}}$. Differently, they can decay to DM and another SM meson~(pion, Kaon, D-meson and B-meson).

\renewcommand{\arraystretch}{1.0}
\begin{table}[h!]
\centering
\caption{$R$-hadrons with integer electric charge formed wth the first generation light quark.}
\label{tbl:Rhadron}
\begin{tabular}{c|c}
\hline
$\beta=-\sqrt{3}$ & $(D\bar{u})^+,~(D\bar{d})^{++},~(S\bar{u})^+,~(S\bar{d})^{++},~(T\bar{u})^{--},~(T\bar{d})^-$\\
\hline
$\beta=\sqrt{3}$ & $(D\bar{u})^{--},~(D\bar{d})^-,~(S\bar{u})^{--},~(S\bar{d})^-,~(T\bar{u})^+,~(T\bar{d})^{++}$\\
\hline
$\beta=-\frac{1}{\sqrt{3}}$ & $(D\bar{u})^0,~(D\bar{d})^+,~(S\bar{u})^0,~(S\bar{d})^+,~(T\bar{u})^-,~(T\bar{d})^0$\\
\hline
$\beta=\frac{1}{\sqrt{3}}$ & $(D\bar{u})^-,~(D\bar{d})^0,~(S\bar{u})^-,~(S\bar{d})^0,~(T\bar{u})^0,~(T\bar{d})^+$\\
\hline
\end{tabular}
\end{table}

\section{Higgs boson Phenomenology}\label{sec:sec3}

There are three scalar triplets responsible for spontaneously symmetry braking in the 331 model. In general, there will be tree-level FCNC and the couplings of the 125~GeV Higgs boson to the fermions and gauge bosons are also modified. New neutral and charged scalars appears after symmetry breaking and exhibit collider signatures very different from other NP models~\cite{Okada:2016whh}. In this work, we focus on the deviation of 125 GeV Higgs boson couplings to the fermions and gauge bosons and further assume the FCNC via the Higgs boson is negligible. We first use the recent measurements of the Higgs boson couplings at the LHC to constrain the parameter space of the 331 model and then discuss how much the production $pp\to h\to Z\gamma$ and di-Higgs production $pp\to hh$ will be affected in the allowed parameter space. 

\subsection{Constraints from the single Higgs boson production}

In the 331 model, the scalar mixing modifies the couplings of the Higgs boson to the SM fermions and gauge bosons. The loop-induced couplings will also be affected by new particles. Below we first review the modification of Higgs couplings in the model and then explore their impacts on the Higgs boson production at the LHC.

As the three generations of fermions are in different representations of $SU(3)_L$ group, the Higgs couplings to the SM fermions are different from those in the SM. For example, the masses of the first and second generations and the third generation arises from different origins. Therefore, the Yukawa interaction and the mass matrix of the quark sector can not be diagonalized simultaneously, inevitably leading to FCNC process. We assume no mixing between the first two generations and the third generation to forbid the FCNC processes that are severely constrained by low energy data. The Yukawa interaction related to the Higgs boson and the mass of up-type quarks is given by
\be
\mathcal{L}_{Yuk}^h=\frac{1}{\sqrt{2}}\overline
{(\begin{array}{ccc}
u & c & t
\end{array}
)}_L 
\left(\begin{array}{ccccc}
y_{11}^u(U_{21}h+v_2) \quad & ~&y_{12}^u(U_{21}h+v_2) \quad & ~&0\\
y_{21}^u(U_{21}h+v_2) \quad &~& y_{22}^u(U_{21}h+v_2) \quad &~& 0\\
0 \quad &~& 0 \quad &~& y_{33}^u(U_{11}h+v_1)
\end{array}\right)
\left(
\begin{array}{c}
u\\
c\\
t
\end{array}\right)_R.
\ee
The Yukawa couplings are now easy to read
\be
y_{huu}=\frac{U_{21}v}{v_2}\left(\frac{m_u}{v}\right),\quad\quad y_{hcc}=\frac{U_{21}v}{v_2}\left(\frac{m_c}{v}\right),\quad\quad y_{htt}=\frac{U_{11}v}{v_1}\left(\frac{m_t}{v}\right).\label{yukh}
\ee
In the decoupling limit,  
\be
U_{21}=\frac{v_2}{v},\qquad U_{11}=\frac{v_1}{v}.
\ee
Therefore, the Yukawa couplings are exactly the same as the SM values, e.g. $m_q/v$. The same statement also works for the down-type quarks. For the lepton Yukawa couplings, there is no need to make assumptions on the mixing pattern simply because all the leptons are in the same representation of $SU(3)_L$ group. The Yukawa couplings of the down type quarks and leptons read as follows:
\begin{align}
&y_{hdd}=\frac{U_{11}v}{v_1}\left(\frac{m_d}{v}\right), &&y_{hss}=\frac{U_{11}v}{v_1}\left(\frac{m_s}{v}\right), &&y_{hbb}=\frac{U_{21}v}{v_2}\left(\frac{m_b}{v}\right),\nn\\
&y_{hee}=\frac{U_{21}v}{v_2}\left(\frac{m_e}{v}\right), &&y_{h\mu\mu}=\frac{U_{21}v}{v_2}\left(\frac{m_\mu}{v}\right), &&y_{h\tau\tau}=\frac{U_{21}v}{v_2}\left(\frac{m_\tau}{v}\right).\nn
\end{align}
Again, as expected, the couplings shown above approach to the SM values in the decoupling limit.

It is more straightforward to obtain the couplings of the Higgs boson to the SM gauge bosons. From Eq.~\eqref{gaugemass}, we obtain
\bea
\mathcal{L}_{gauge}^h&=&\frac{g^2}{4}W_\mu^+ W^{\mu,-}\left[\left(U_{11}h+v_1\right)^2+\left(U_{21}h+v_2\right)^2\right]\nn\\
&+&\frac{g^2}{8c_W}Z_\mu Z^\mu\left[\left(U_{11}h+v_1\right)^2+\left(U_{21}h+v_2\right)^2\right].
\eea
It gives rise to the $HWW$ and $HZZ$ couplings,
\be
g_{hWW}=\frac{U_{11}v_1+U_{21}v_2}{v}\left(\frac{2 m_W^2}{v}g_{\mu\nu}\right),\qquad g_{hZZ}=\frac{U_{11}v_1+U_{21}v_2}{v}\left(\frac{2 m_Z^2}{v}g_{\mu\nu}\right),
\ee
which approach the SM coupling $\dfrac{2m^2}{v}  g_{\mu\nu}$ in the decoupling limit. 

In addition to above tree-level Higgs couplings, there are three loop-induced Higgs couplings ($g_{hgg}$, $g_{h\gamma\gamma}$ and $g_{hZ\gamma}$) that play an important role in the Higgs-boson phenomenology. They will be affected by the mixing of the scalars and new particles inside the loop.

First, consider the Higgs-gluon-gluon anomalous coupling $g_{hgg}$. In addition to the top-quark loop, the $g_{hgg}$ coupling receives additional contributions from heavy quark ($J=D,S,T$) loops. It yields 
\be
g_{hgg}=\frac{\omega_{hgg}}{\dfrac{1}{v} A_{1/2}^h(\tau_t)}g_{hgg}^{\rm SM},
\ee
where $g_{hgg}^{\rm SM}$ denotes the $hgg$ anomalous coupling in the SM and 
\be
\omega_{hgg}=\frac{U_{11}}{v_1} A_{1/2}^h(\tau_t)
+\sum\limits_{J=D,S,T}\frac{U_{31}}{v_3}A_{1/2}^h(\tau_J),
\ee
with $\tau_i = m_h^2/4m_i^2$ where $m_i$ is the mass of the particle propagating inside the triangle loop. The function $A_{1/2}^h(\tau)$ is given in Appendix~\ref{app_c}.

Next, consider the $h\gamma\gamma$ and $hZ\gamma$ anomalous couplings. Both couplings are induced by the loop effects of heavy quarks ($J$), heavy leptons ($E=E_e,E_{\mu},E_{\tau}$), charged gauge bosons ($V$ and $Y$) and charged scalars ($H^\pm$, $H^{Q_V}$, $H^{Q_Y}$). The contribution from the charged scalars will be highly suppressed by its mass and is neglected in our work. The $g_{h\gamma\gamma}$ couplings is
\be
g_{h\gamma\gamma}=\frac{\omega_{h\gamma\gamma}}{3\left[\dfrac{4}{9v}A_{1/2}^h(\tau_t)\right]+\dfrac{1}{v}A_{1}^h(\tau_W)}g_{h\gamma\gamma}^{SM},
\ee
where $g_{h\gamma\gamma}^{\rm SM}$ denotes the $h\gamma\gamma$ anomalous coupling in the SM and 
\bea
\omega_{h\gamma\gamma}&=&3\bigg[\frac{4U_{11}}{9v_1}A_{1/2}^h(\tau_t)+\sum\limits_{J=D,S,T}\frac{Q_J^2 U_{31}}{v_3}A_{1/2}^h(\tau_J)\bigg]\\
&+&\sum\limits_{E=E_e,E_\mu,E_\tau}\frac{Q_{E}^2U_{31}}{v_3}A_{1/2}^h(\tau_{E})+\frac{v_1 U_{11}+v_2 U_{21}}{v^2}A_{1}^h(\tau_W)\nn\\
&+&\frac{Q_Y^2 (v_2 U_{21}+v_3 U_{31})}{v_2^2+v_3^2}A_{1}^h(\tau_Y)+\frac{Q_V^2 (v_1 U_{11}+v_3 U_{31})}{v_1^2+v_3^2}A_{1}^h(\tau_V).
\label{eq:haa}
\eea
The function $A_{1}^h(\tau)$ can be found in Appendix~\ref{app_c}.
The $hZ\gamma$ anomalous coupling is
\bea
g_{hZ\gamma}=\frac{\omega_{hZ\gamma}}{3\bigg[\dfrac{2 \hat{v}_t}{3 v c_W}A_{1/2}^h(\tau_t,\lambda_t)\bigg]+\dfrac{1}{v}A_1^h(\tau_W,\lambda_W)}g_{hZ\gamma}^{\rm SM},
\eea
where $g_{hZ\gamma}^{\rm SM}$ denotes the $hZ\gamma$ anomalous coupling in the SM  and
\bea
\omega_{hZ\gamma}&=&3\bigg[\frac{2 U_{11}\hat{v}_t}{3 v_1 c_W}A_{1/2}^h(\tau_t,\lambda_t)+\sum\limits_{J=D,S,T}\frac{Q_J U_{31}\hat{v}_J}{ v_3 c_W}A_{1/2}^h(\tau_J,\lambda_J)\bigg]\\
&+&\sum\limits_{E=E_e,E_\mu,E_\tau}\frac{Q_E U_{31}\hat{v}_E}{ v_3 c_W}A_{1/2}^h(\tau_E,\lambda_E)+\frac{v_1 U_{11}+v_2 U_{21}}{v^2}A_1^h(\tau_W,\lambda_W)\nn\\
&+&\frac{Q_Y (v_2 U_{21}+v_3 U_{31})}{v_2^2+v_3^2}\frac{1-(\sqrt{3}\beta+1)s_W^2}{2c_W^2}A_{1}^h(\tau_Y,\lambda_Y)\nn\\
&+&\frac{Q_V (v_1 U_{11}+v_3 U_{31})}{v_1^2+v_3^2}\frac{(1-\sqrt{3}\beta)s_W^2-1}{2c_W^2}A_{1}^h(\tau_V,\lambda_V),\label{whzr}
\eea
with $\hat{v}_f=2I_f^3-4Q_f s_W^2$ as usual. The functions $A_{1/2}^h(\tau,\lambda)$ and $A_{1}^h(\tau,\lambda)$ can be found in Appendix \ref{app_c}.

Obviously, the deviation of the Higgs boson couplings highly depends on many parameters, e.g. the symmetry breaking scale ($v_{1,2,3}$), the mixing in the scalar sector, and the mass of new particles in the loop.
However, the number of parameters can be greatly reduced when new particles inside the loop are much heavier than the Higgs-boson mass. The loop functions have a nice decoupling behavior when $\tau\to 0$, e.g.  
\beq
A_{1/2}^h(\tau)\to 4/3,\qquad A_{1}^h(\tau)\to -7.
\eeq
Thus, when the new particles inside the loop are very heavy, the loop functions do not depend on the masses of new particles. As the masses of new particles are quite heavy in the decoupling limit $v_3\gg v_{1,2}$ of our interest, one can approximate the full loop functions by their constance limits. Besides, from the unitarity of the mixing matrix $U$ and the $W$ boson mass, we obtain two conditions: $\sum_{i=1,2,3}U_{i1}^2=1$ and $\sum_{i}v_i^2=v^2$. Therefore, we end up with only four independent parameters which are chosen to be $v_1$, $v_3$, $U_{11}$ and $U_{21}$. Apparently, the constraints on the parameters will be weakened when $v_3$ increases. Below we fix a few values of $v_3$ and then scan other three parameters with respect to recent data from the ATLAS and CMS collaborations~\cite{ATLAS-CONF-2015-044}. The ATLAS and CMS collaborations have combined their global fitting results of the Higgs boson effective couplings divided by the SM value which are parameterized as $\kappa_i$
\begin{align}
&\kappa_W=0.91^{+0.10}_{-0.10}\quad &&\kappa_Z=1.03^{+0.11}_{-0.11}\quad &&\kappa_t=1.43^{+0.23}_{-0.22}\quad &&\kappa_\tau=0.88^{+0.13}_{-0.12}\nn\\
&\kappa_b=0.60^{+0.18}_{-0.18}\quad &&\kappa_g=0.81^{+0.11}_{-0.10}\quad &&\kappa_\gamma=0.92^{+0.11}_{-0.10}\nn
\end{align}
Taking the Higgs precision data into account, we perform a global scan to derive the constraints on the parameters of the 331 model.

\begin{figure}
\centering
\includegraphics[width=0.235\textwidth]{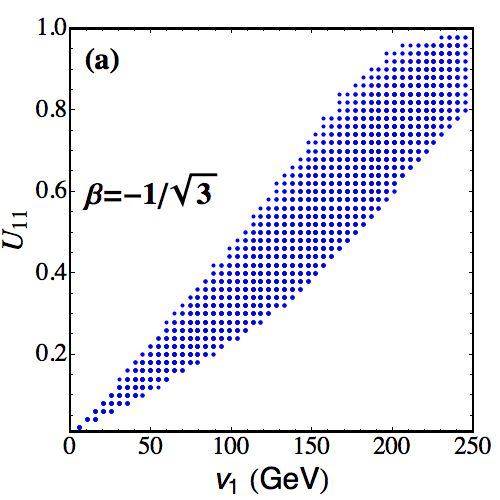}
\includegraphics[width=0.235\textwidth]{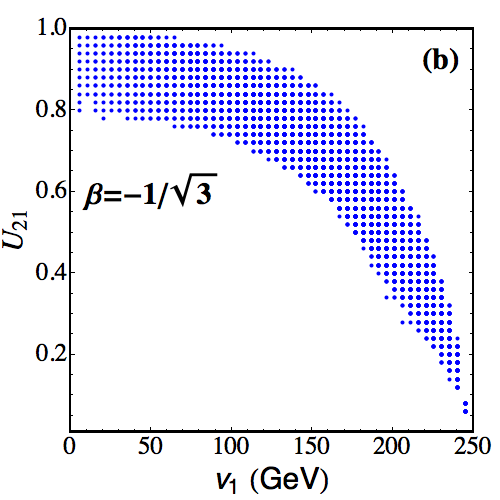}
\includegraphics[width=0.235\textwidth]{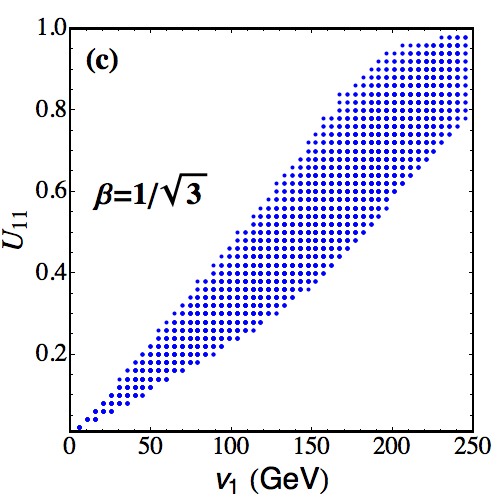}
\includegraphics[width=0.235\textwidth]{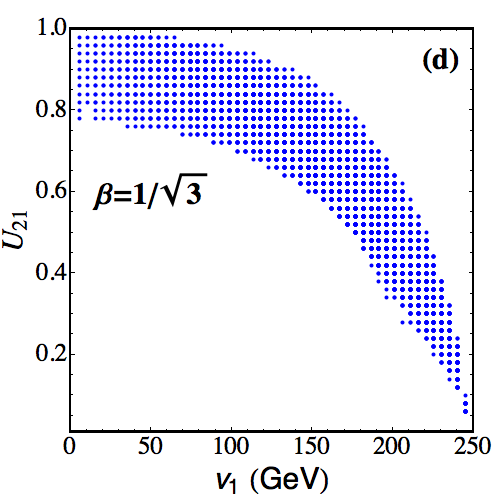}\\
\includegraphics[width=0.235\textwidth]{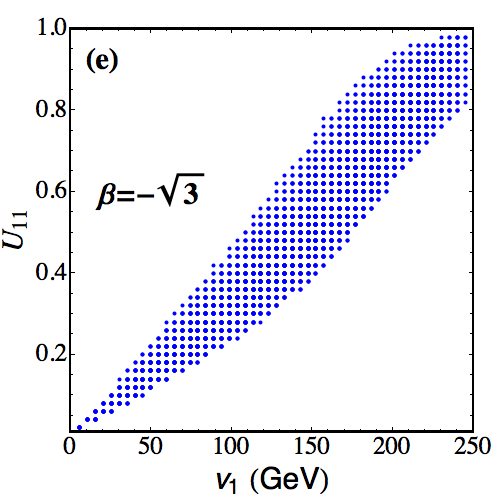}
\includegraphics[width=0.235\textwidth]{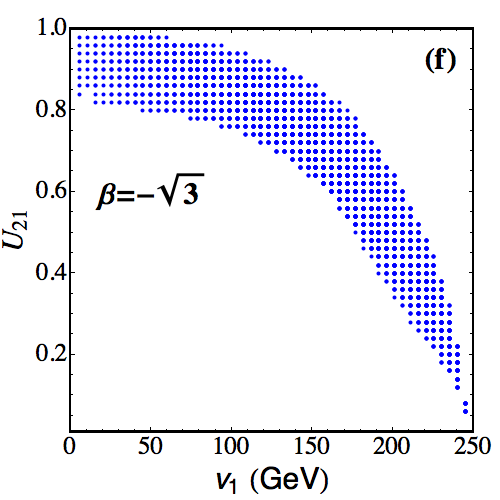}
\includegraphics[width=0.235\textwidth]{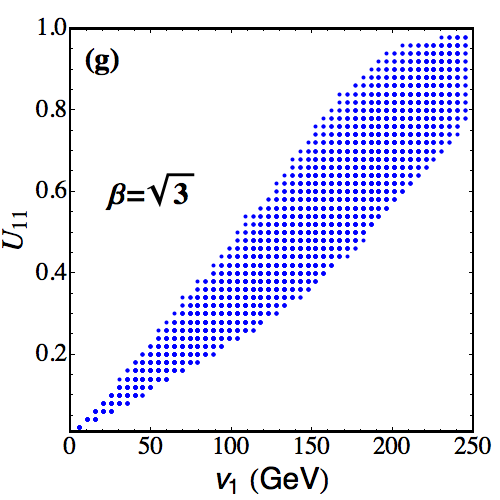}
\includegraphics[width=0.235\textwidth]{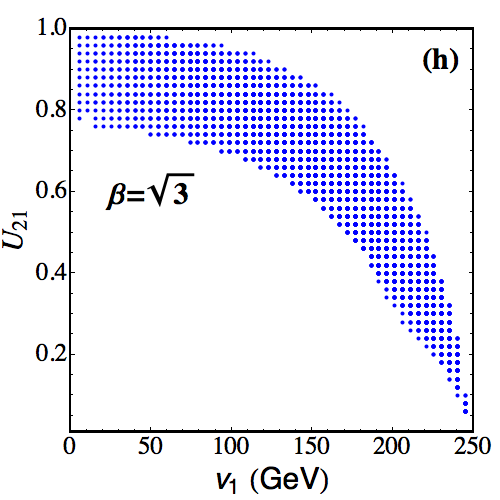}
\caption{\it The allowed parameter space in the plane of $v_1-U_{11}$ and $v_1-U_{21}$ for $v_3 = 2~{\rm TeV}$ at the {\rm 95\%} confidence level: (a,b) $\beta=-1/\sqrt{3}$, (c,d) $\beta=+1/\sqrt{3}$, (e,f) $\beta=-\sqrt{3}$, and (g,h) $\beta=+\sqrt{3}$.
}
\label{chi2_1a}
\end{figure}

Figure~$\ref{chi2_1a}$ displays the 95\% confidence level allowed regions for $v_3=2{~\rm TeV}$, respectively, with different choices of $\beta$: (a,b) $\beta=-1/\sqrt{3}$, (c,d) $\beta=+1/\sqrt{3}$, (e,f) $\beta=-\sqrt{3}$, and (g,h) $\beta=+\sqrt{3}$. First, we note that the allowed parameter space is not sensitive to $\beta$. Even though the electronic charges of new resonances depend on the value of $\beta$, the loop effects of those new resonances are suppressed by the large value of $v_3$, yielding a less sensitive dependence on $\beta$. 

Second, the $U_{11}$ parameter depends mainly linearly on $v_1$ in the decoupling limit, i.e. $U_{11}\simeq v_1/v$; see Eq.~\eqref{U}. The dependence of $U_{21}$ on $v_1$ is in the form of $\sqrt{1-v_1^2/v^2}$; see Fig.~\ref{chi2_1a}. The bands of allowed parameter space are mainly from the $\kappa_\gamma$ data, of which the $W$-boson contribution dominates. The $W$-boson contribution to the $H\gamma\gamma$ anomalous coupling, $\frac{v_1 U_{11}+v_2 U_{21}}{v^2}A_{1}^h(\tau_W)$, is close to the SM value, i.e. $ (v_1 U_{11}+v_2 U_{21})/v^2 \sim 1$. Together with the condition $v_1^2 + v_2^2=v^2$, it yields the bands shown in Fig.~\ref{chi2_1a}, which clearly demonstrates the competition between $v_1$ and $v_2$.

\subsection{$pp\to h\to Z\gamma$ channel}

With the parameter regions allowed by the Higgs boson signal strength, we can give the prediction of $\sigma(pp\to h\to Z\gamma)$ in the 331 model. The $Z\gamma$ decay mode is complementary to the $\gamma\gamma$ mode as the $Z\gamma$ is sensitive to the weak eigenstate particles inside the loop while the $\gamma\gamma$ mode just probes charged particles inside the loop. Combining both $Z\gamma$ and $\gamma\gamma$ modes would help deciphering the nature of the Higgs boson~\cite{Cao:2015scs,Cao:2015pto}. 

\begin{figure}\centering
\includegraphics[width=0.24\textwidth]{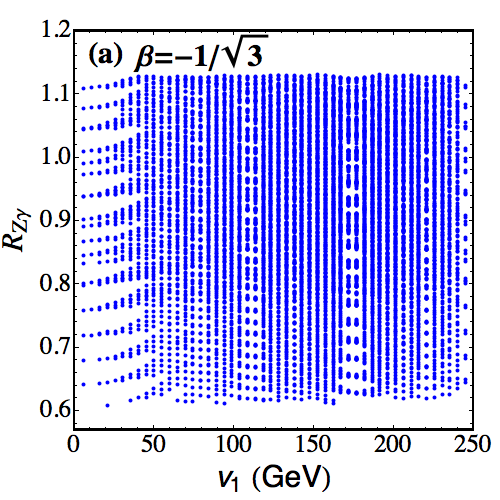}
\includegraphics[width=0.24\textwidth]{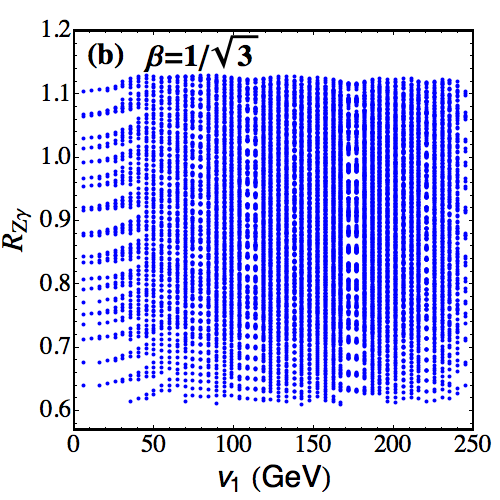}
\includegraphics[width=0.24\textwidth]{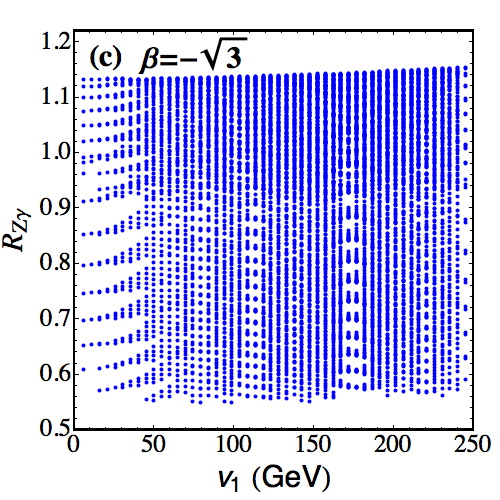}
\includegraphics[width=0.24\textwidth]{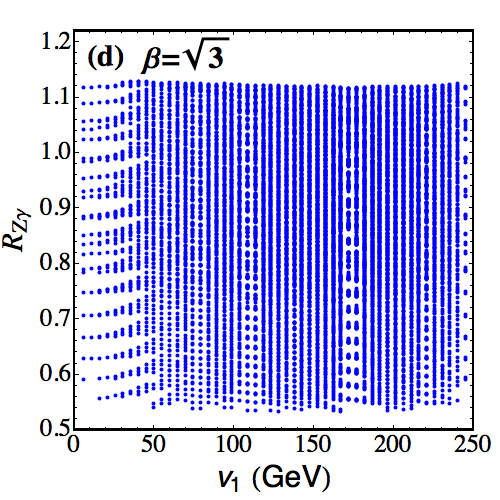}
\caption{\it The signal strength $R_{Z\gamma}$ as a function of $v_1$ for different choices of $\beta$ and $v_3=2{~\rm TeV}$ in the parameter space allowed by single Higgs boson measurements at {\rm 95\%} confidence level.
}\label{Rzr}
\end{figure}

The signal strength relative to the SM expectation, 
\be
R_{Z\gamma}\equiv\frac{\sigma(pp\to h\to Z\gamma)}{\sigma(pp\to h\to Z\gamma)_{\rm SM}}=\left(\frac{g_{hZ\gamma}}{g_{hZ\gamma}^{\rm SM}}\right)^2,
\ee
is shown in Fig.~\ref{Rzr} as a function of $v_1$ with fixed $v_3=2{\rm~TeV}$. The signal strength $R_{Z\gamma}$ is not sensitive to either $\beta$ or $v_1$. 
The $R_{Z\gamma}$ can be enhanced as large as $\sim 1.2$. The cancellation between the contribution of the new fermions and gauge bosons can also reduce the signal strength. For example, we observe that the signal strength are typically larger than $\sim 0.6$ for $\beta=\pm 1/\sqrt{3}$ while larger than $\sim 0.5$ for $\beta=\pm \sqrt{3}$.  If the $Z\gamma$ mode is found to be suppressed sizably, say $R_{Z\gamma}<0.6$, then one can exclude the option of $\beta=\pm 1/\sqrt{3}$.

\subsection{Higgs-boson pair production}

\begin{figure}[b]
\centering
\includegraphics[width=0.3\textwidth]{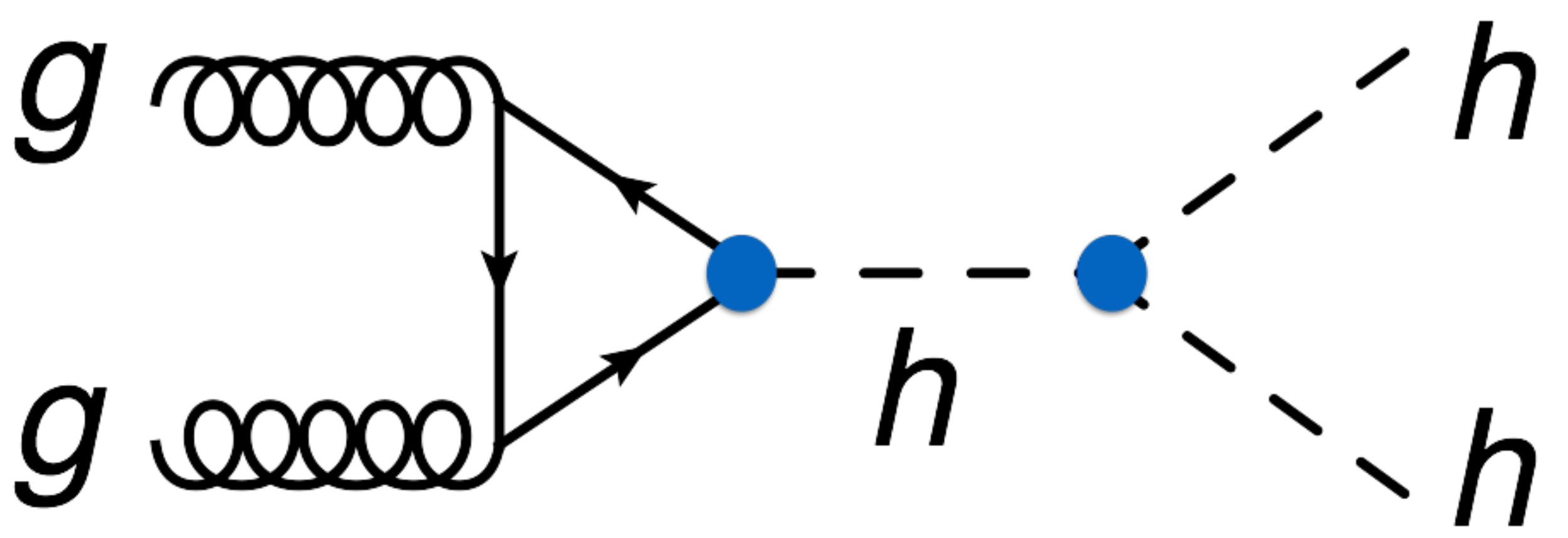}\quad
\includegraphics[width=0.26\textwidth]{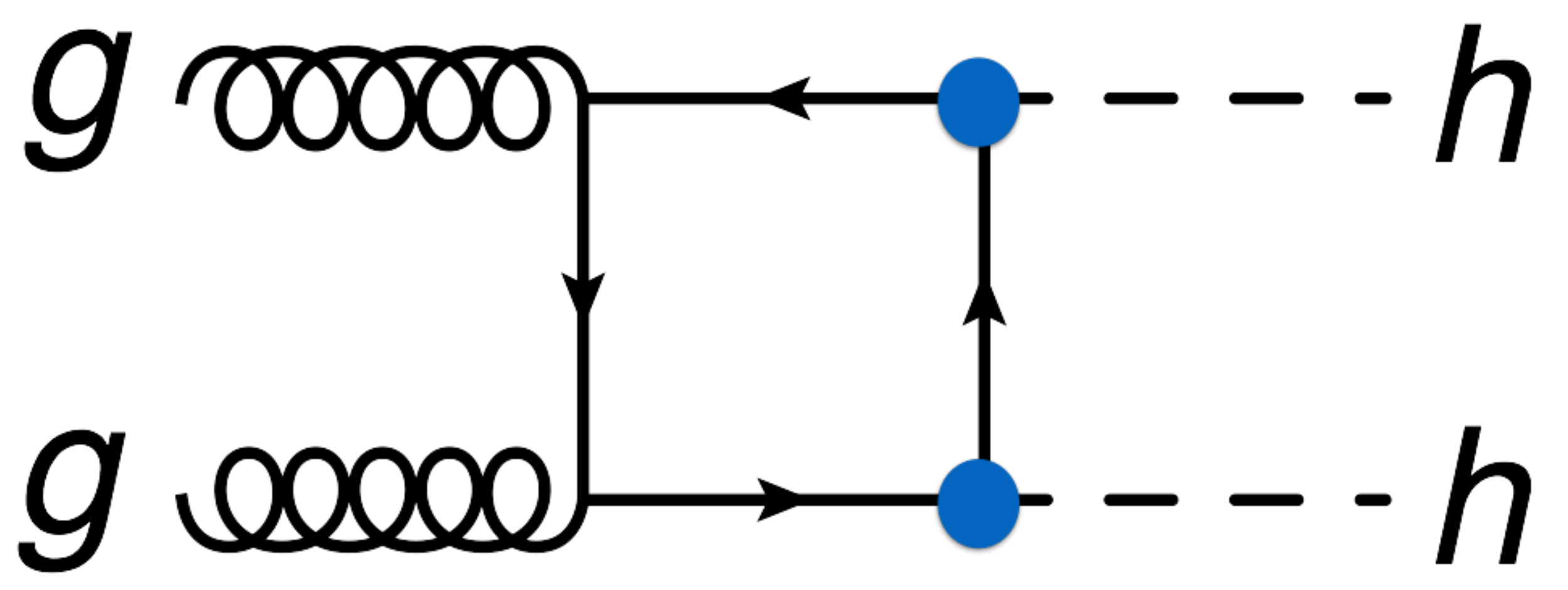}
\caption{\it The triangle and box Feynman diagram of double Higgs boson production in the SM.
}\label{hh}
\end{figure}

The double Higgs boson production has drawn a lot of attentions because it is the golden channel to directly measure the triple Higgs-boson self-interaction in the SM, and therefore, tests the electroweak symmetry breaking mechanism. As the Higgs boson does not carry any color, they are produced in pair through the triangle loop and box loop shown in Fig.~$\ref{hh}$. The production rate in the SM is small mainly due to the large cancellation between the triangle and box diagrams which can be easily understood from the low energy theorem~\cite{Shifman:1979eb,Kniehl:1995tn}. At the LHC with an center of mass energy of $14{~\rm TeV}$, the production cross section is about $40~{\rm fb}$, which cannot be measured owing to the small branching ratio of the Higgs boson decay and large SM backgrounds~\cite{Shao:2013bz}. However, in new physics models, the Higgs trilinear coupling can significantly deviate from the SM value. It then could enhance the di-Higgs production make it testable at the LHC. Therefore, it is important to study how large can the cross section of the double Higgs boson production be considering all the constraints from the single Higgs boson measurements.

The squared amplitude of $gg\to hh$ in the SM is given by~\cite{Dawson:2015oha}
\bea
|{\cal M}|^2_{\rm SM} = \frac{\alpha_s^2 \hat{s}^2}{\pi^2 v^4}
\bigg[ \bigg|  {3m_h^2\over {\hat s}-m_h^2} F_\bigtriangleup
+ F_\Box \bigg|^2+ \left| G_\Box\right|^2\bigg],
 \label{gg2hh_SM}
\eea
where $F_\bigtriangleup\equiv F_\bigtriangleup({\hat s},{\hat t},m_h^2, m_t^2)$, 
$F_\Box\equiv F_\Box({\hat s}, {\hat t}, m_h^2, m_t^2)$ and 
$G_\Box\equiv G_\Box({\hat s},{\hat t}, m_h^2, m_t^2)$ are the 
form factors~\cite{Plehn:1996wb} with $\hat{s}$ and 
$\hat{t}$ the canonical Mandelstam 
variables. $G_\Box$ represents the $d$-wave contribution, 
which is negligible~\cite{Dawson:2012mk}. In the large $m_t$ limit, $F_\bigtriangleup\to +2/3$ and $F_\Box\to -2/3$, therefore,  they tend to cancel around the energy threshold of Higgs boson pairs, say $\sqrt{\hat s}\sim 4m_h^2$.
In the 331 model, the top Yukawa coupling and Higgs trilinear coupling are 
\bea
y_{htt}&=&\frac{U_{11}v}{v_1}\left(\frac{m_t}{v}\right)\equiv c_t\left(\frac{m_t}{v}\right)\equiv c_t y_{htt}^{\rm SM},\\
\lambda_{hhh}&=&\left(1+\frac{kv_1v_2U_{31}^2}{m_h^2}\right)\frac{3m_h^2}{v}\equiv c_3\frac{3m_h^2}{v}= c_3 \lambda_{hhh}^{\rm SM}.
\eea
Therefore, the squared amplitude of $gg\to hh$ is 
\bea
|{\cal M}|^2 = \frac{\alpha_s^2 \hat{s}^2}{\pi^2 v^4}
\bigg[ \bigg|  {3m_h^2\over {\hat s}-m_h^2} c_3c_t F_\bigtriangleup
+ c_t^2F_\Box \bigg|^2+ \left| c_t^2G_\Box\right|^2\bigg].
 \label{gg2hh_331}
\eea

\begin{figure}[h!]
\centering
\includegraphics[width=0.24\textwidth]{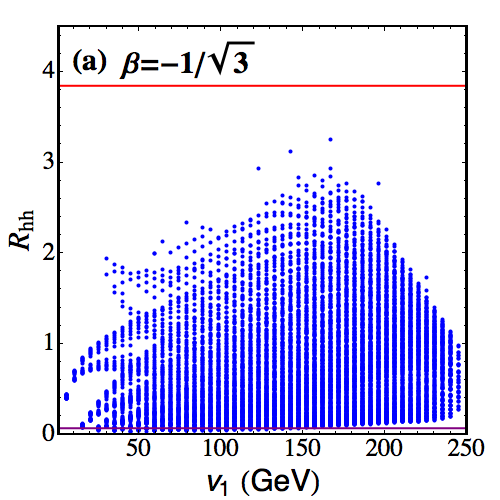}
\includegraphics[width=0.24\textwidth]{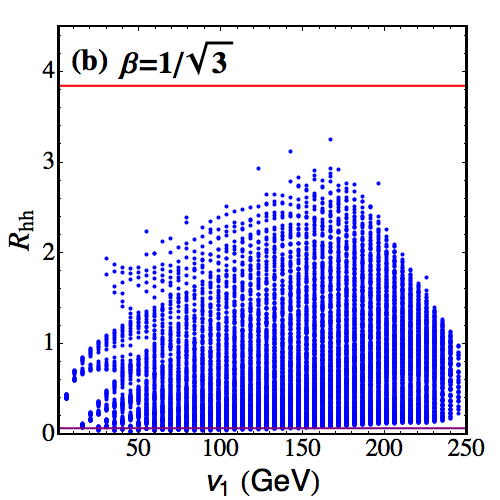}
\includegraphics[width=0.24\textwidth]{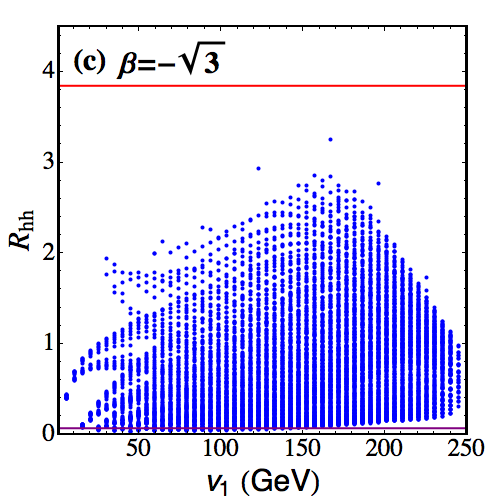}
\includegraphics[width=0.24\textwidth]{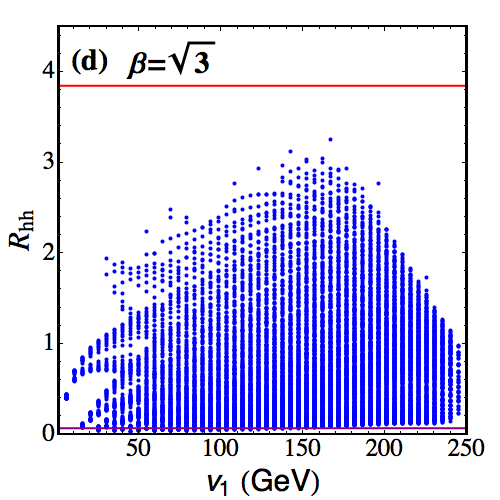}
\caption{\it The signal strength $R_{hh}$ as a function of $v_1$ for different choices of $\beta$ and $v_3=2{~\rm TeV}$ in the parameter space allowed by single Higgs boson measurements at the {\rm 95\%} confidence level. The red line is the 5$\sigma$ discovery limit for the HL-LHC. While the purple line shows the discovery potential of the di-Higgs signal at the 100 TeV FCC-hh or SppC with an integrated luminosity 6 and 30~{\rm ab}$^{-1}$, respectively.
}\label{diHiggs}
\end{figure}

Again, we consider the signal strength relative to the SM expectation $R_{hh}$ defined as following: 
\be
R_{hh}=\frac{\sigma(pp\to hh)}{\sigma(pp\to hh)_{SM}}.
\ee
Figure~\ref{diHiggs} displays the signal strength $R_{hh}$ as a function of $v_1$ in the parameter space allowed by the single Higgs production measurement at the 95\% confidence level. Similar to the case of $R_{Z\gamma}$, the signal strength $R_{hh}$ is not sensitive to $\beta$. That is simply because the di-Higgs production is mainly from the QCD, Yukawa and Higgs trilinear coupling, which is not sensitive to $\beta$ at all. The loop corrections from new quarks inside the triangle and box diagrams could enhance $R_{hh}$ and the maximum of $R_{hh}$ is around 3. Unfortunately, it is still not enough to observe the $hh$ pair production with such an enhancement at the high luminosity LHC with an integrate luminosity of $3000{\rm~fb^{-1}}$~\cite{ATL-PHYS-PUB-2014-019}; for example, the red line in Fig.~\ref{diHiggs} represents the $5\sigma$ discovery potential of the HL-LHC. The future hadron-hadron circular collider (FCC-hh) or the super proton-proton collider (SppC), designed to operate at the energy of 100~TeV, can easily probe most of the parameter space through the $hh$ pair production~\cite{Gomez-Ceballos:2013zzn,CEPC-SPPCStudyGroup:2015csa}. It is shown that the di-Higgs signal can be discovered if $R_{hh}>0.07$,  which covers also entire parameter space of the 331 model. 

\section{Phenomenology of $Z^\prime$}\label{sec:sec4}

The 331 model consists of new fermions, gauge bosons and scalars, which yields very rich collider phenomenologies. In this section we first examine the phenomenology of new neural gauge boson $Z^\prime$ at the LHC and then discuss other resonances after.  

\subsection{Production of the $Z^\prime$ boson}

At the LHC the $Z^\prime$ boson is produced singly through quark annihilation processes. The interaction of the $Z^\prime$ boson to the fermion $f$ is 
\bea
\mathcal{L}_{f}=Z^\prime_{\mu}\,\bar{f}\,\gamma^{\mu}(g_L P_L+g_R P_R)f,
\label{eq:effcoup}
\eea
where $P_{L,R}=(1\mp\gamma_{5})/2$ are the usual chirality projectors.
Without loss of generality we assume $V_L=V_{\rm CKM}$ and the rotation matrix of up-type quark $U_L$ being the Identity matrix. Therefore, the couplings of $Z^\prime$-boson to fermions, $g_L$ and $g_R$, depends only on $\beta$. The relevant Feynman rules are given in Appendix \ref{app_a}. 

At the LHC, the cross section of $pp\to Z^\prime \to XY$ is
\begin{equation}
\sigma_{pp\to V'\to\bar{f}f'}=\sum_{\{ij\}}\int_{\tau_{0}}^{1}
\frac{d\,\tau}{\tau}\cdot\frac{1}{s}\frac{d\,\mathcal{L}_{ij}}{d\,\tau}\cdot[\hat{s}\,\hat{\sigma}_{ij\to V'\to\bar{f}f'}(\hat{s})]\,,
\end{equation}
where $X$ and $Y$ denote the decay products of the $Z^\prime$ boson,
$\sqrt{s}$ is the total energy of the incoming proton-proton beam,
$\sqrt{\hat{s}}$ is the partonic center-of-mass (c.m.) energy and $\tau\equiv \hat{s}/s$.
The lower limit of $\tau$ variable is determined by the kinematics threshold of 
the $Z^\prime$ production, i.e. $\tau_{0}=M_{Z^\prime}^2/s$. 
The parton luminosity $\frac{1}{s}\frac{d\,\mathcal{L}_{ij}}{d\,\tau}$
is defined as
\begin{equation}
\frac{1}{s}\frac{d\,\mathcal{L}_{ij}}{d\,\tau}=\frac{1}{1+\delta_{ij}}
\frac{1}{s}\int_{\tau}^{1}\frac{d\, x}{x}
[f_{i}^{(a)}(x)f_{j}^{(b)}(\tau/x)+f_{j}^{(a)}(x)f_{i}^{(b)}(\tau/x)]\,,
\end{equation}
where $i$ and $j$ denote the initial state partons and $f_i^{(a)}(x)$ is the 
parton distribution of the parton $i$ inside the hadron $a$ with a momentum
fraction of $x=p_i/p_a$. 
Using the narrow width approximation (NWA) one can factorize 
the $pp\to Z^\prime \to XY$ process into 
the $Z^\prime$ production and the $Z^\prime$ decay,
\begin{equation}
\sigma_{pp\to Z^\prime\to XY}=\left(\sum_{\{ij\}}\int_{\tau_{0}}^{1}
\frac{d\,\tau}{\tau}\cdot\frac{1}{s}\frac{d\,\mathcal{L}_{ij}}{d\,\tau}\cdot[\hat{s}\,
\hat{\sigma}_{ij\to Z^\prime}(\hat{s})]\right)\,\times {\rm Br}(Z^\prime \to XY),
\label{eq:nlo_vprime}
\end{equation}
where the branching ratio (Br) is defined as 
$
{\rm Br}(Z^\prime\to XY) = \Gamma(Z^\prime \to XY)
/\Gamma_{\rm tot}.
$
As to be shown later, the decay width of $Z^\prime$ boson in most of the allowed parameter space  
are much smaller than their masses, which validates the NWA adapted in this work. 
The partonic cross section of the $Z^\prime$ production is 
\be
\hat{\sigma}_{ij\to V'}(\hat{s})=
\frac{\pi}{6\hat{s}}(g_L^2+g_R^2)\delta(1-\hat{\tau}),
\ee
where $\hat{\tau}\equiv M_{Z^\prime}^2/\hat{s}$.

\begin{figure}[h!]
\begin{center}
\includegraphics[width = 0.6\textwidth]{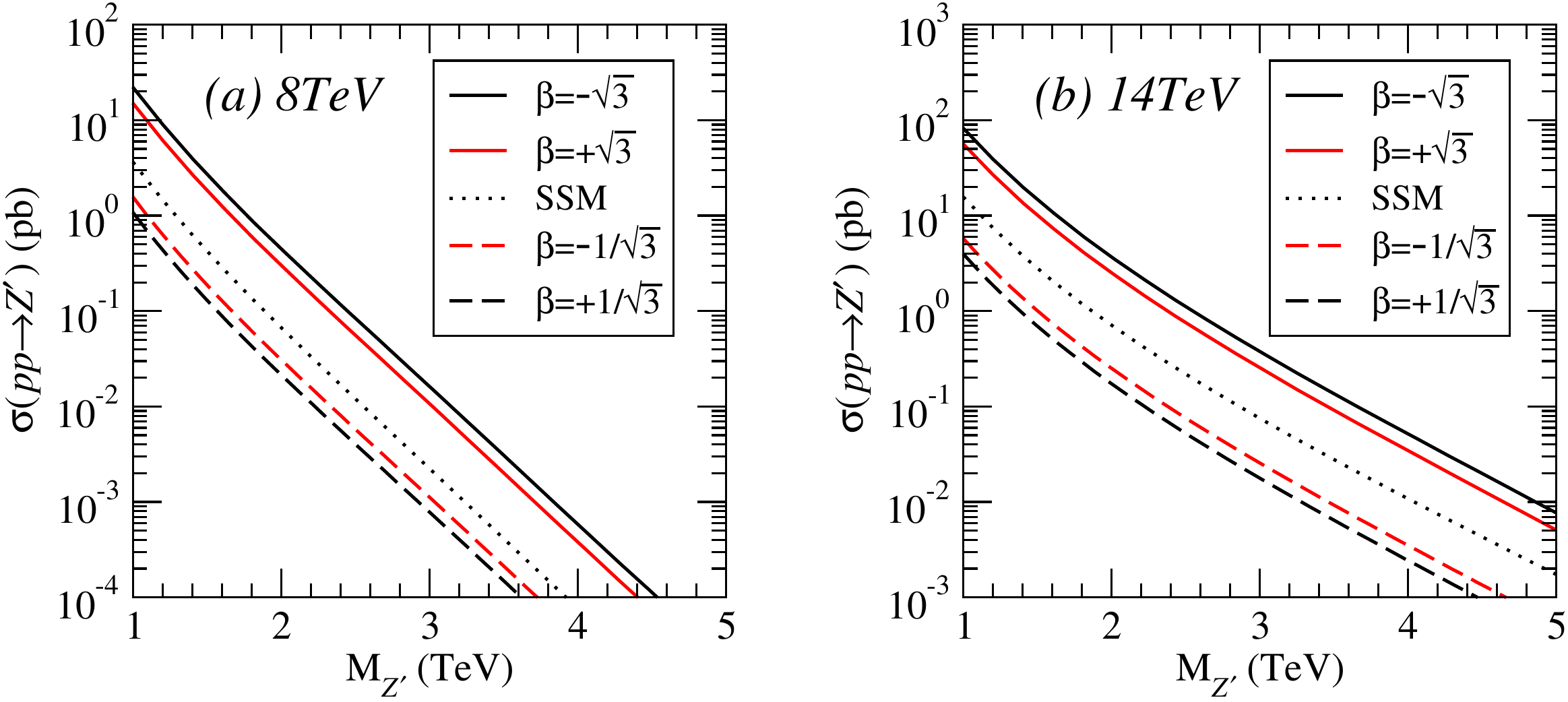}
\caption{\it The production cross sections of the $Z^\prime$ boson as a function of $M_{Z^\prime}$ for different choices of $\beta$ at the 8~TeV LHC (a) and the 14~TeV LHC (b). For comparison we also plots the production cross section of a sequential $Z^\prime$ boson (SSM).}
\label{ppzp}
\end{center}
\end{figure}

We use MadGraph5\_aMC@NLO~\cite{Alwall:2014hca} to calculate the $Z^\prime$ production cross section. The FeynRules \cite{Alloul:2013bka} package is used to generate the 331 model files. Figure~$\ref{ppzp}$ displays the $Z^\prime$ production cross sections as a function of $M_{Z^\prime}$ for different choices of $\beta$. We also plot the $Z^\prime$ production cross section in the Sequential Standard Model (SSM) as a reference. The $Z^\prime$ boson can be copiously produced at the 8 TeV and 14~TeV LHC. Note that the cross sections of $\beta=\pm \sqrt{3}$ are about one order of magnitude larger than the cross sections of $\beta=\pm 1/\sqrt{3}$. That is due to the large enhancements of the $g_L$ and $g_R$ couplings for $\beta=\pm \sqrt{3}$, e.g.
\bea
\beta=+\sqrt{3}&: & g_L^{u,d} \sim -0.37~,~g_R^u \sim +0.59~,~g_R^d=-0.30~,\\
\beta=-\sqrt{3} &: & g_L^{u,d} \sim -0.66~,~g_R^u \sim -0.59~,~g_R^d=+0.30~,\\
\beta=+\tfrac{1}{\sqrt{3}} &:& g_L^{u,d} \sim -0.18~,~ g_R^u \sim +0.08~,~g_R^d=-0.04~,\\
\beta=-\tfrac{1}{\sqrt{3}} &:& g_L^{u,d} \sim -0.22~,~ g_R^u \sim -0.08~,~g_R^d=+0.04~.
\label{Zpuu}
\eea
That leads to an enhancement factor $\sim 10$ in the $Z^\prime$ production cross section in the case of $\beta=\pm \sqrt{3}$.

\subsection{Decay of the $Z^\prime$ boson}

Now consider the $Z^\prime$ boson decay. The decay channels can be classified into four categories. First, it can decay to a pair of fermion and anti-fermion. Second, it can decay into a pair of gauge boson, owing to non-abelian interaction. To be more specific, it could decay to $YY$ and $VV$ pairs, but cannot decay into a pair of $WW$ bosons. It is due to $SU(3)$ gauge group, whose structure constant $f^{128}=0$. The decay mode $Z^\prime\to ZZ$ is also forbidden by CP symmetry of the 331 model. Third, it may decay into a pair of Higgs bosons. The case can be further classified into two subcategories: one is charged Higgs boson pairs in the final states, the other involves one CP-even Higgs boson and one CP-odd Higgs boson. Last, the $Z^\prime$ may decay into a pair of gauge boson and Higgs boson as well. 

\begin{figure}[b]
\centering
\includegraphics[width=0.24\textwidth]{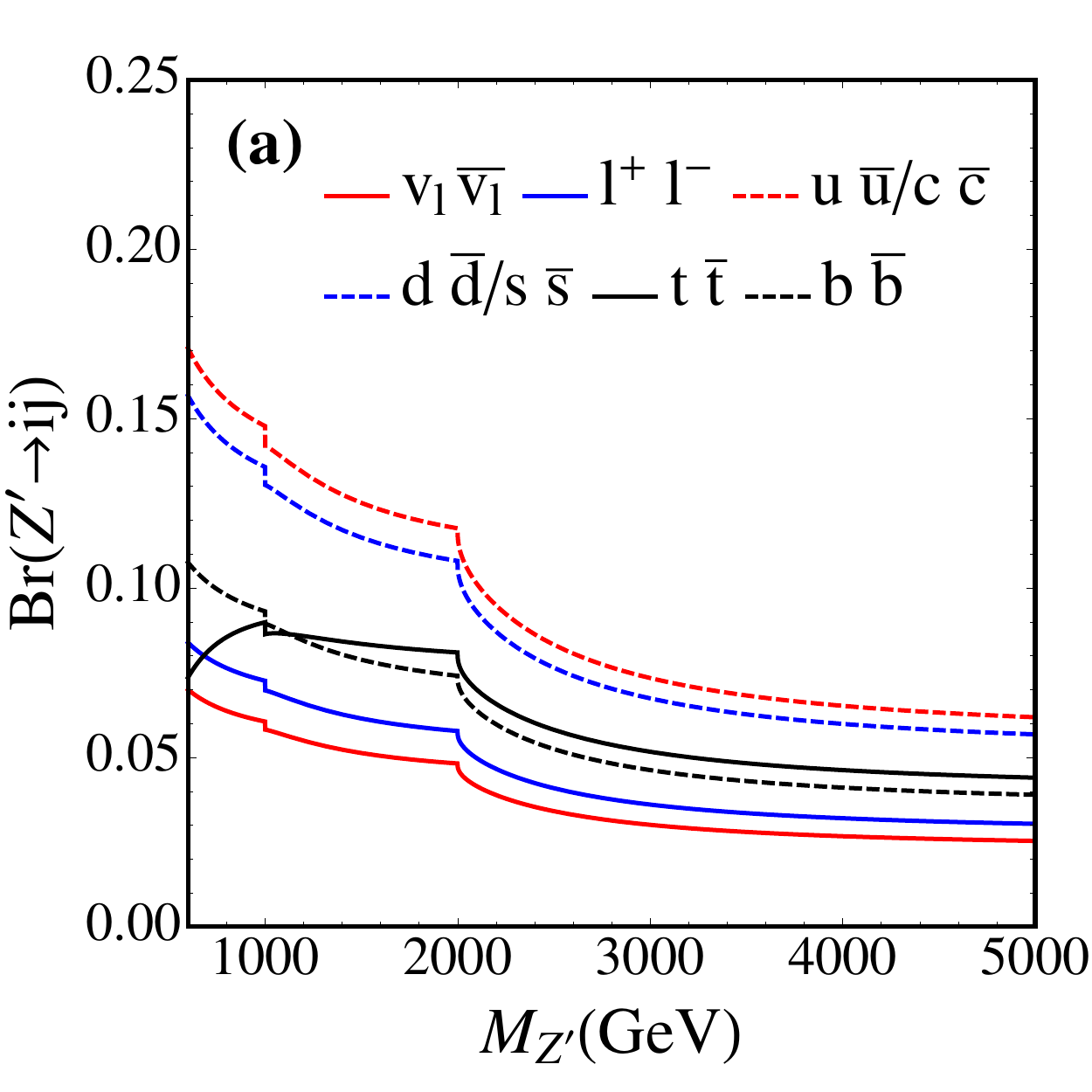}
\includegraphics[width=0.24\textwidth]{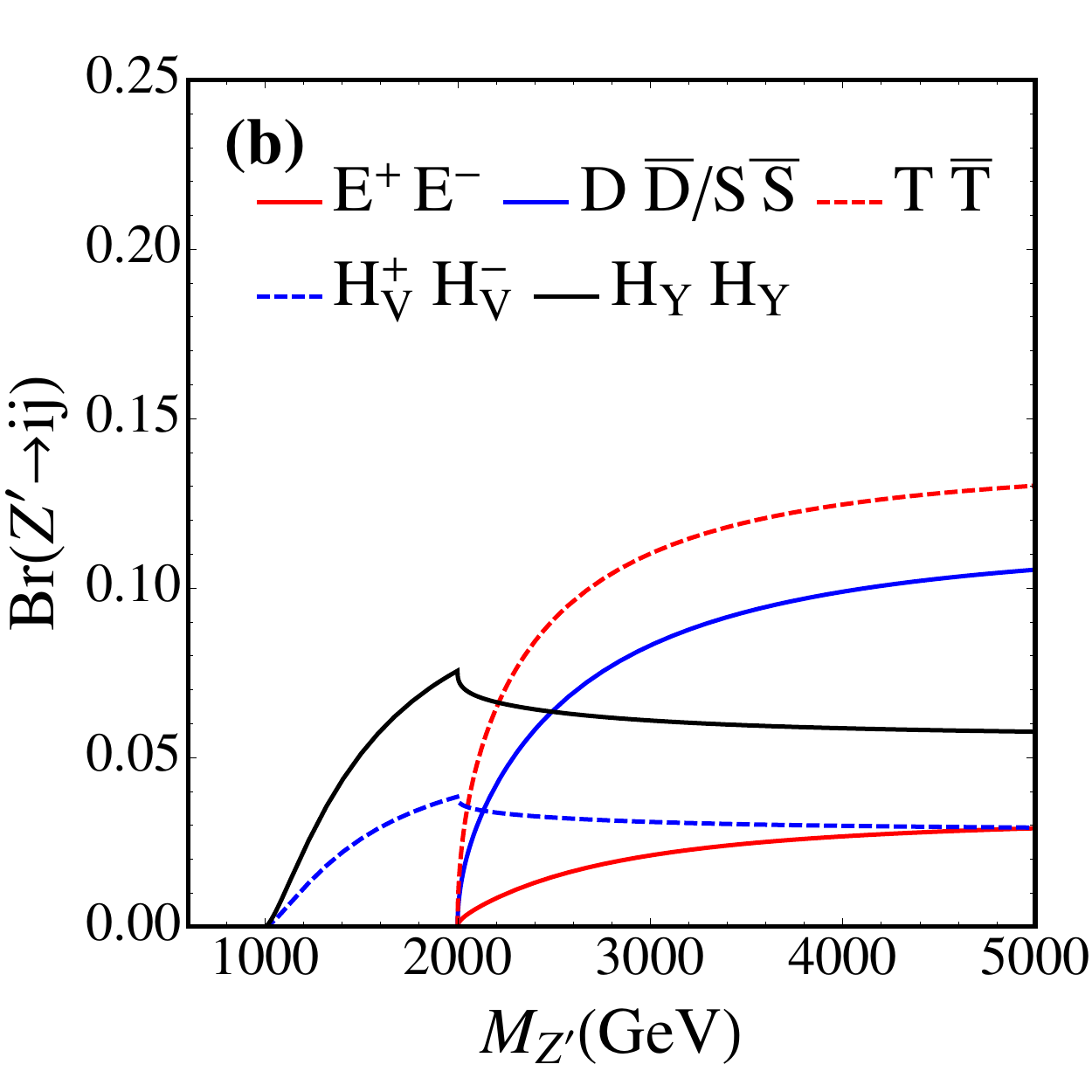}
\includegraphics[width=0.24\textwidth]{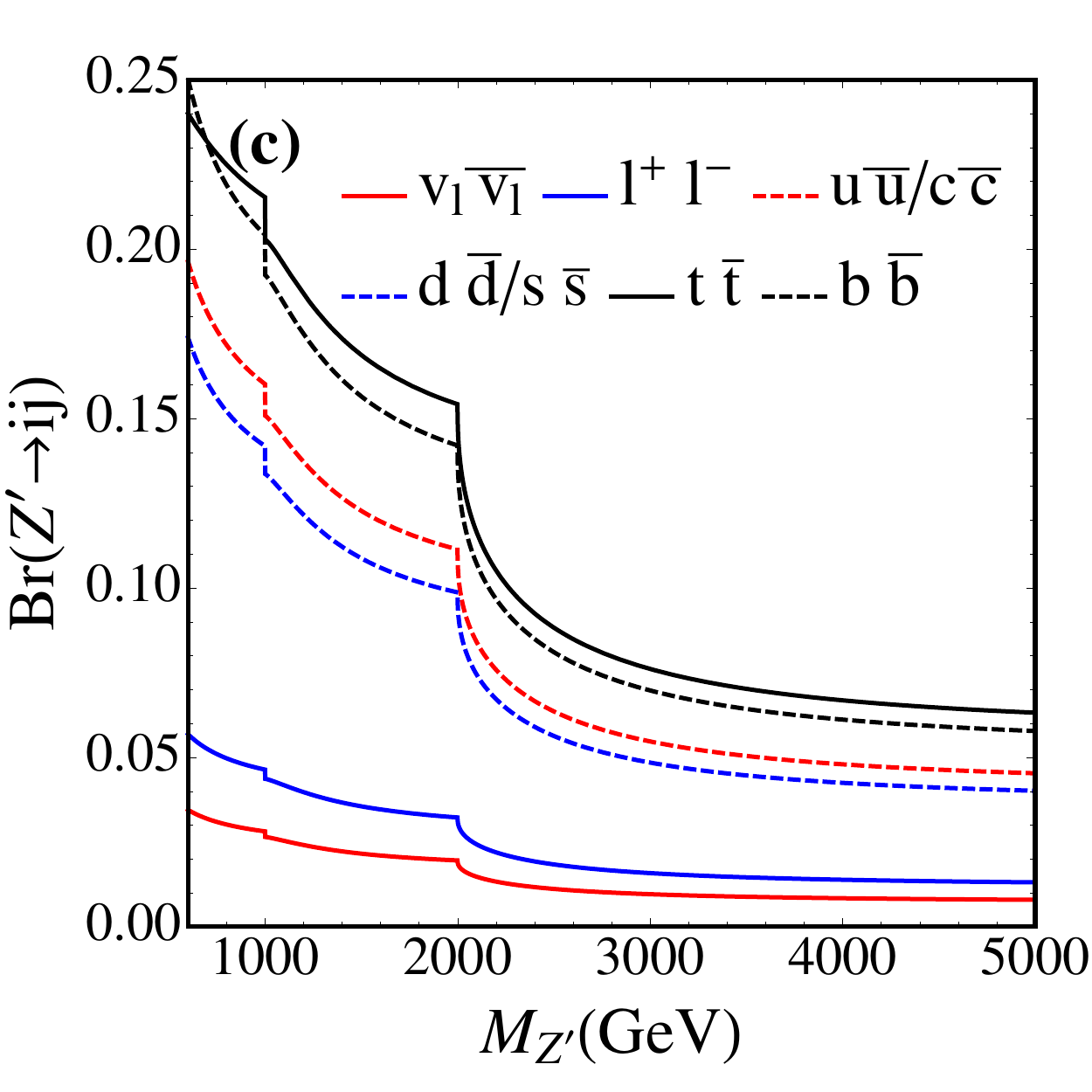}
\includegraphics[width=0.24\textwidth]{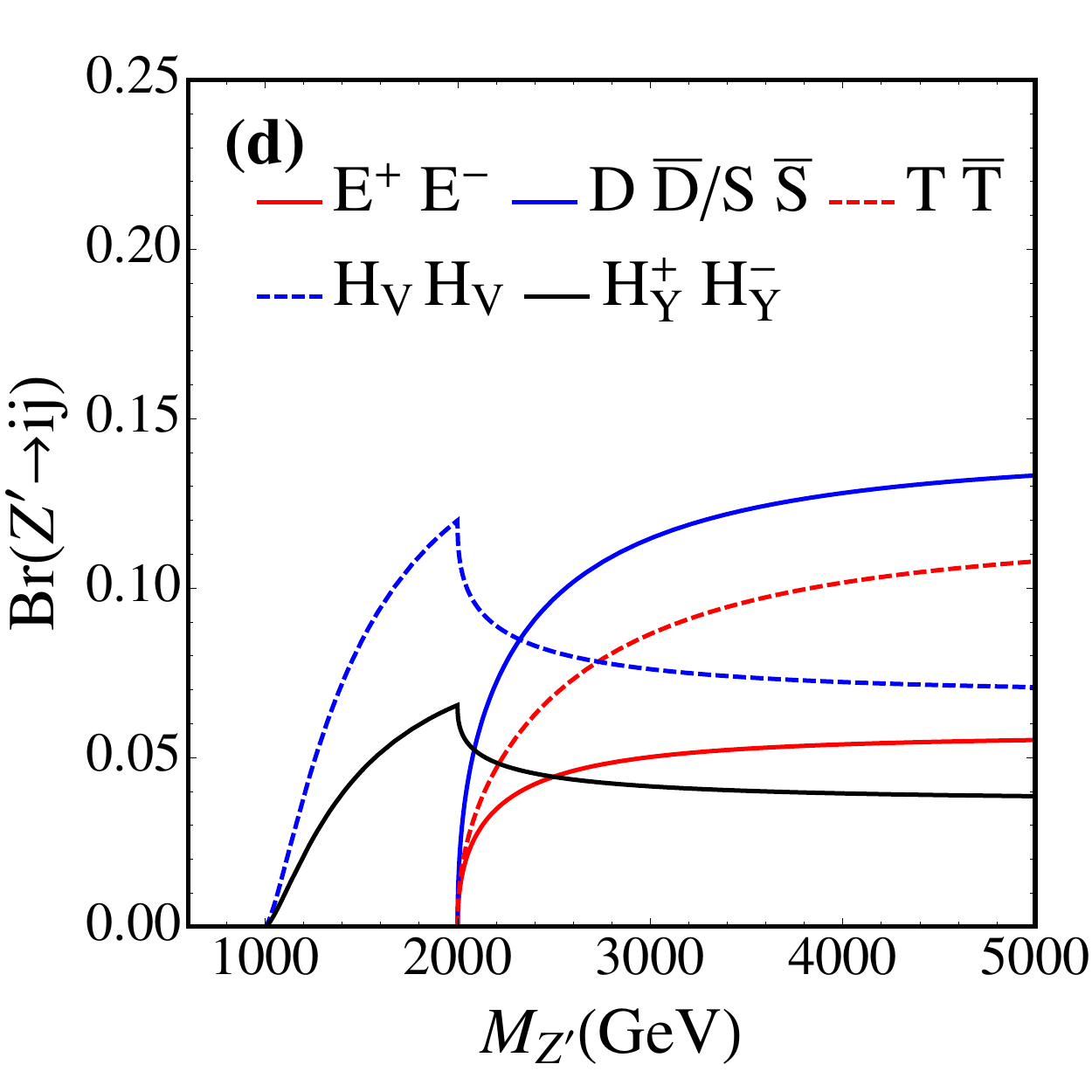}
\caption{\it Decay branching ratios of $Z^\prime$ with respect to $M_{Z^\prime}$. (a) and (b) correspond to $\beta=-1/\sqrt{3}$, while
(c) and (d) correspond to $\beta=+1/\sqrt{3}$. (a) and (c) include the SM particles, while (b) and (d) include the 331 model particles. Here we use a different notation of the charged scalar, where the electric charge is expressed apparently and a subscript is added to represent the corresponding charged scalar. The heavy fermion masses are chosen to be $1000\gev$, and the charged Higgs boson masses are chosen to be $500\gev$.
}\label{zpbr1}
\end{figure}

\begin{figure}\centering
\includegraphics[width=0.24\textwidth]{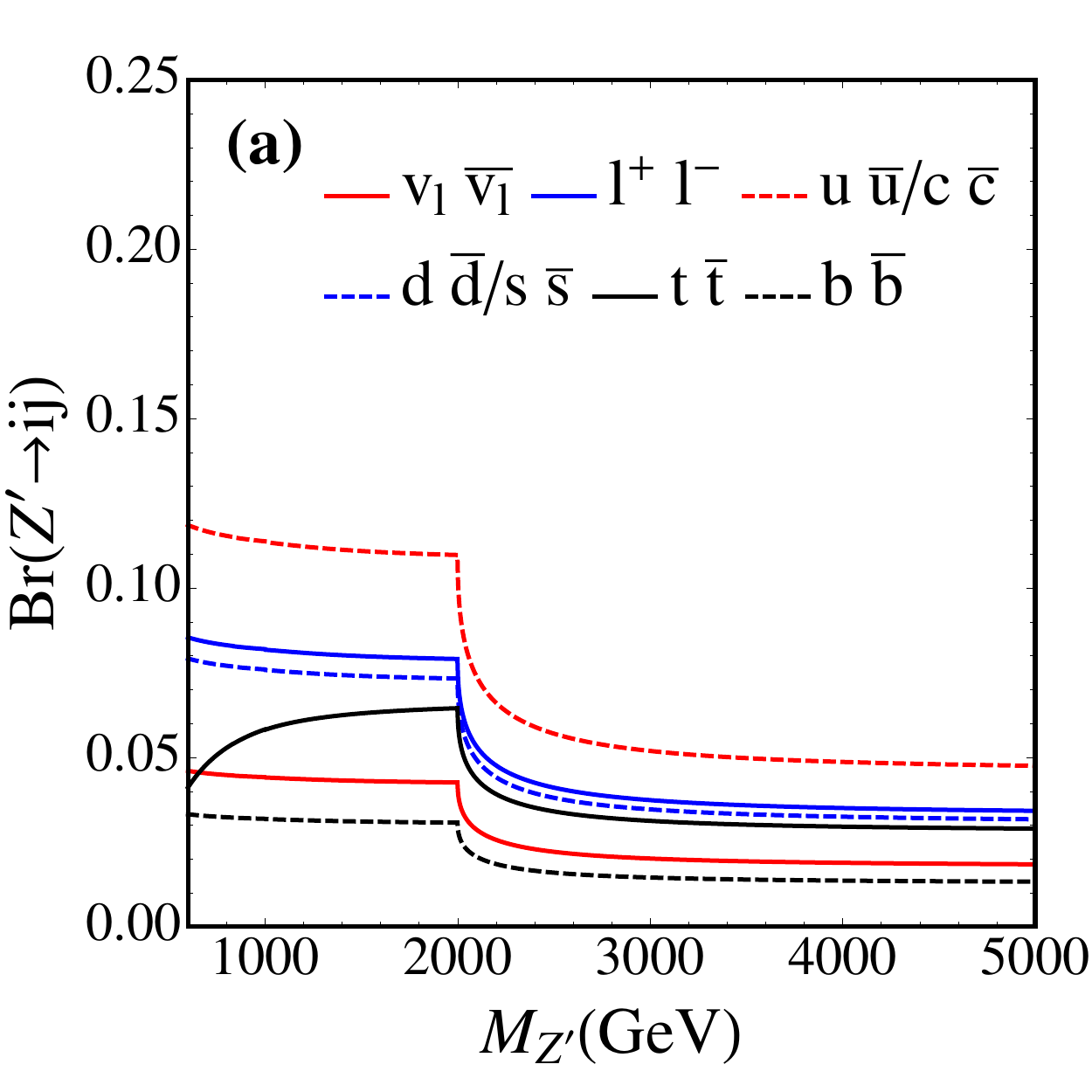}
\includegraphics[width=0.24\textwidth]{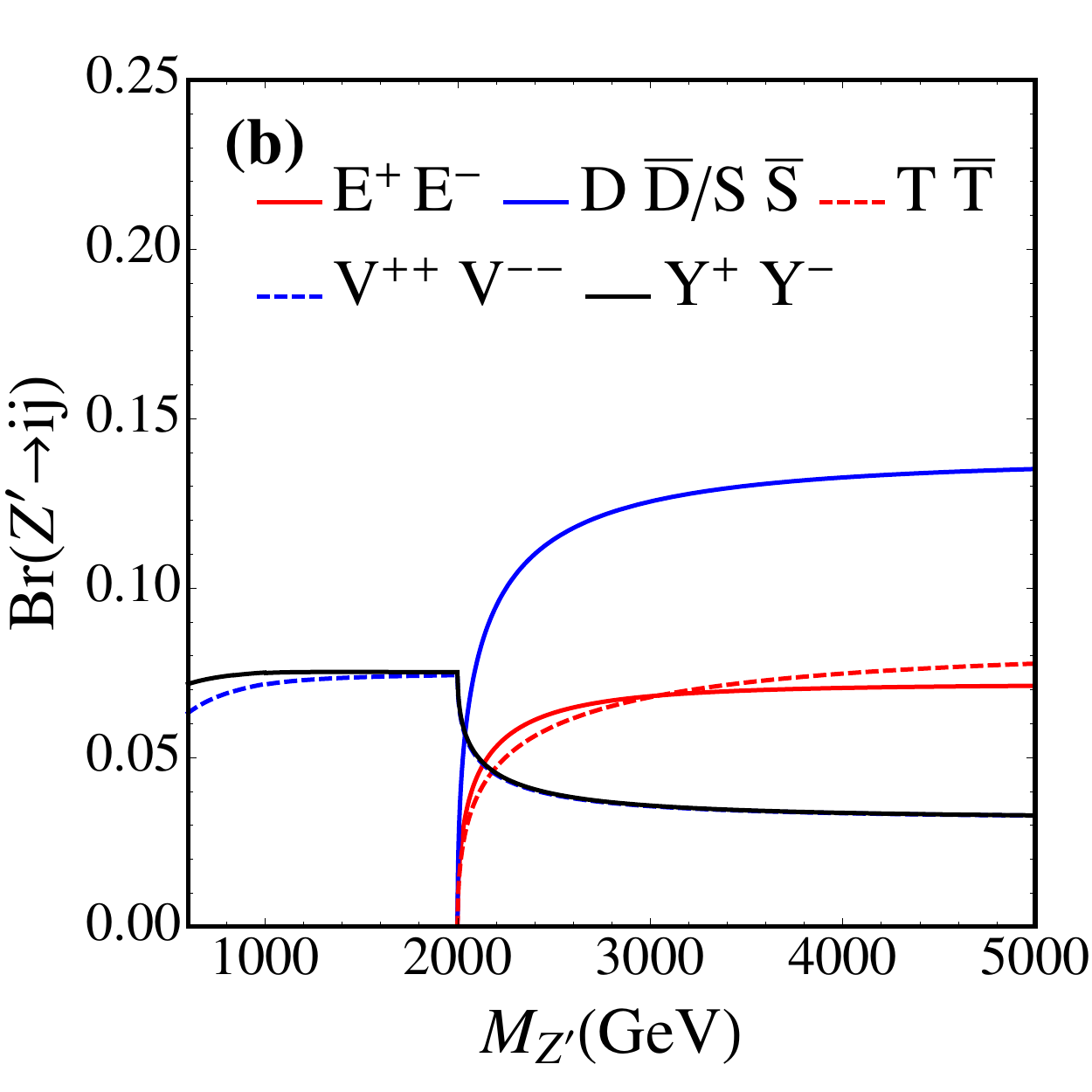}
\includegraphics[width=0.24\textwidth]{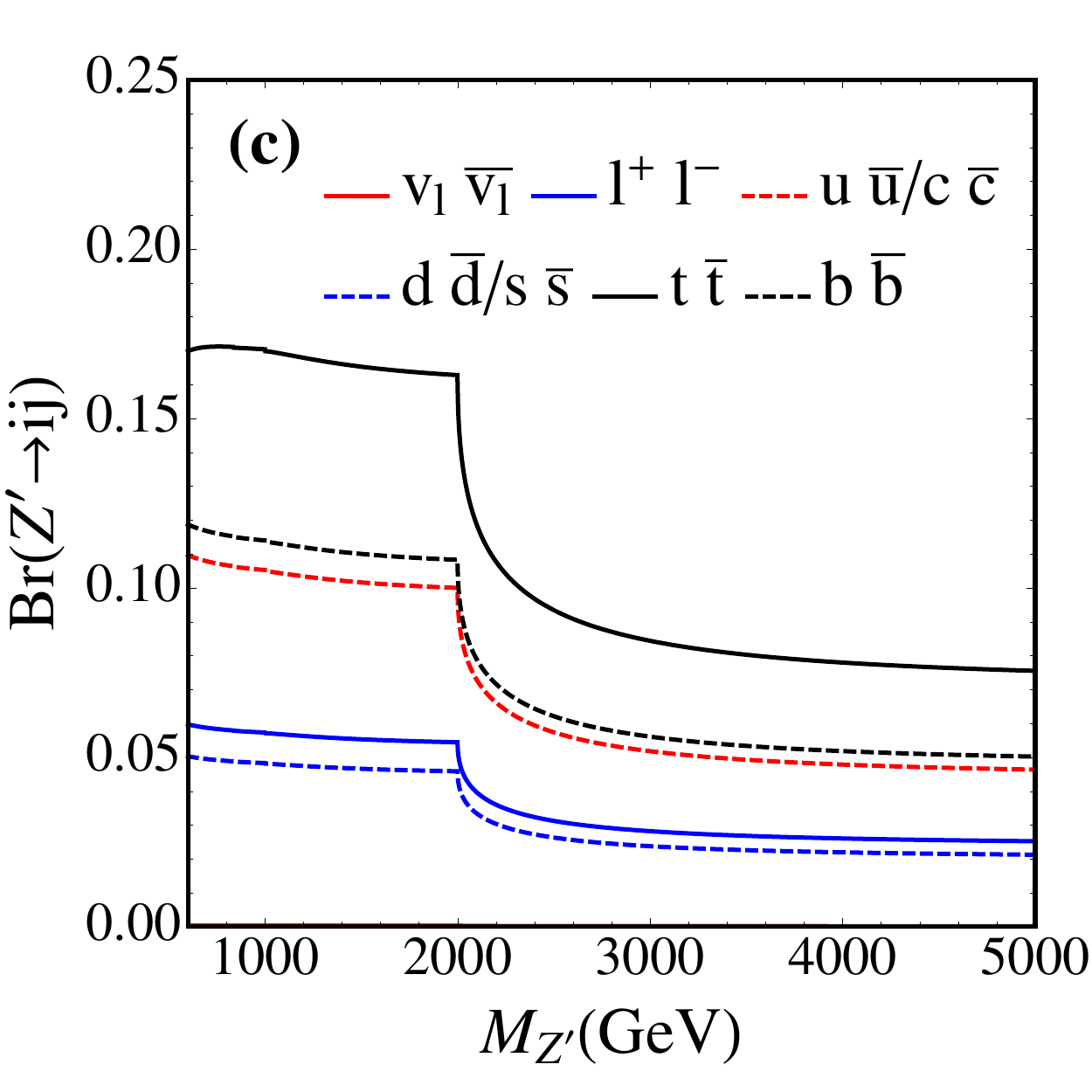}
\includegraphics[width=0.24\textwidth]{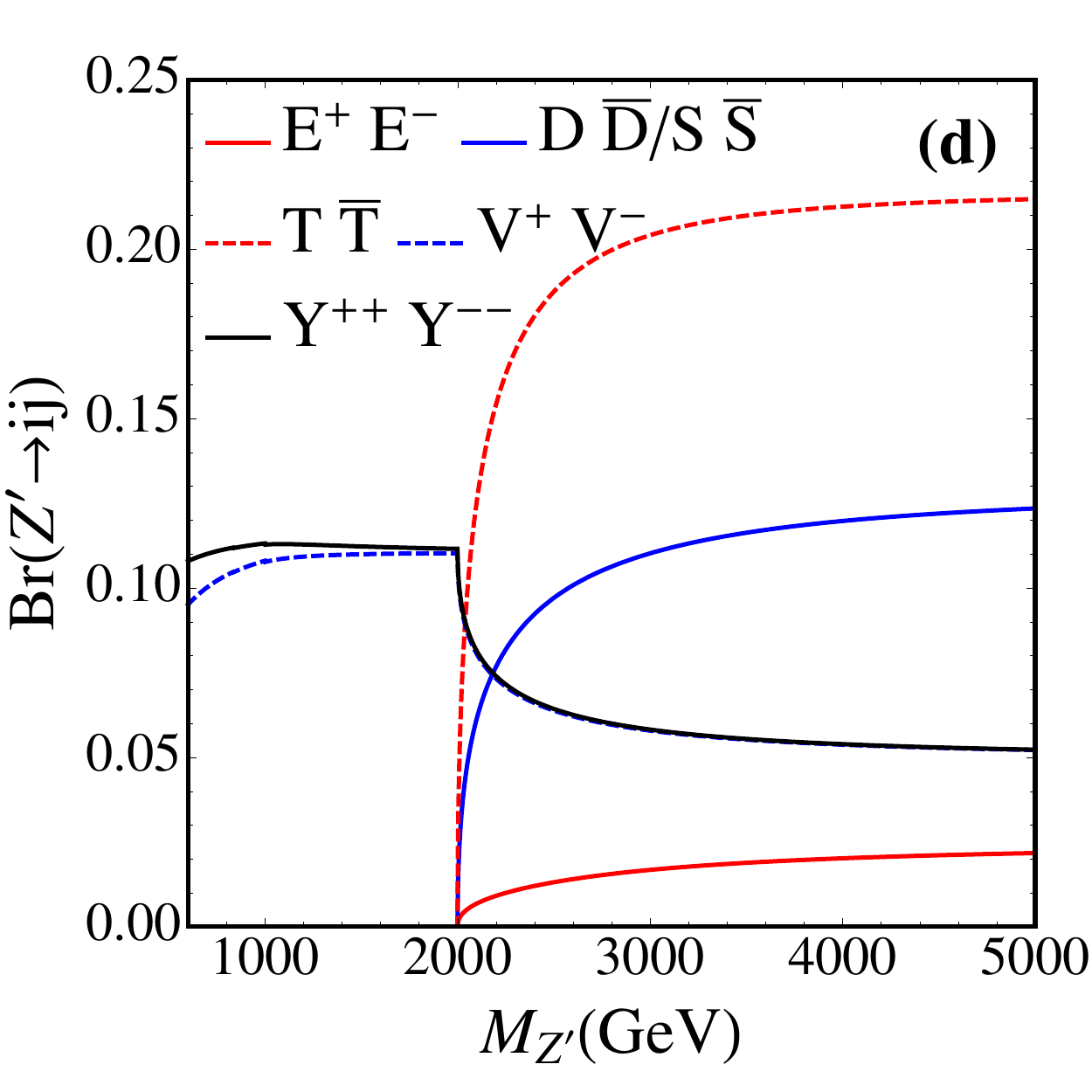}
\caption{\it Decay branching ratios of $Z^\prime$ with respect to $M_{Z^\prime}$. (a) and (b) correspond to $\beta=-\sqrt{3}$, while (c) and (d) correspond to $\beta=\sqrt{3}$. (a) and (c) include the SM particles, while (b) and (d) include the 331 model particles. The heavy fermions' masses are chosen to be $1000\gev$, and the charged Gauge bosons' masses are determined by $Z^\prime$ mass~\eqref{zpvy}.
}\label{zpbr2}
\end{figure}

Figures~$\ref{zpbr1}$ and $\ref{zpbr2}$ display the branching ratio of different decay modes of the $Z^\prime$ boson as a function of $M_{Z^\prime}$ in different 331 models. The branching ratio depends on many parameters. A global fitting has to be carried out to fully understand the model parameter space, but for illustration, we choose $v_1=200\gev$ and fix $M_{H^{\pm Q_V}}=1.3\tev$ and $M_{H^{\pm Q_Y}}=1.4\tev$ in this work by tuning $\lambda_{13}$ and $\lambda_{23}$ for different $v_1$ and $v_3$. 
We drop those decay modes with a branching ratio less than 1 percent. For example, the decay modes involving the $H^\pm$, $H_2$ and $H_0$ scalars will be highly suppressed, e.g. $Z^\prime\to H^+ H^-$, $Z^\prime\to hH_0$, $Z^\prime\to H_3H_0$, $Z^\prime\to H_2H_0$, and $Z^\prime\to ZH_2$, etc. That is owing to the fact that the nearly degenerate masses of $H^\pm$, $H_2$ and $H_0$ is slightly smaller or even larger than $M_{Z^\prime}$, which leads to a large suppression from the phase space. Also, the decay modes involving one $Z$ boson and one neutral Higgs boson are highly suppressed by the small coupling, e.g.  $Z^\prime\to Zh$ and $Z^\prime\to ZH_3$, etc. 

Note that the specific value of $v_1$ is not important. The $v_1$ will affect both the mass of $V$ or $Y$ and the couplings of $ZH^{+Q_V}H^{-Q_V}$ and $ZH^{+Q_Y}H^{-QY}$, however, its effect is overwhelmed by $v_3$. As a consequence, the branching ratios alter less than 1 percent while varying $v_1$, and therefore, we just consider one value of $v_1$ throughout this work.

Different 331 models share a few common features of the branching ratio of $Z^\prime$-boson decay, as depicted in Figs. \ref{zpbr1} and \ref{zpbr2}. First, the decay of $Z^\prime$ boson into a pair of the SM fermions dominate for a light $Z^\prime$ boson. It is worth mentioning that the decay of the $Z^\prime$ boson into a pair of neutrinos, $Z^\prime\to \nu_l \bar{\nu}_l$, is absent in the case of $\beta=\sqrt{3}$ as the coupling is zero. 
Second, for a $Z^\prime$ boson in the medium mass region, the decay modes of the $Z^\prime$ boson into heavy leptons, heavy quarks and heavy charged Higgs bosons, if allowed kinematically, will contribute sizably and compete with those light SM fermion modes. Third, when $M_{Z^\prime}$ is very large, we can ignore the masses of decay products. The heavy quarks is the predominant decay channel of the $Z^\prime$ boson. All the decay branching ratios of $Z^\prime\to E_\ell E_\ell$ ($\ell=e,\mu,\tau$) are  identical, and so do the $Z^\prime\to DD$ and $Z^\prime\to SS$ channels. The decay of $Z^\prime\to TT$ deviates from the $Z^\prime\to DD/SS$ channels since the $T$ quark lives in the anti-triplet representation of the $SU(3)_L$ group. Owing to the same reason, the branching ratio of the $Z^\prime\to t\bar{t}$ channel is not identical to that of $Z^\prime\to u\bar{u}(c\bar{c})$. Similarly, the decay of $Z^\prime\to b\bar{b}$ is also not identical to the decay of $Z^\prime\to d\bar{d}(s\bar{s})$.

Note that $M_{Z^\prime}> 2M_{V(Y)}$ for $\beta=\pm \sqrt{3}$ and $M_{Z^\prime}< 2M_{V(Y)}$ for $\beta=\pm 1/\sqrt{3}$. As a consequence, the $Z^\prime\to VV(YY)$ channel only opens for $\beta=\pm \sqrt{3}$. We plot the branching ratio of the $Z^\prime\to H^{+Q_{V(Y)}} H^{-Q_{V(Y)}}$ mode for $\beta=\pm 1/\sqrt{3}$ but not for $\beta=\pm \sqrt{3}$; see the blue and red dotted curves in Fig. \ref{zpbr1}. Note that it is not because the partial decay width of $Z^\prime\to H^{+Q_{V(Y)}} H^{-Q_{V(Y)}}$ is small or absent for $\beta=\pm\sqrt{3}$; on the contrary, the partial width is larger in the case of $\beta=\pm\sqrt{3}$ than in the case of $\beta=\pm1/\sqrt{3}$.
However, the total width of the $Z^\prime$ also increases dramatically for $\beta=\pm\sqrt{3}$ such that the branching ratio of $Z^\prime\to H^{+Q_{V(Y)}} H^{-Q_{V(Y)}}$ mode is highly suppressed. As a consequence, it is difficult to detect the $Z^\prime$ through a pair of $H^{\pm Q_{V(Y)}}$ scalar in the case of $\beta=\pm \sqrt{3}$.

\begin{figure}\centering
\includegraphics[width=0.3\textwidth]{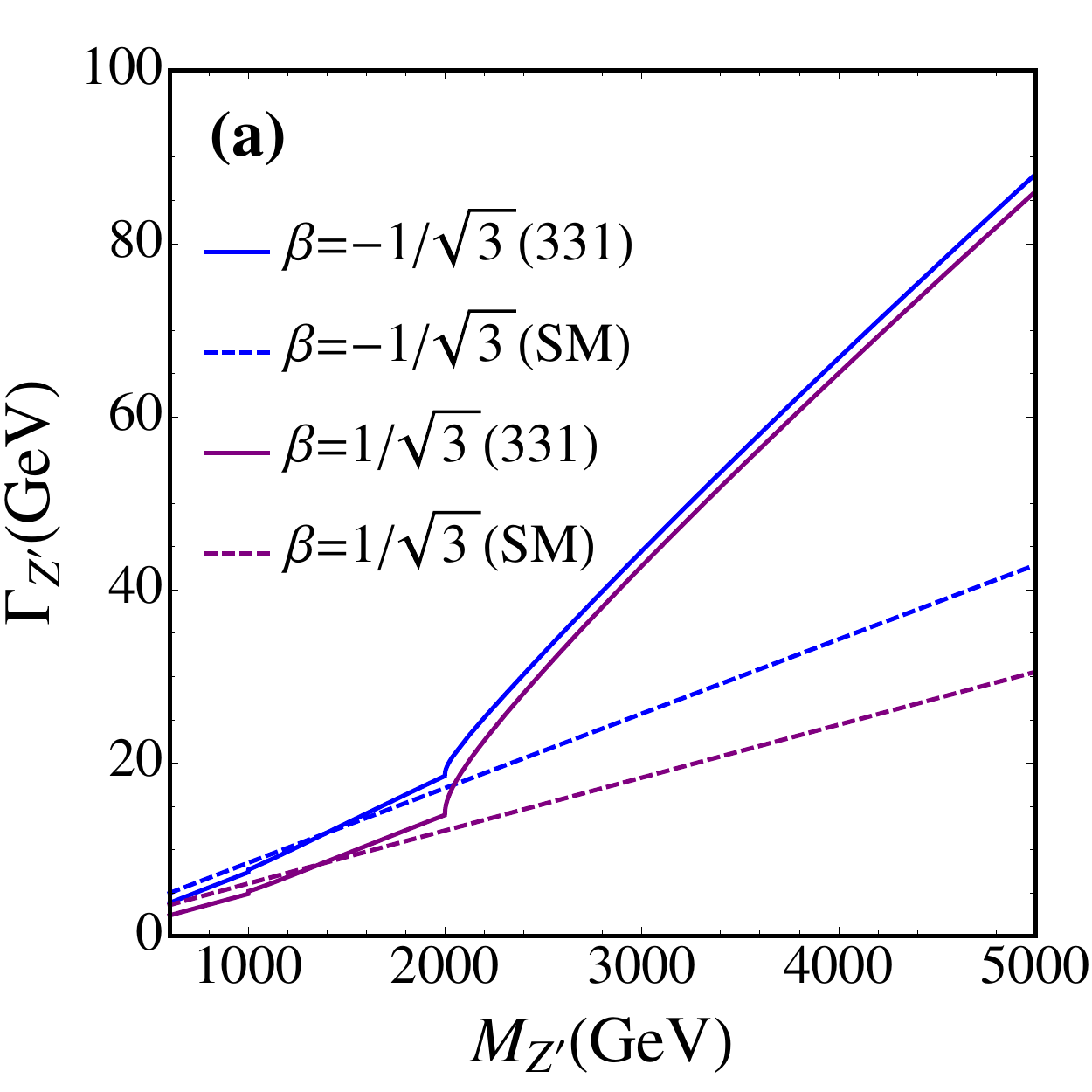}
\includegraphics[width=0.3\textwidth]{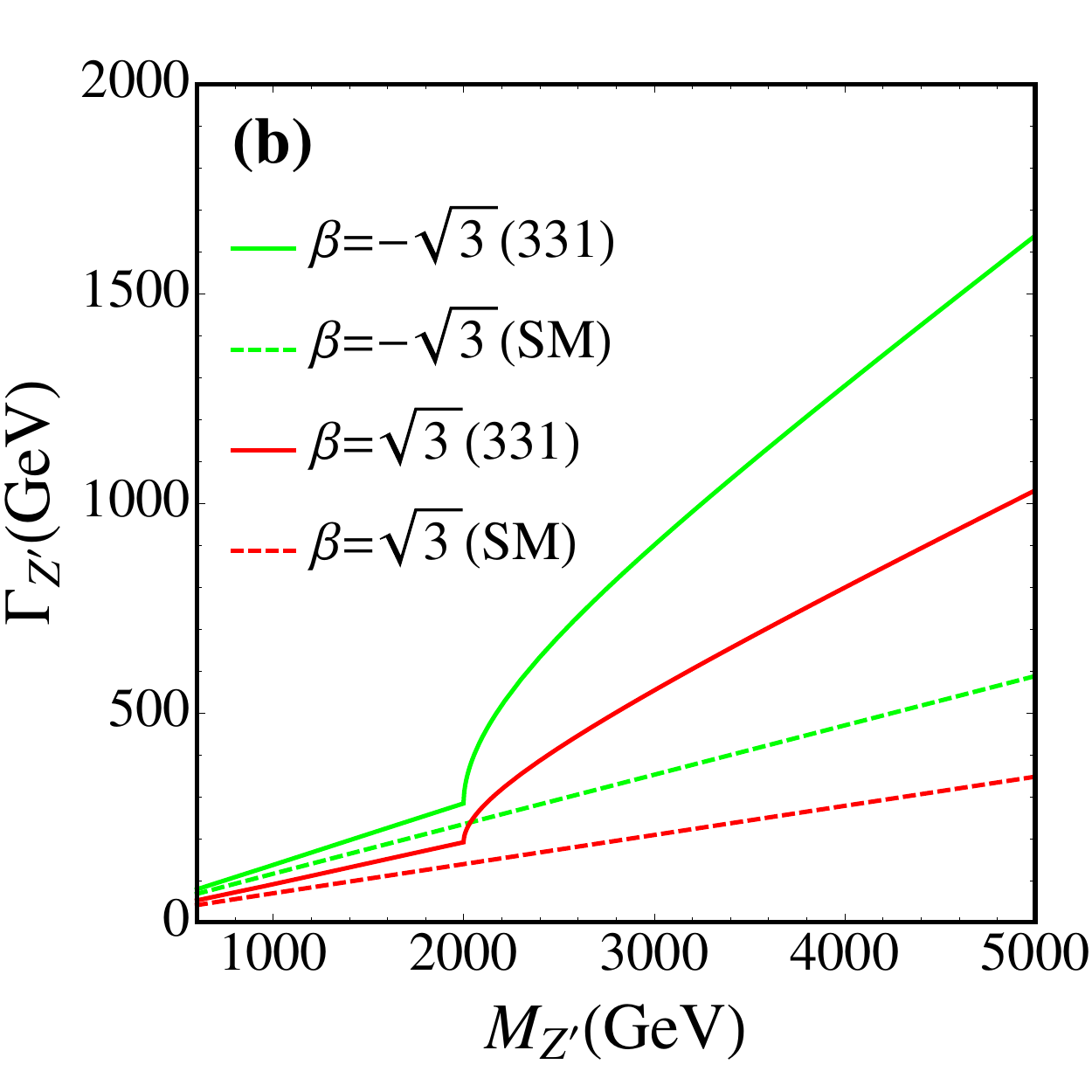}
\caption{\it Total width of $Z^\prime$ with respect to $M_{Z^\prime}$ in different versions of the 331 model. (a) and (b) correspond to $\beta=\pm1/\sqrt{3}$ and $\beta=\pm\sqrt{3}$ respectively.
The solid lines represent the total width including all the decay modes. While the dashed lines stand for the total width that the $Z^\prime$ boson decays only the SM fermions, which is a lowest limit of the real total width.
}\label{zpwidth}
\end{figure}

Figure \ref{zpwidth} plots the total width of the $Z^\prime$ as a function of $M_{Z^\prime}$ in different 331 models with the masses of new physics particles specified above.  The total width of $Z^\prime$ boson varies greatly for different $\beta$'s.  For example, the $Z^\prime$ boson exhibits a narrow width for $\beta=\pm 1/\sqrt{3}$; while the width of $Z^\prime$ boson for $\beta=\pm \sqrt{3}$ is much broader than the case of $\beta=\pm 1/\sqrt{3}$. Of course the decay width depends on whether or not the decay channels involving new physics particles. For illustration, we also plot the sum of partial widths of the $Z^\prime$ decaying into a pair of SM fermions in Fig. \ref{zpwidth}, depicted by the dashed curves. One should treat the dashed curve as the lower limit of the $Z^\prime$-boson width. It is clear that, owing to the large couplings of $Z^\prime$ boson to SM fermions for $\beta=\pm \sqrt{3}$, the $Z^\prime$ boson in the $\beta=\pm \sqrt{3}$ models exhibit much larger width than in the $\beta=\pm 1/\sqrt{3}$ models.

\subsection{$M_{Z^\prime}$ constraint from the dilepton search at the LHC}

\begin{figure}
\centering
\includegraphics[width=0.3\textwidth]{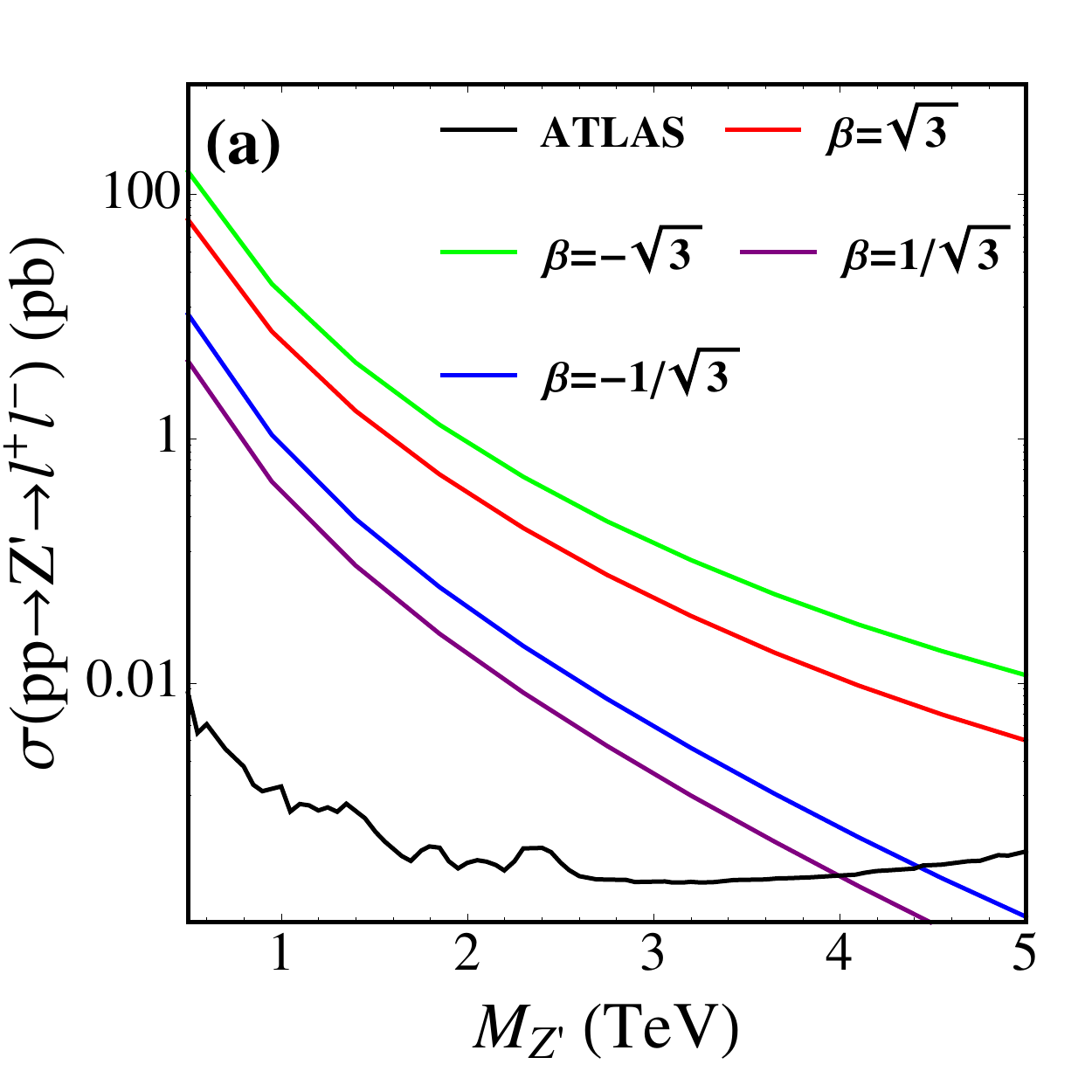}
\includegraphics[width=0.3\textwidth]{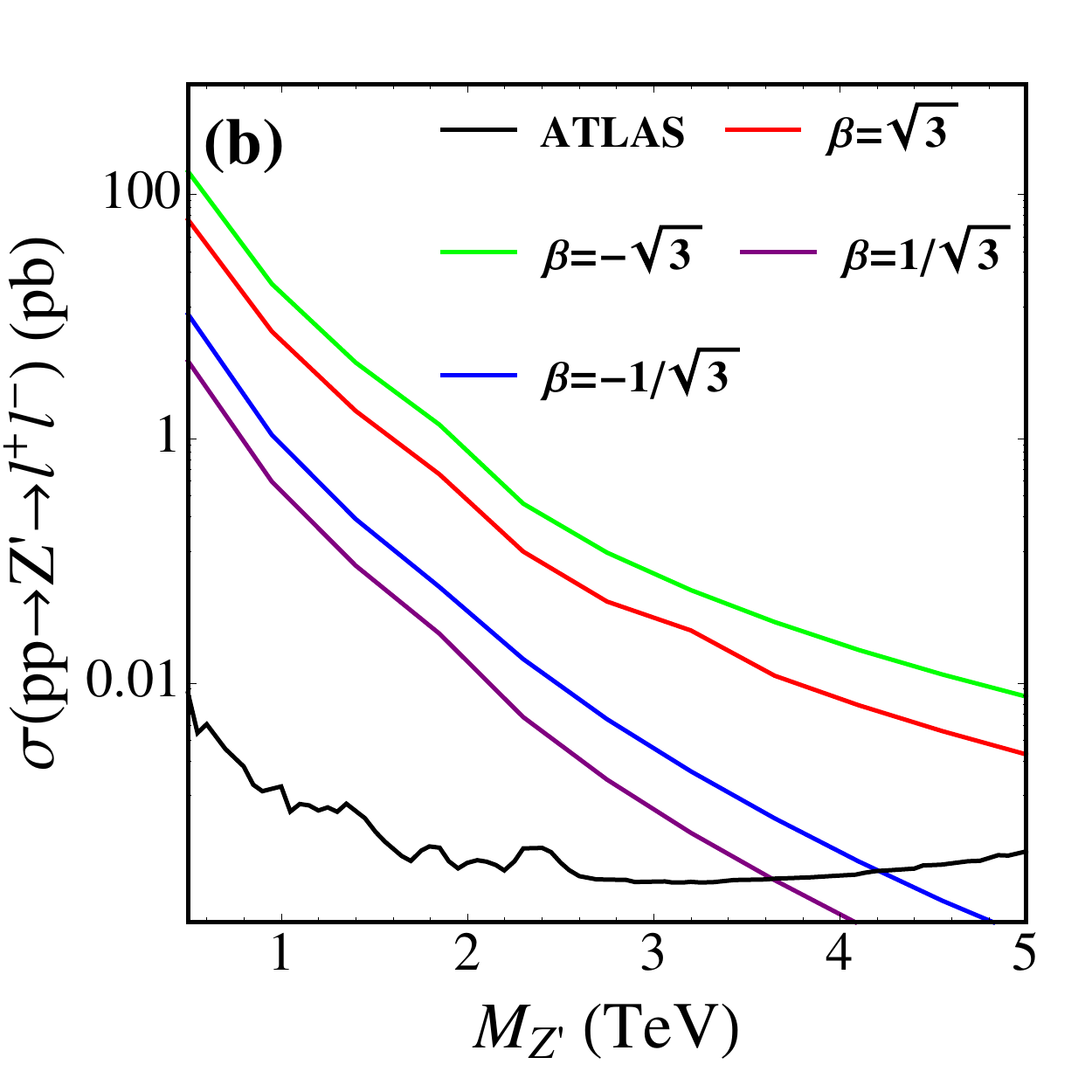}
\caption{\it Constraint on $M_{Z^\prime}$ in different versions of 331 model, where the black line shows the 95\% C.L. limit on the $Z^\prime$ production cross section times branching ratio to two leptons as a
function of $M_{Z^\prime}$ from~\cite{ATLAS:2016cyf}. (a) is assuming $Z^\prime$ only decays to SM fermions, while (b) assumes $Z^\prime$ can also decay to the heavy fermions with fermion's masses all set to be 1 TeV.}\label{Fig:zpll}
\end{figure}

The constraint on $M_{Z^\prime}$ is obtained from the Drell-Yan channel $pp\to Z^\prime \to e^+e^-/\mu^+\mu^-$ at the LHC. Ref.~\cite{Salazar:2015gxa} studied the constraint on $M_{Z^\prime}$ in different 331 models from the dilepton signal using the 8 TeV LHC data~\cite{Aad:2014cka} and obtained a typical bound of $M_{Z^\prime}\gtrsim 3$ TeV. In this work we consider the recent negative result of dilepton searches at the 14 TeV LHC~\cite{ATLAS:2016cyf} to set a constraint on $M_{Z^\prime}$ in the 331 models. The cross sections of $pp\to Z^\prime \to \ell^+\ell^-$ ($\ell=e,\mu$) are plotted in Fig.~\ref{Fig:zpll} as a function of $M_{Z^\prime}$. The exclusion limit on $\sigma(pp\to Z^\prime\to \ell^+\ell^-)$ is also shown; see the black curve. Two scenarios of $Z^\prime$ decay is considered. First, we consider the case that the $Z^\prime$ boson decays only into a pair of SM fermions, therefore, $\sigma(pp\to Z^\prime \to \ell^+\ell^-)$ is maximumly enhanced and that leads to stronger constraints on $M_{Z^\prime}$. For example, it yields $M_{Z^\prime}\gtrsim 4.4$ TeV for $\beta=-1/\sqrt{3}$ and $M_{Z^\prime}\gtrsim 4$ TeV for $\beta=+1/\sqrt{3}$, respectively; see Fig.~\ref{Fig:zpll}(a). For $\beta=\pm \sqrt{3}$, the $Z^\prime$ boson with a mass smaller than 5 TeV is completely excluded. 
Second, we allow the $Z^\prime$ boson can also decay into a pair of heavy fermions and fix all the heavy fermion masses to be 1 TeV. The cross section $\sigma(pp\to Z^\prime\to \ell^+\ell^-)$ is slightly smaller than the first case; see Fig.~\ref{Fig:zpll} (b). That yields a slightly weakened bound on $M_{Z^\prime}$: $M_{Z^\prime}\gtrsim 4.2~{\rm TeV}$ for $\beta=-1/\sqrt{3}$ and $M_{Z^\prime}\gtrsim 3.6~{\rm TeV}$ for $\beta=+1/\sqrt{3}$, respectively. Again, a $Z^\prime$ boson with mass less than 5~TeV in the $\beta=\pm \sqrt{3}$ models is already ruled out. 

We choose two benchmark masses of the $Z^\prime$ boson in the study: $M_{Z^\prime}=4~{\rm TeV}$ in the $\beta=\pm 1/\sqrt{3}$ models and $6~{\rm TeV}$ in the $\beta=\pm \sqrt{3}$ models, to respect the current LHC bounds. For simplicity, we consider the $Z^\prime$ boson decaying only into the SM fermions. 

\subsection{Search for $Z^\prime$ through $A_{FB}$ at the LHC}

\begin{figure}\centering
\includegraphics[width=0.4\textwidth]{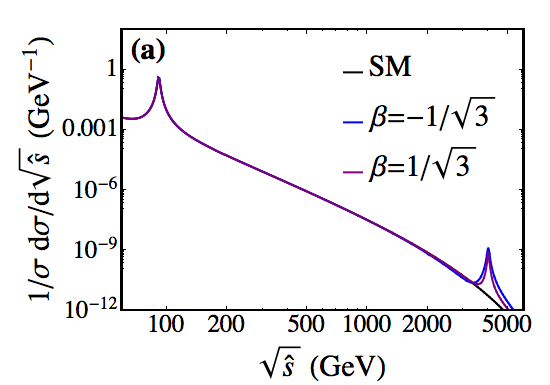}
\includegraphics[width=0.4\textwidth]{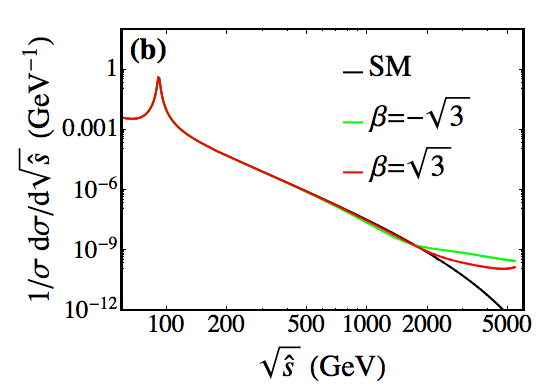}
\caption{\it Invariant mass ($m_{\ell^+\ell^-}$) distribution of lepton pair in the process of $pp\to \gamma/Z/Z^\prime \to \ell^+\ell^-$ at the 14~TeV LHC in the SM (black) and various 331 model: (a) $M_{Z^\prime}=4\tev$ and (b) $M_{Z^\prime}=6\tev$. The $Z^\prime$ boson decaying into a pair of the SM fermions is understood.}
\label{crosc}
\end{figure}

The width of $Z^\prime$ boson is very broad when $\beta=\pm \sqrt{3}$ due to its large couplings with the fermions. It is hard to probe such a broad resonance through the usual bump search. For illustration, Fig.~$\ref{crosc}$ plots the invariant mass distribution of the charged-lepton pair $m_{\ell\ell}$, which is equal to the center of mass energy of the hard scattering $\sqrt{\hat{s}}$ at the tree level, in the process of $pp\to \gamma/Z/Z^\prime\to \ell^+\ell^-$ for both $\beta=\pm\sqrt{3}$ and $\beta=\pm 1/\sqrt{3}$ models at the 14~TeV LHC: (a) $M_{Z^\prime}=4~{\rm TeV}$ and (b) $M_{Z^\prime}=6~{\rm TeV}$. The $Z^\prime$ bosons in the $\beta=\pm 1/\sqrt{3}$ models (purple and blue curves) appear as a peak on top of the SM background (black curve), but the $Z^\prime$ bosons in the $\beta=\pm \sqrt{3}$ models (green and red curves) display a very broad bump on top of the SM background. It is very hard to detect such a non-resonance shape experimentally.  In addition, the distribution will be distorted by the detector effects which make the detection more challenging. 

Fortunately, the so-called Forward-Backward Asymmetry ($\AFB$) of the charged lepton provides us a powerful tool to probe a ``fat" $Z^\prime$ boson at the LHC~\cite{Accomando:2015cfa}.  The $\AFB$ of the charged lepton is defined as the difference of the fractions that the charged lepton appears in the forward and backward regimes of the detector, i.e. 
\be
A_{FB}(\hat{s})=\frac{\sigma_{F}(\hat{s})-\sigma_{B}(\hat{s})}{\sigma_{F}(\hat{s})+\sigma_{B}(\hat{s})},
\ee
where $\hat{s}$ is the lepton pair invariant mass squared, $\sigma_{F}$ and $\sigma_{B}$ are the cross section of the charged lepton in the forward and backward regimes, respectively. We call the charged lepton in the forward region when its polar angle with respect to the quark direction is from 0 to $\pi/2$ in the center-of-mass-frame of the di-lepton system, and in the backward region when the angle is from $\pi/2$ to $\pi$.

At the partonic level the $\AFB$ is given by
\be
A_{FB}(\hat{s})=\frac{3}{4}\times \frac{\sum_{i,j=1}^N C_A^{ij}P_{ij}(\hat{s})}{\sum_{i,j=1}^N C_S^{ij}P_{ij}(\hat{s})},
\ee
where $N$ stands for the numbers of gauge bosons involved involved in the process, $C_A^{ij}$ ($C_S^{ij}$) denotes the parity antisymmetric (symmetric) coefficient, respectively. Both coefficients are a function of the chiral gauge couplings of the $i$-boson to the quark ($q_{L/R}^i$ ) and lepton ($l_{L/R}^i$), which are listed as follows:
\bea
C_A^{ij}&=&(q_L^i q_L^j-q_R^i q_R^j)(l_L^i l_L^j-l_R^i l_R^j),\\
C_S^{ij}&=&(q_L^i q_L^j+q_R^i q_R^j)(l_L^i l_L^j+l_R^i l_R^j).
\eea
The $\hat{s}$ dependent variable $P_{ij}(\hat{s})$ arises from the propagators of gauge bosons involved in the process
\be
P_{ij}(\hat{s})=\frac{(\hat{s}-M_i^2)(\hat{s}-M_j^2)+M_i M_j \Gamma_i \Gamma_j}{[(\hat{s}-M_i^2)^2+M_i^2\Gamma_i^2][(\hat{s}-M_j^2)^2+M_j^2\Gamma_j^2]},
\ee
where $M_i$ and $\Gamma_i$ is the mass and width of the $i$-boson, respectively.

\begin{figure}\centering
\includegraphics[width=0.45\textwidth]{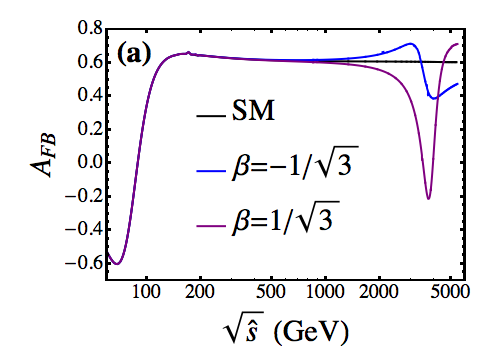}
\includegraphics[width=0.45\textwidth]{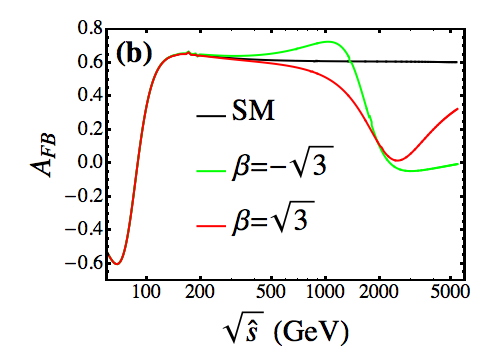}
\caption{\it $A_{FB}$ distribution versus the lepton pair invariant mass $\sqrt{\hat{s}}=m_{\ell^+\ell^-}$ in the SM and various 331 models at the 14 TeV LHC for (a) $M_{Z^\prime}=4\tev$ and (b) $M_{Z^\prime}=6\tev$ respectively. The width of $Z^\prime$ is chosen to be the lowest limit shown in Fig.~\ref{zpwidth}.}
\label{AFB}
\end{figure}

Figure~$\ref{AFB}$ shows the $\AFB$ distribution as a function of $\sqrt{\hat{s}}$ in the SM and the two benchmark 331 models: (a) $M_{Z^\prime}=4~{\rm TeV}$ in the $\beta=\pm 1/\sqrt{3}$ models and (b) 6~TeV in the $\beta=\sqrt{3}$ models. The black (red, green, purple, blue) curve denotes the SM ($\beta=+\sqrt{3}$, $\beta=-\sqrt{3}$, $\beta=+1/\sqrt{3}$, $\beta=-1/\sqrt{3}$), respectively. When $\sqrt{\hat{s}}$ is larger than 200~GeV,  the $\AFB$ approaches 0.6 in the SM as there is no other heavier gauge bosons. But if there is a $Z^\prime$ boson, the $\AFB$ can deviate from the SM prediction. The deviation depends on the $Z^\prime$ boson mass, width and its couplings to the SM fermions.

In the $\beta=\pm 1/\sqrt{3}$, owing to the narrow width of the $Z^\prime$ boson, large deviations occur around $M_{Z^\prime}=4~{\rm TeV}$; see Fig.~\ref{AFB}(a). Thanks to the large width of $Z^\prime$ in the $\beta=\pm \sqrt{3}$ models, the $\AFB$ can be affected at a energy scale far below $M_{Z^\prime}$; see Fig.~\ref{AFB}(b). The large deviations occur around half of $M_{Z^\prime}$. The $A_{FB}$ distribution has the advantage searching for or excluding a broad resonance even in the case that the collider energy is not enough to see those ``fat" resonances directly. We note that, owing to the interference effect between the $Z$ boson and $Z^\prime$, the sign of $\beta$ plays an important role in the $\AFB$ distributions, which can be used to distinguish different 331 models when collecting enough data.

\section{Phenomenology of the $V$ and $Y$ bosons}\label{sec:sec5}

In this section we study the phenomenology of two unconventional gauge bosons in the 331 models, $V$ and $Y$, at the LHC.  

\begin{figure}[b]
\centering
\includegraphics[width=0.23\textwidth]{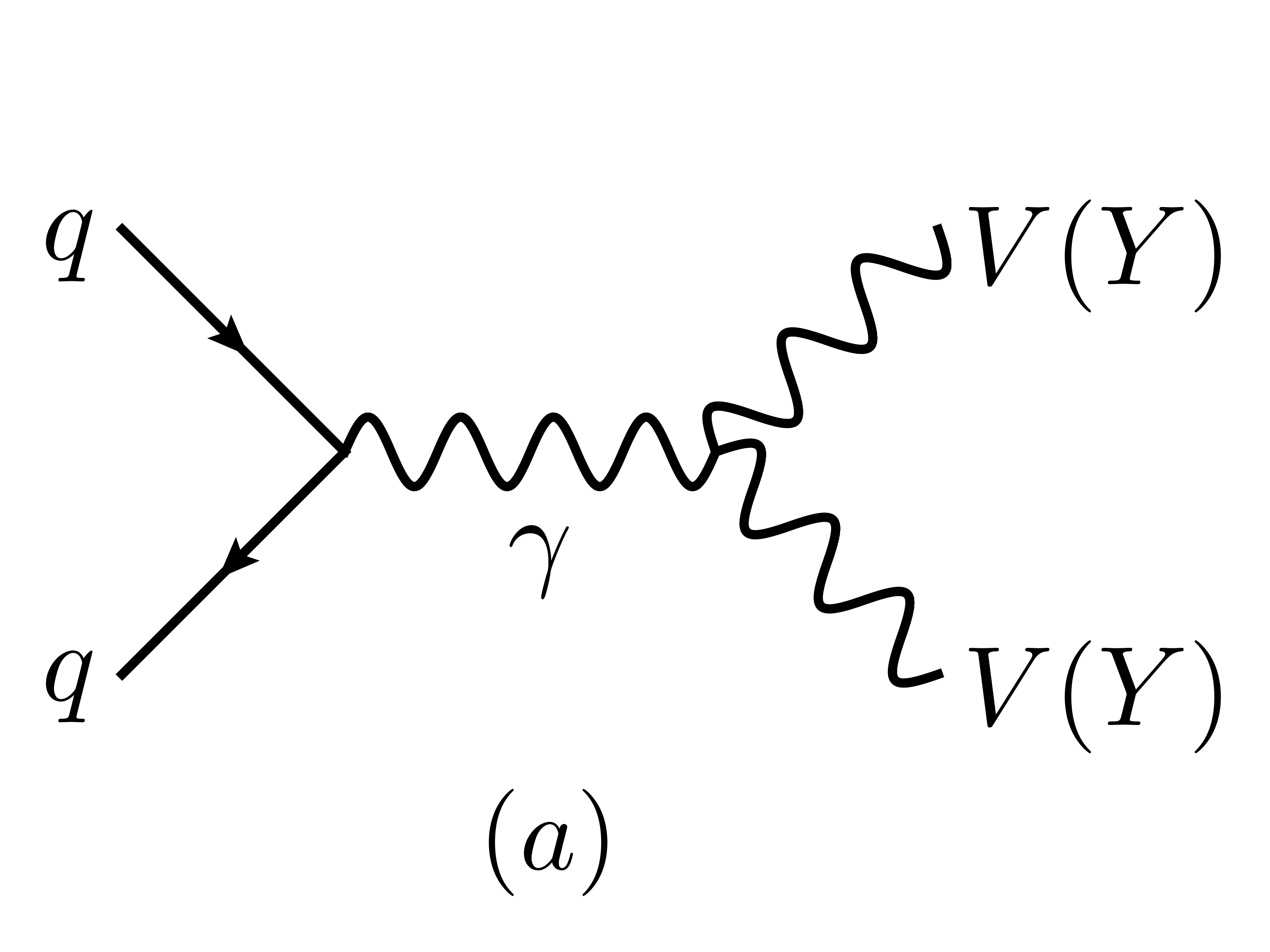}
\includegraphics[width=0.23\textwidth]{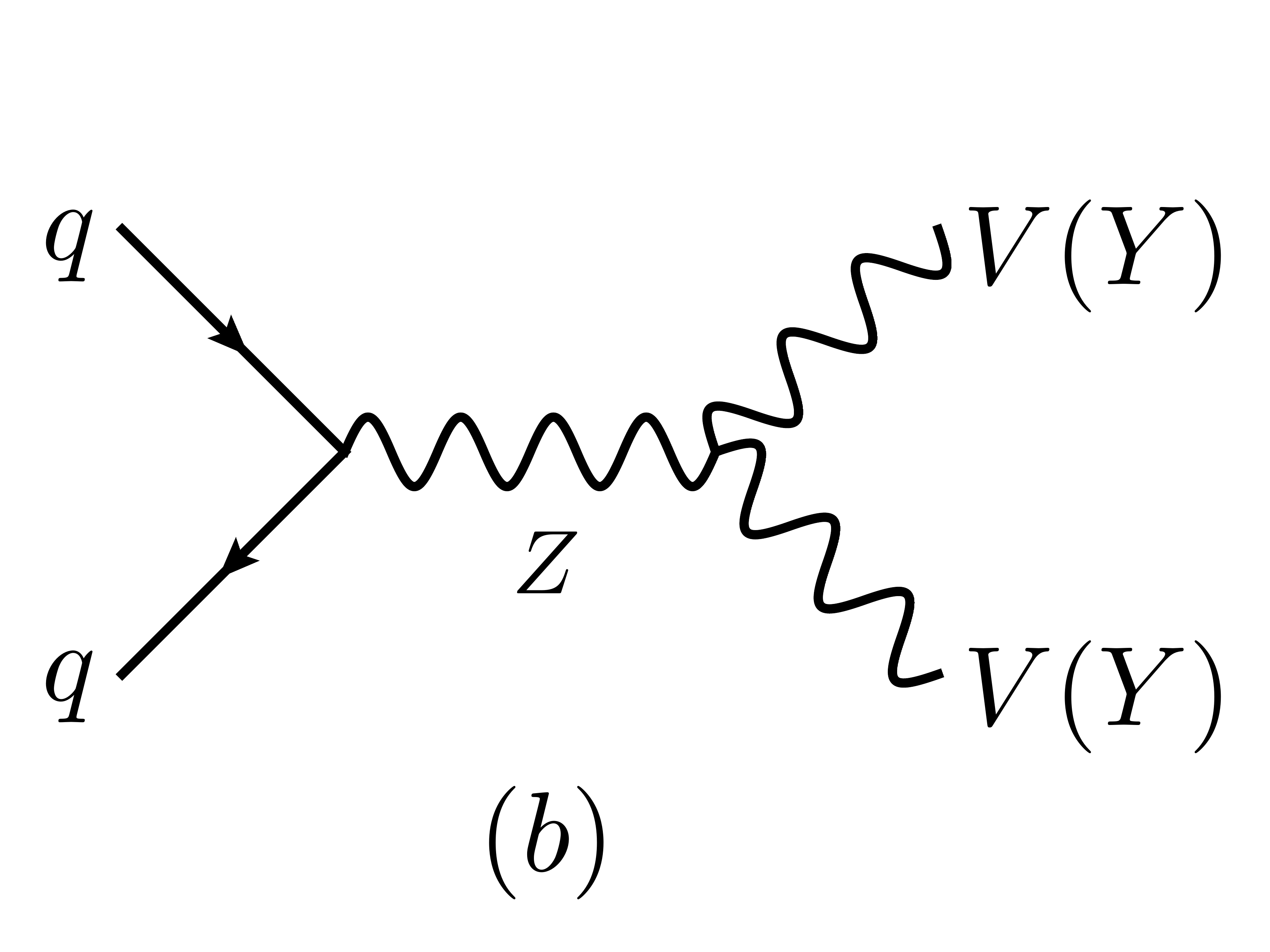}
\includegraphics[width=0.23\textwidth]{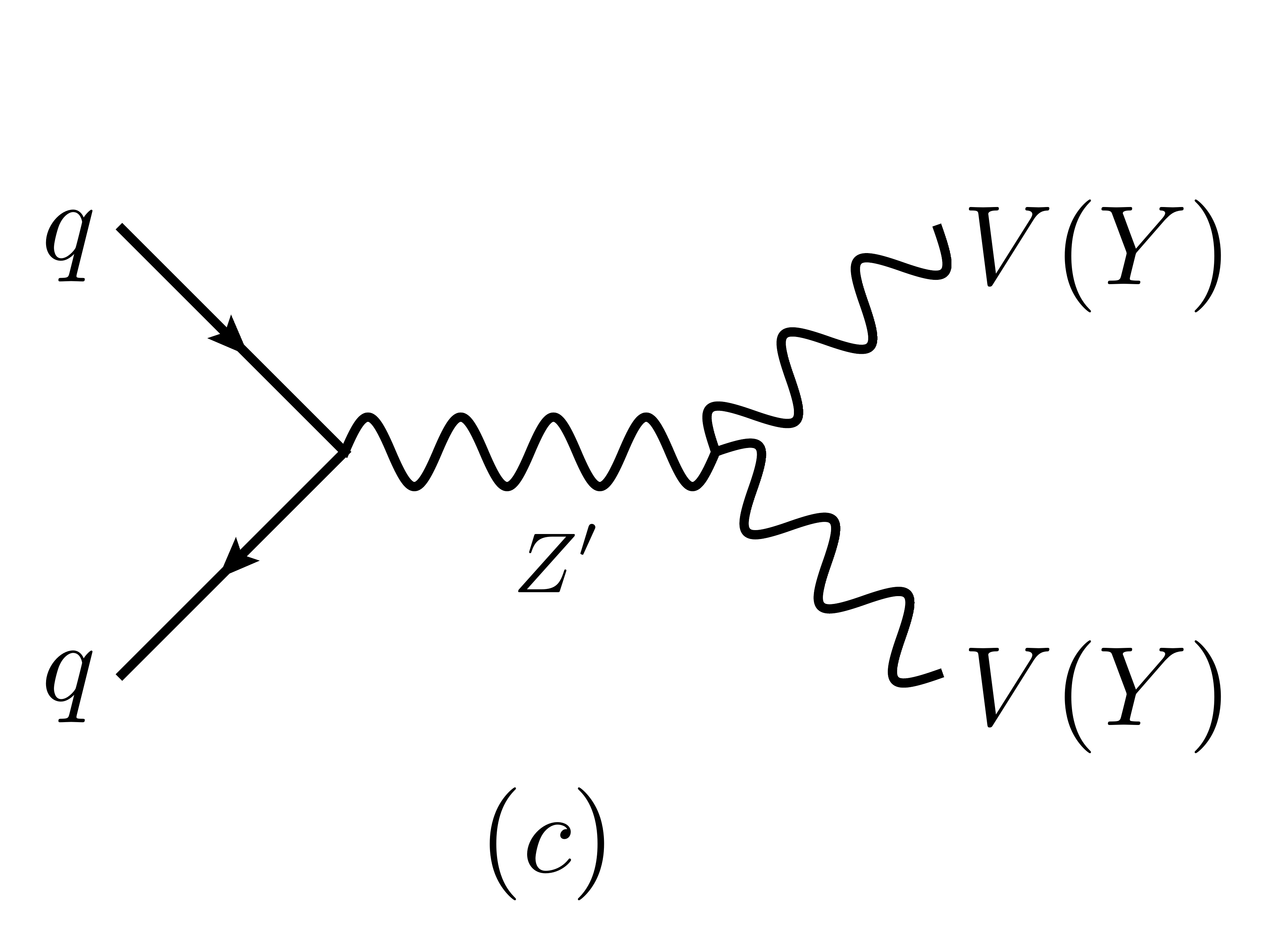}
\includegraphics[width=0.23\textwidth]{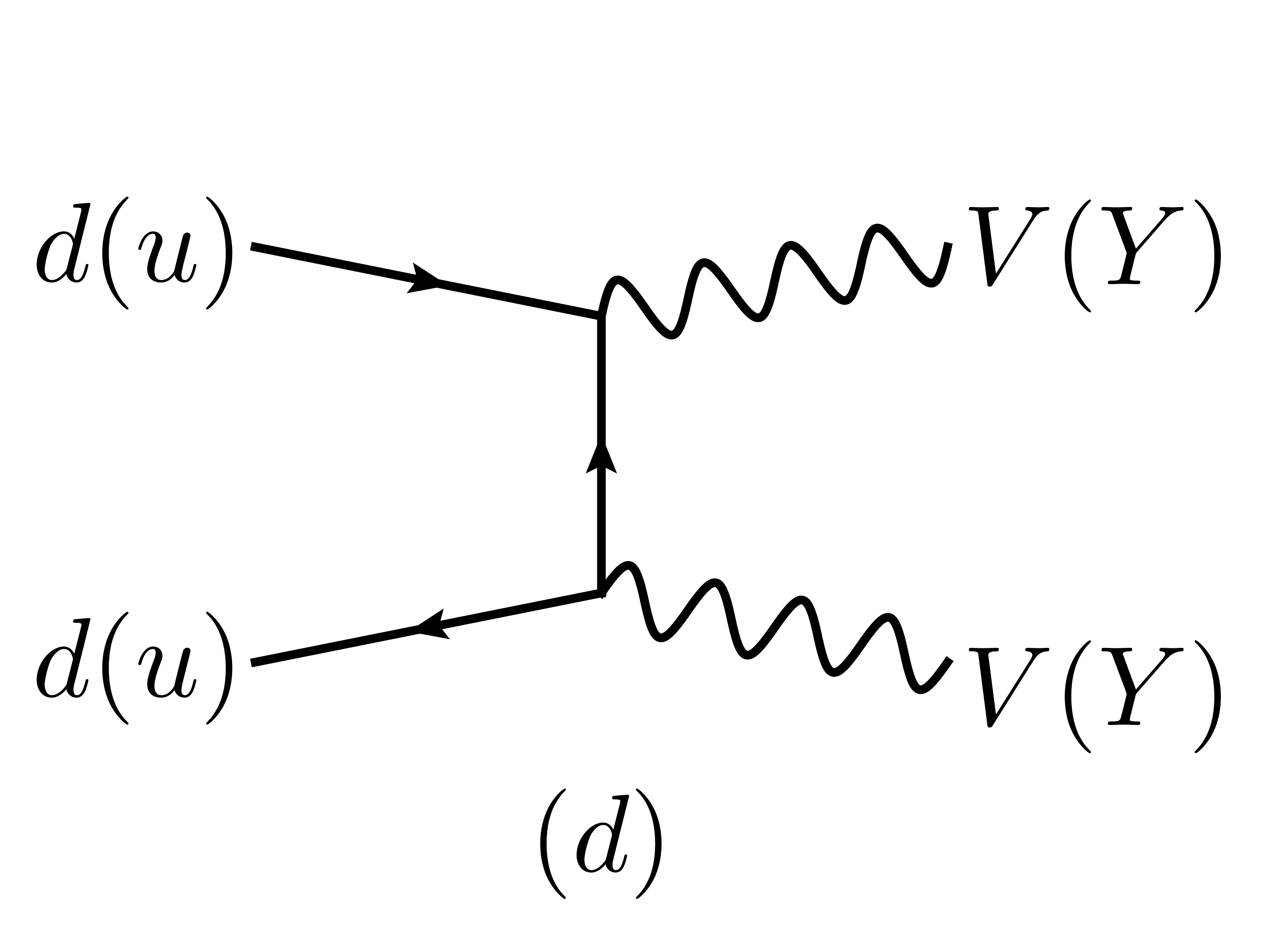}
\caption{\it Feynman diagrams for $V(Y)$-pair production, where $q$ stands for SM quarks. In the t-channel diagram $(d)$, we have neglected the flavor mixing in the quark sector and only considered the dominant contribution from $u$ and $d$ quark.}\label{vypair}
\end{figure}

\subsection{Production of $V$ and $Y$ bosons}
As shown in Fig.~\ref{fig:gaugeint}, the $V$ and $Y$ bosons connect the SM fermions and the exotic fermions that are added to fill the $SU(3)$ (anti)fundamental representation, therefore, the $V$ and $Y$ bosons cannot be produced singly through the Drell-Yan channel as the $Z^\prime$ boson. Figure~$\ref{vypair}$ displays a few production channels of $V$ or $Y$ bosons. They can be produced in pair through $V$-$V$-$\gamma/Z/Z^\prime$ and  $Y$-$Y$-$\gamma/Z/Z^\prime$ triple gauge boson interactions (a,b,c) and through the $t$-channel diagram involving a heavy quark (d).  They can also be produced in association with a heavy quark production or association with a heavy charged scalar production, e.g. $d g \to D V$ or $u g \to D Y$. In the work we focus on the $VV$ and $YY$ pair productions and will study the associated production elsewhere~\cite{Cao:2017abc}.

\begin{figure}[t]
\begin{center}
\includegraphics[width = 0.24\textwidth]{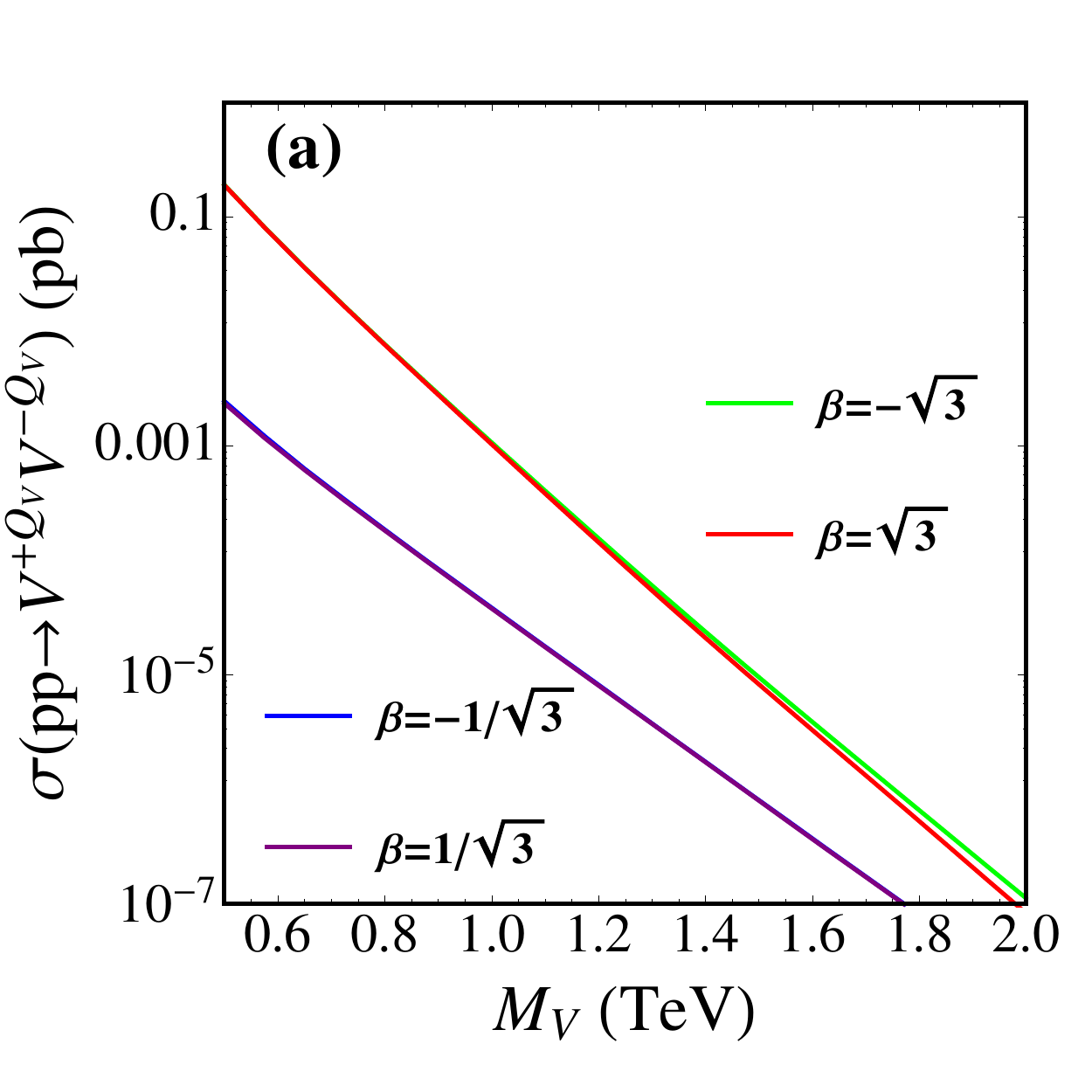}
\includegraphics[width = 0.24\textwidth]{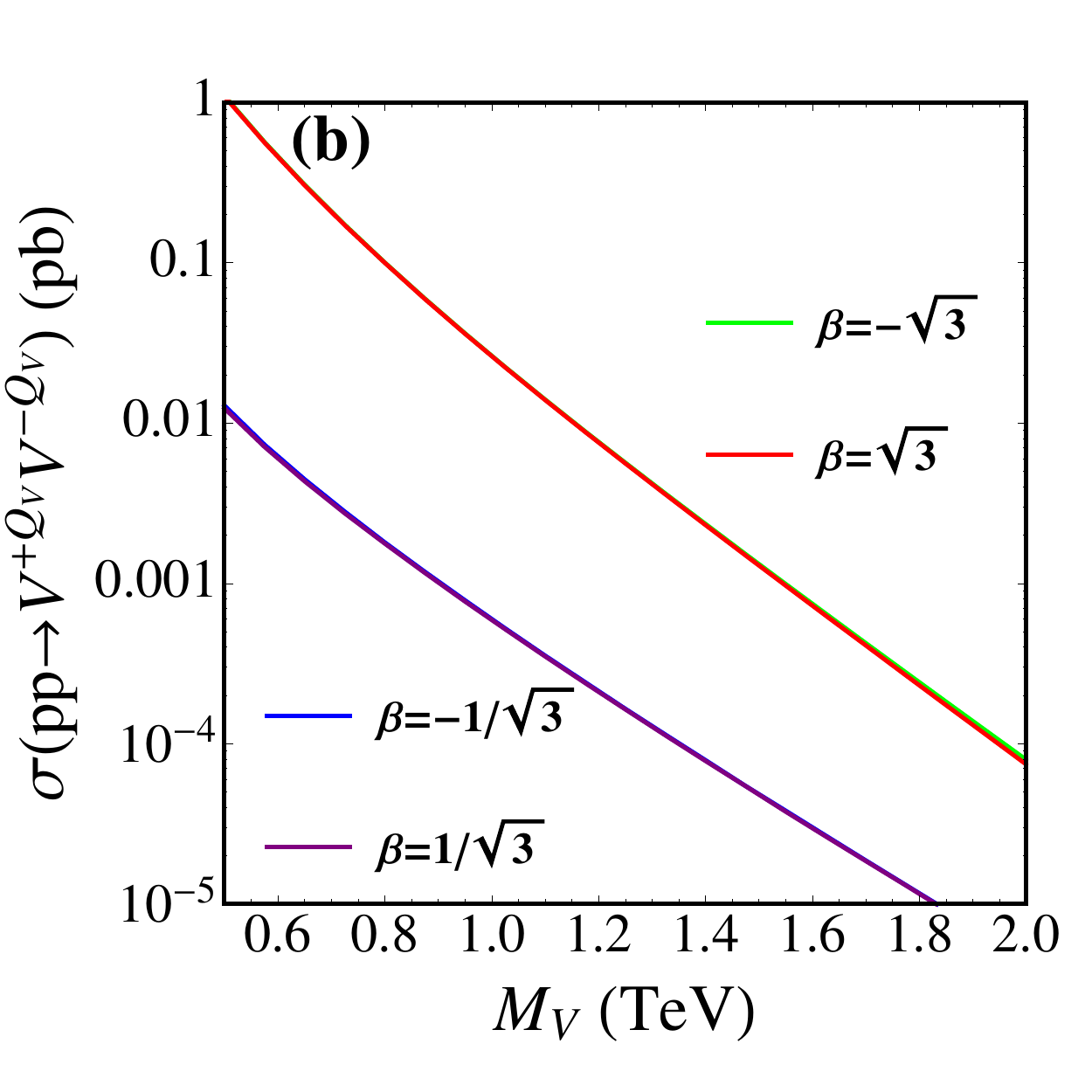}
\includegraphics[width = 0.24\textwidth]{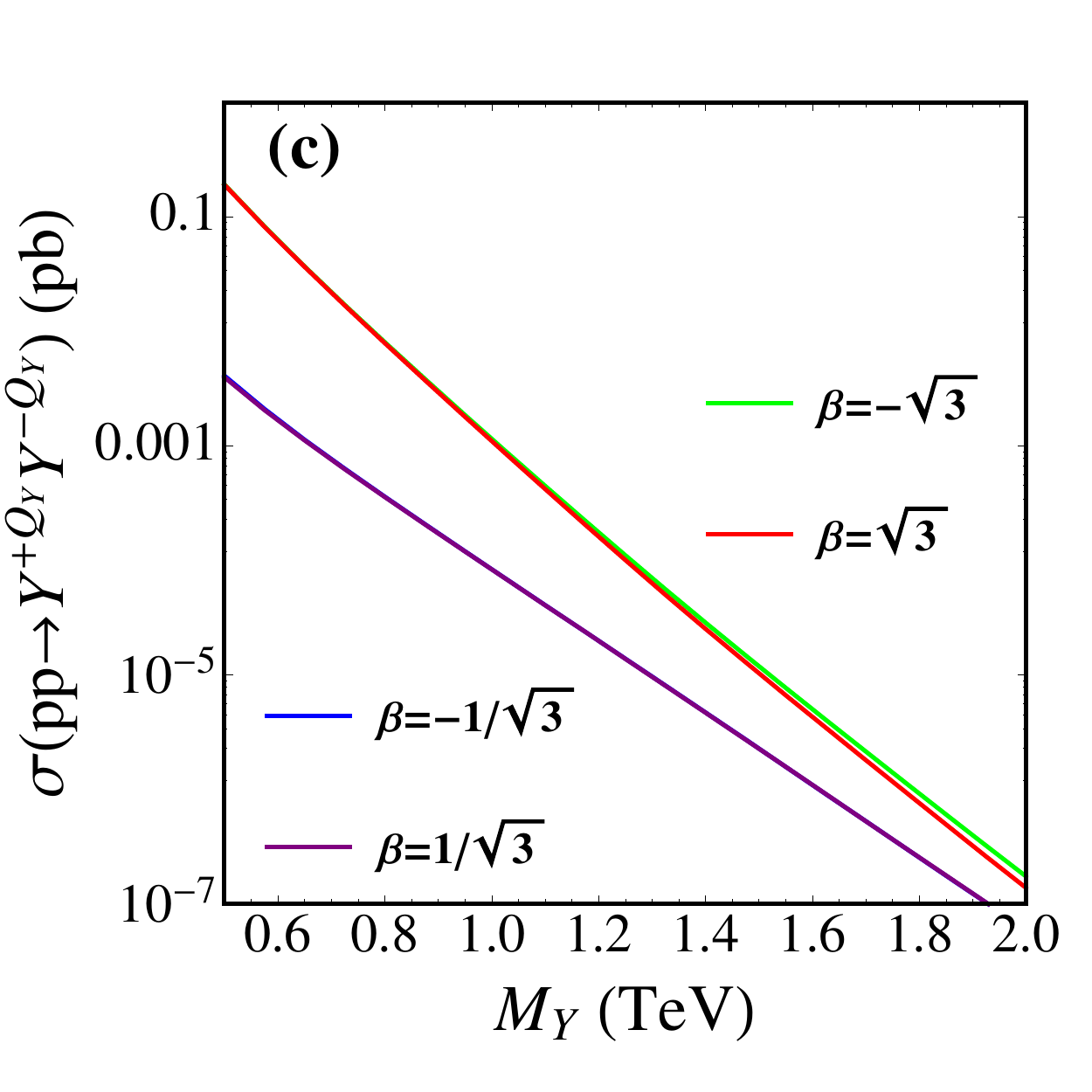}
\includegraphics[width = 0.24\textwidth]{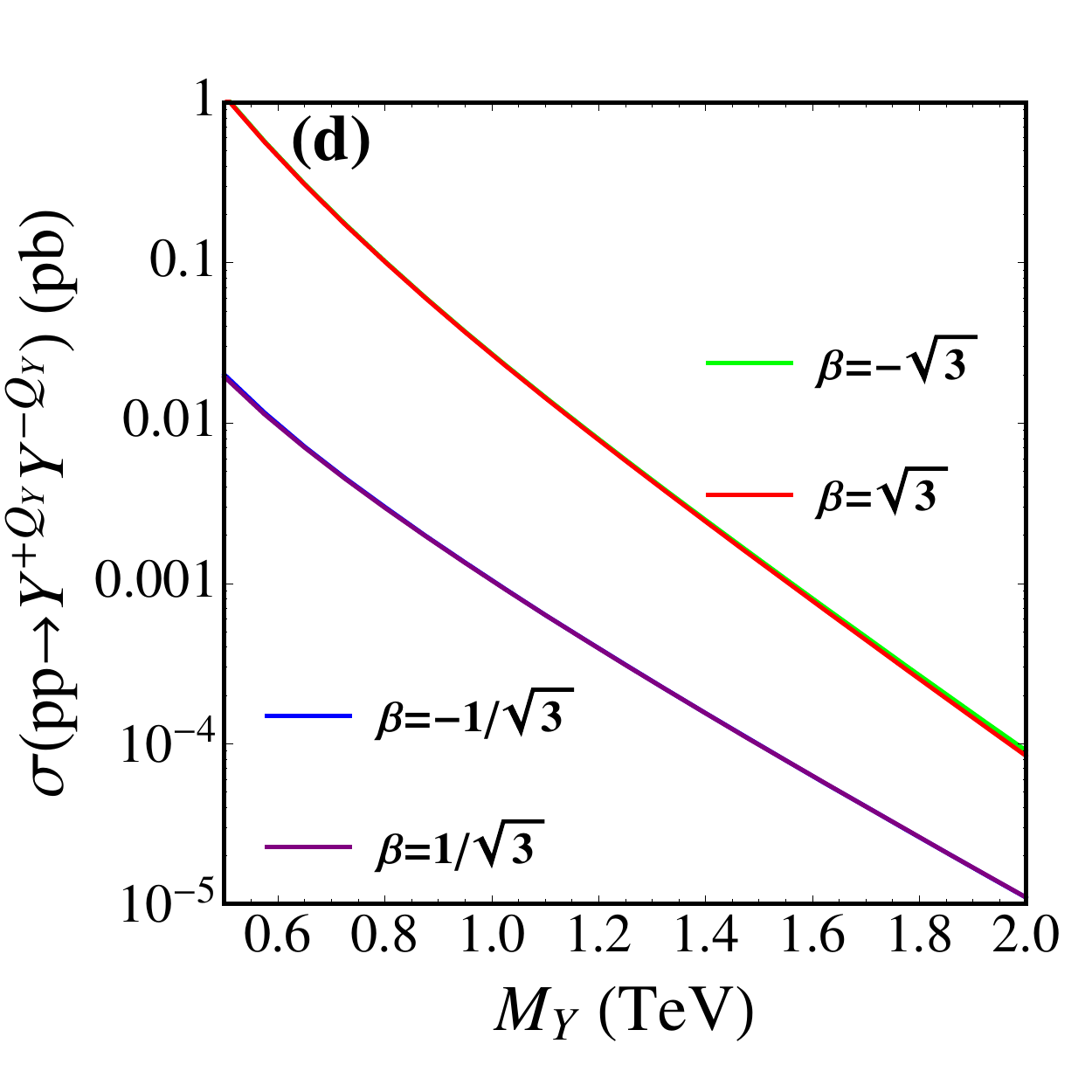}
\caption{\it Inclusive cross sections of $VV$ and $YY$ pair productions as a function of  $M_{V}$ and $M_Y$ in the 331 models at the 8~TeV LHC (a, c) and 14~TeV LHC (b, d), respectively, where we have set all the heavy quark masses to be 500{~\rm GeV}. The width of $Z^\prime$ is chosen to be the lowest limit shown in Fig.~\ref{zpwidth}.}\label{ppvv}
\end{center}
\end{figure}

Figure~\ref{ppvv} displays the cross section of $VV$ pair production at the 8~TeV (a) and 14~TeV LHC (b) and $YY$ pair production at the 8~TeV (c) and 14~TeV (d). The cross section of $VV(YY)$ productions depends mainly on the absolute value of $\beta$. The cross sections in the $\beta=\pm \sqrt{3}$ models are larger than those in the $\beta=\pm 1/\sqrt{3}$ models. As mentioned above, the gauge couplings of $Z^\prime$ to the SM quarks in the $\beta=\pm \sqrt{3}$ model is larger than the couplings in the $\beta=\pm 1/\sqrt{3}$ models. That yields an enhancement factor of $\sim 10$ to the $VV$ pair production in the case of $\beta=\pm \sqrt{3}$. 
Another enhancement factor originates from the Breit-Wigner resonance effects. Since $M_{Z^\prime}\sim 3 M_{V(Y)}$ for $\beta=\sqrt{3}$, the $Z^\prime$ propagator in Fig.~\ref{vypair}(c) can be produced on mass shell and then exhibit a large Breit-Wigner enhancement. On the other hand, $M_{Z^\prime}\sim 1.2 M_{V(Y)}$ for $\beta=\pm 1/\sqrt{3}$, therefore, the $Z^\prime$ propagator can never be shell, yielding a big suppression in the cross section.

The $YY$ pair production is very similar to the $VV$ pair production. Its production rate is slightly more  than the rate of $VV$ pair for the same value of $\beta$.  The difference arises from Fig.~\ref{vypair}(d), where $V$-pair can only be produced from $d$ quark initial state, while $Y$-pair from $u$ quark initial state due to the $SU(3)$ gauge symmetry. As the parton distribution function of $u$-quarks is larger than that of $d$-quarks, it yields $\sigma(pp\to YY)\gtrsim \sigma(pp\to VV)$.

\subsubsection{Unitarity analysis of $u\bar{u} \to YY$ scattering}

As a gauge invariant set, the unitarity of the $VV$ or $YY$ pair production process must be maintained considering all the four diagrams. Since the $\beta$ dependence comes only from Fig.~\ref{vypair} (a)-(c). The bad high energy behavior that depends on $\beta$ must disappear within Fig.~\ref{vypair} (a)-(c). A simple example is to consider the amplitude of $u(p_1)\bar{u}(p_2)\to Y(p_3)Y(p_4)$. We can expand the amplitude into four terms $A^\gamma$, $A^Z$, $A^{Z^\prime}$ and $A^D$ given by
\begin{eqnarray*}
A^{\gamma} & = & \frac{g_{\gamma YY}g_{\gamma uu}}{s}\bar{v}(p_{2})\{(-\not\not{p}_{3}+\not\not{p_{4}})
\varepsilon^{\ast}(p_{3})\cdot\varepsilon^{\ast}(p_{4})
\\
&  & 
-2p_{4}\cdot\varepsilon^{\ast}(p_{3})\not{\varepsilon^{\ast}(p_{4})}
+2p_{3}\cdot\varepsilon^{\ast}(p_{4})\not{\varepsilon^{\ast}(p_{3})}\} u(p_{1})
\, ,\\
A^{Z} & = & \frac{g_{ZYY}}{s-m_{Z}^{2}}\bar{v}(p_{2})
\{(-\not\not{p_{3}}+\not\not{p_{4}})\varepsilon^{\ast}(p_{3})\cdot\varepsilon^{\ast}(p_{4})\\
&  & -2p_{4}\cdot\varepsilon^{\ast}(p_{3})\not{\varepsilon^{\ast}(p_{4})}+2p_{3}
\cdot\varepsilon^{\ast}(p_{4})\not{\varepsilon^{\ast}(p_{3})}\}(g_L^{Zuu}+g_R^{Zuu})u(p_{1})
\, ,\\
A^{Z^\prime} & = & \frac{g_{Z^\prime YY}}{s-m_{Z^\prime}^{2}}\bar{v}(p_{2})
\{(-\not\not{p_{3}}+\not\not{p_{4}})\varepsilon^{\ast}(p_{3})\cdot\varepsilon^{\ast}(p_{4})\\
&  & -2p_{4}\cdot\varepsilon^{\ast}(p_{3})\not{\varepsilon^{\ast}(p_{4})}+2p_{3}
\cdot\varepsilon^{\ast}(p_{4})\not{\varepsilon^{\ast}(p_{3})}\}(g_L^{Z^\prime uu}+g_R^{Z^\prime uu})u(p_{1})
\, ,\\
A^{D} & = & -\frac{g_{YDu}}{t-M_{D}^{2}}\bar{v}(p_{2})\not{\varepsilon}^{\ast}(p_{4})
P_{L}(\not\not{t}-M_{D})\not{\epsilon^{\ast}(p_{3})}P_{L}u(p_{1}) \, ,
\end{eqnarray*}
where the related couplings can be found in Appendix~\ref{app_a}.
In the center-of-mass frame of $YY$,
the 4-momenta of the particles can be chosen to be
\begin{eqnarray*}
p_{1}  &=&  (E,0,0,E),~~
p_{2}  =  (E,0,0,-E),\\
p_{3}  &=&  (E,\, p\sin\theta,0,p\cos\theta),~~
p_{4}  =  (E,-p\sin\theta,0,-p\cos\theta) \, ,
\end{eqnarray*}
where $E$ is the energy of incoming and outgoing particles, $p$
is the momentum of outgoing heavy gauge bosons and $\theta$ is the scattering
angle. In order to check its high energy behavior, we consider 
the case that both of the outgoing heavy gauge boson $Y$ 
are longitudinally polarized. Since the incoming fermion $u$
and anti-fermion $\bar{u}$ have opposite helicities, the helicity
amplitudes of s-channel and t-channel processes can be easily found
to be
\begin{eqnarray*}
A^{\gamma}(-+) & = & g_{\gamma YY}g_{\gamma uu} \frac{4Ep(p^2-3E^2)}{sM_{V}^{2}}\sin\theta
\, , \\
A^{Z}(-+) & = & g_{ZYY}g_{Zuu}
\frac{4Ep(p^{2}-3E^{2})}{(s-m_{Z}^{2}) M_{V}^{2}}\sin\theta
\, , \\
A^{Z^\prime}(-+) & = & g_{Z^\prime YY}g_{Z^\prime uu}
\frac{4Ep(p^{2}-3E^{2})}{(s-M_{Z^\prime}^{2}) M_{V}^{2}}\sin\theta
\, , \\
A^{D}(-+) & = & g_{YDu}^2 \frac{2E(2E^{3}\cos\theta+p^{3}
-3pE^{2})}{(t-M_{D}^{2})M_{V}^{2}}\sin\theta
\, ,
\end{eqnarray*}
where $(-+)$ are the helicities of
$(u\,\bar{u})$, the Mandelstam variables $s\equiv(p_{1}+p_{2})^{2}$ and
$t\equiv(p_{1}-p_{3})^{2}$. As we take the high energy limit, i.e. $\sqrt{s}\gg M_{X}$
($X=Z$, $Z^\prime$ and
$D$), each amplitude behaves as follows:
\begin{eqnarray}
A^{\gamma}(-+)  &=&  \frac{8(\sqrt{3}\beta+1)e^2}{3M_V^2}s,~~\label{eq:amplitude1}\\
A^{Z}(-+)  &=& \frac{2[1-(\sqrt{3}\beta+1)s_W^2](3-4s_W^2)e^2}{3s_W^2 c_W^2 M_V^2}s,~~\label{eq:amplitude2}
\\
A^{Z^\prime}(-+)  &=& -\frac{2\sqrt{3}[(\beta+\sqrt{3})s_W^2-\sqrt{3}]e^2}{3s_W^2 c_W^2 M_V^2}s,~~\label{eq:amplitude3}
\\
A^{D}(-+) &=& -\frac{4e^2}{s_W^2 M_V^2}s\label{eq:amplitude4}.
\end{eqnarray}
It turns out by summing over the Eq.~\eqref{eq:amplitude1}-\eqref{eq:amplitude3}, one cancels out all the $\beta$ dependent terms and leaves only a term $4e^2s/s_W^2 M_V^2$ which exactly cancels Eq.~\eqref{eq:amplitude4}.

\subsection{Decay of the $V$ and $Y$ bosons}

Now we firstly consider the $V$ boson and study its decay branching ratio. The result can be easily applied to the $Y$ boson. The decay modes of the $V$ bosons can be classified into three categories according to the final state particles. In the first class the $V$ boson decays into a SM fermion and a new heavy fermion. In the second class the $V$  boson decays into a SM gauge boson and a new charged scalar. In the third class the $V$ boson decays into a pair of two scalars.

\begin{figure}
\begin{center}
\includegraphics[width = 0.3\textwidth]{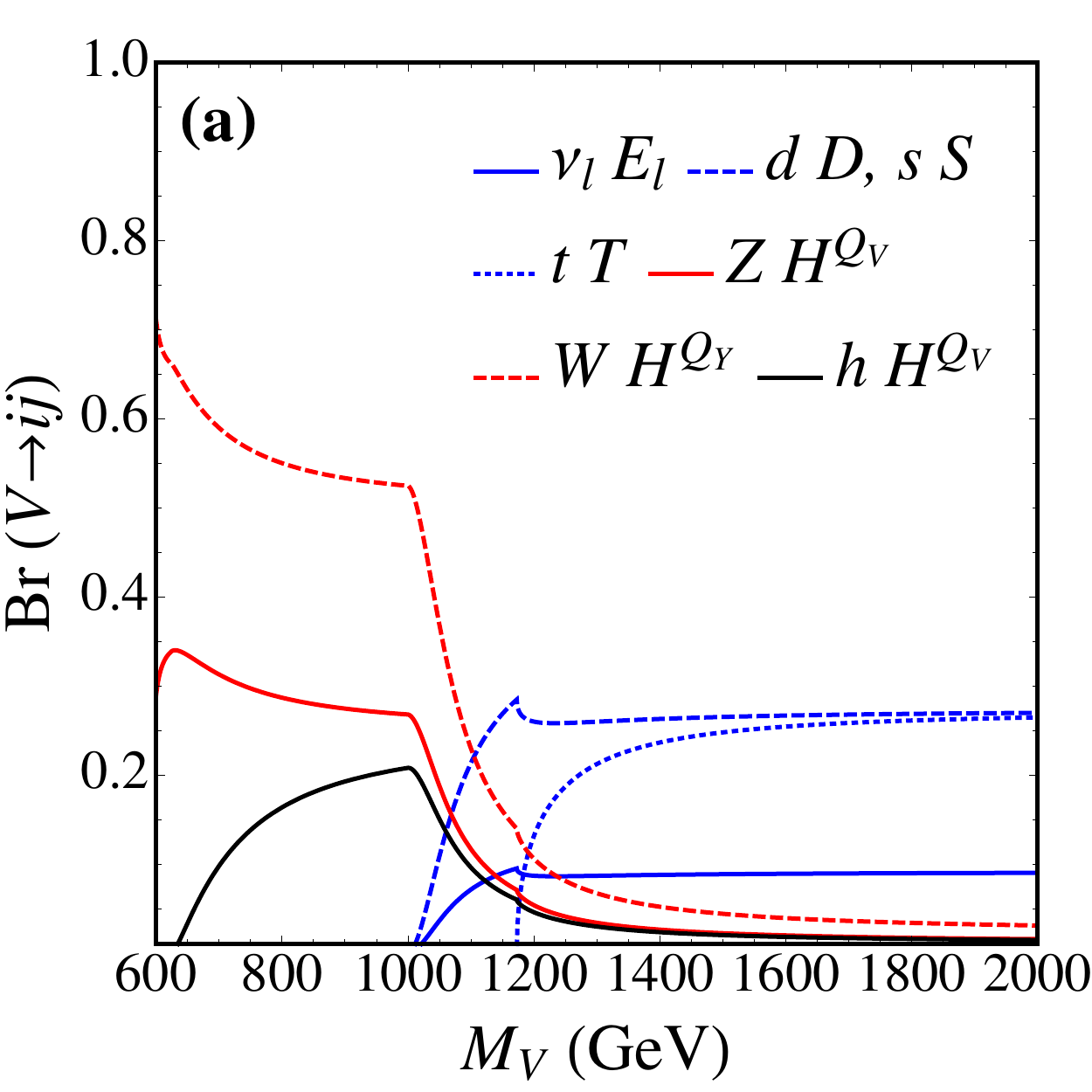}
\includegraphics[width = 0.3\textwidth]{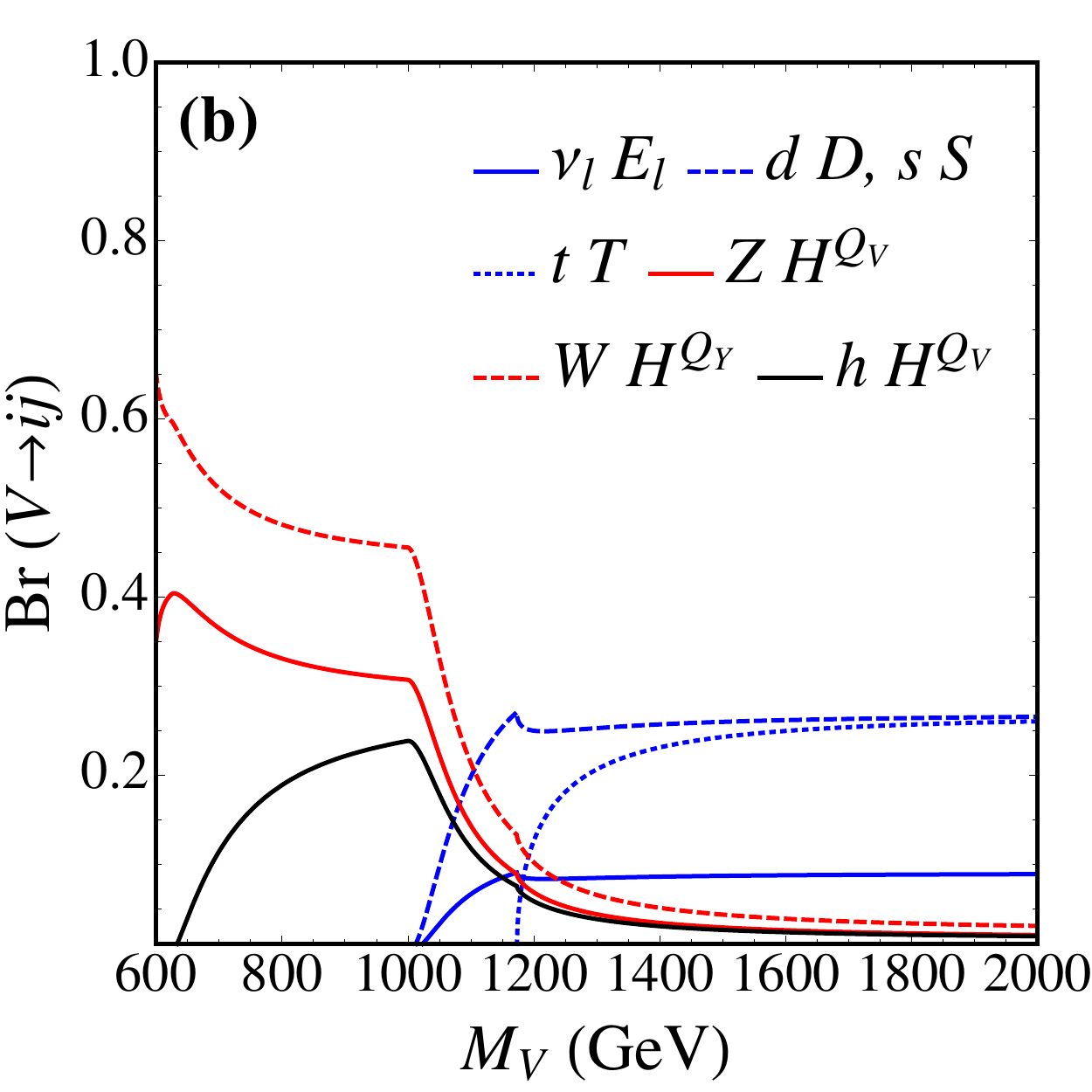}\\
\includegraphics[width = 0.3\textwidth]{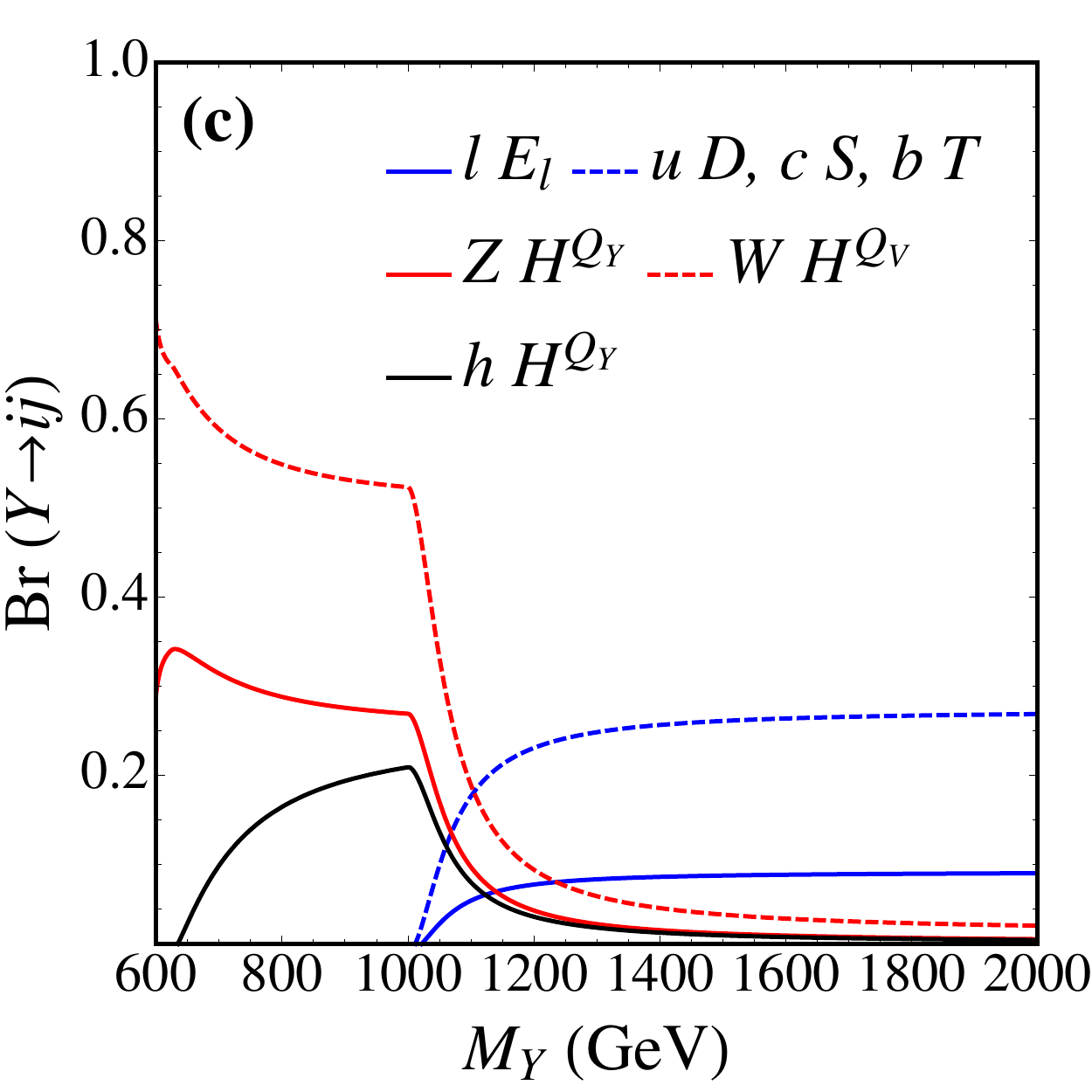}
\includegraphics[width = 0.3\textwidth]{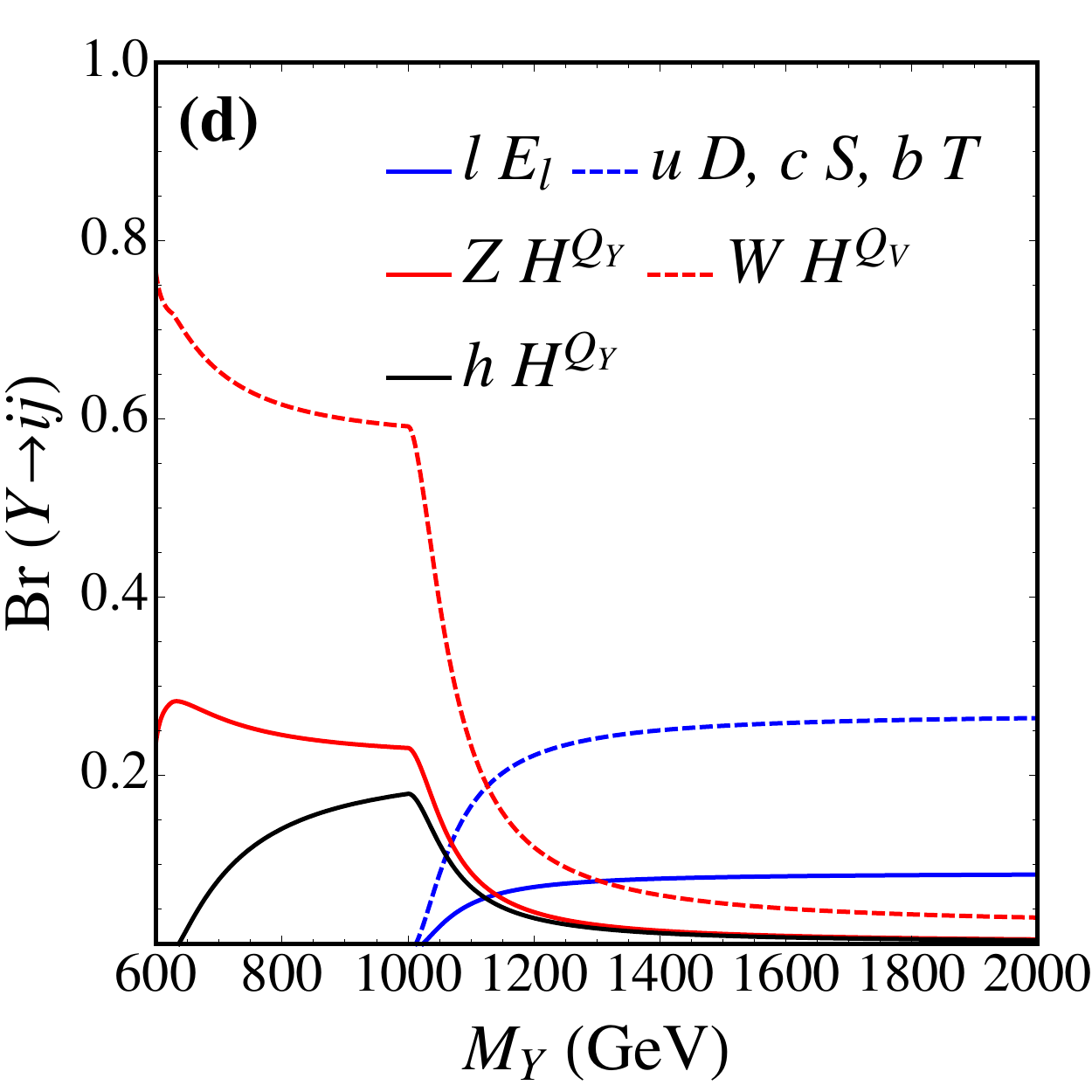}

\caption{\it Decay branching ratios of $V(Y)$ boson versus $M_{V}(M_Y)$ for different choices of $v_{1,2}$ parameters:  (a,c) $v_1=v_2=174{~\rm GeV}$; (b,d) $v_1=200{~\rm GeV}$ and $v_2=143~{\rm GeV}$. The mass of charged scalars is chosen to be {\rm 500~GeV} and the mass of heavy quarks is {\rm 1000~GeV}. }
\label{vbr}
\end{center}
\end{figure}

The decay width of the $V$ boson into a pair of fermions is
\bea
\Gamma(V\to fF)&=&\frac{N_c g^2}{96 \pi  M_V^5}\Big(2 M_V^4-M_F^4-m_f^4- M_V^2m_f^2-M_V^2M_F^2+2 M_F^2 m_f^2 \Big)\nn\\
&\times& \lambda^{1/2}\left(M_V^2, m_f^2, M_F^2\right)
\label{vff}
\eea
where $f$ and $F$ represent the SM and 331 fermion, respectively, $N_c=1$ for leptons and $N_c=3$ for quarks, and the kinematical function $\lambda$ is defined as $\lambda(x,y,z)=(x-y-z)^2-4yz$. Note that the anomaly cancellation condition requires $V\to t T$, therefore, the $tT$ mode is different from the $dD$ and $sS$ modes. 
Figures~\ref{vbr}(a) and \ref{vbr}(b) display the branching ratio of the $V$ boson as a function of $M_V$ for different choices of $v_1$ and $v_2$ parameters. We choose the masses of charged scalars to be 500~GeV while the masses of heavy quarks as 1000~GeV. 

The decay width of the $V$ boson into a SM gauge boson and a new charged scalar is 
\bea
\Gamma(V\to U_i H^{Q_i})&=&\frac{\delta_i g^4 \left(v_3 c_{i 3}+v_i s_{i 3}\right)^2}{3072\pi  M_V^5  c_W^2 m_{U_i}^2}\Big(M_V^4+M_{H^{Q_i}}^4+10 M_V^2 m_{U_i}^2+m_{U_i}^4 \nn\\
&-&2M_V^2 M_{H^{Q_i}}^2-2 M_{H^{Q_i}}^2 m_{U_i}^2\Big)
\times \lambda^{1/2}\left(M_V^2, m_{U_i}^2, M_{H^{Q_i}}^2\right)
\label{vwz}
\eea
where $U_1=Z$, $U_2=W$, $Q_1=Q_V$, $Q_2=Q_Y$, $\delta_1=1$, $\delta_2=2$. We learn from Eqs.~\eqref{sc23} and \eqref{sc13} that $\left(v_3 c_{i 3}+v_i s_{i 3}\right)^2=(v_i v_3)^2/(v_i^2+v_3^2)\sim v_i^2$, therefore, ${\rm Br}(V\to Z H^{Q_V})$ depends on $v_1^2$ while ${\rm Br}(V\to W H^{Q_Y})$ on $2 v_2^2$. Choosing different $v_1$ and $v_2$ would change the breaking ratio pattern dramatically. For example, ${\rm Br}(V\to W H^{Q_Y})\simeq 2{\rm Br}(V\to Z H^{Q_V})$ for choosing $v_1=v_2=174{~\rm GeV}$; see Fig.~$\ref{vbr}$(a). On the other hand, choosing $v_1=200~{\rm GeV}$ and $v_2=143~{\rm GeV}$ yields ${\rm Br}(V\to W H^{Q_Y}) <  2{\rm Br}(V\to Z H^{Q_V})$ shown in Fig.~$\ref{vbr}$(b).

The $V$ boson can also decay into a pair of scalars, e.g. $V\to h H^{Q_V}$, $V\to H_{0,2,3} H^{Q_V}$ and $V\to H^\pm H^{Q_Y}$, if kinematically allowed. If we neglect the small mixing of $\xi_{\chi}$ with $\xi_{\rho}$ and $\xi_{\eta}$, the decay mode $V\to H_3 H^{Q_V}$ is absent. In addition, the decay modes $V\to H_2 H^{Q_V}$, $V\to H_0 H^{Q_V}$ and $V\to H^\pm H^{Q_Y}$ will be suppressed by the phase space factor.  Therefore we only consider $V\to h H^{Q_V}$ for a light $V$ boson. The partial decay width is
\bea
\Gamma(V\to h H^{Q_V})&=&\frac{g^2 U_{11}^2 s_{13}^2}{192 \pi  M_V^5 }\Big(M_V^4+M_{H^{Q_V}}^4+m_h^4-2 M_V^2 M_{H^{Q_V}}^2-2 M_V^2 m_h^2-2 M_{H^{Q_V}}^2 m_h^2\Big)\nn \\ 
&\times& \lambda^{1/2}\left(M_V^2, m_h^2, M_{H^{Q_V}}^2\right), 
\label{vh}
\eea
where the approximation of $U_{11}\sim v_1/v$ is used to simplify the calculation. Comparing Eqs.~\ref{vwz} and \ref{vh}, we note that $\Gamma(V\to h H^{Q_V})\sim \Gamma(V\to Z H^{Q_V})$ when $M_V\gg m_h, m_Z$, which is a consequence of the Equivalence Theorem. While $M_V$ is not too large, the contribution from the transverse mode of $Z$ can be important and leads to $ \Gamma(V\to Z H^{Q_V})> \Gamma(V\to h H^{Q_V})$.

Figures~\ref{vbr}(c) and \ref{vbr}(d) shows the decay branching ratio of the $Y$ boson, which is similar to that of $V$ with the only difference is about the $v_1$ dependence of the second and third category.

\subsection{Collider signatures of the $V$ and $Y$ bosons}

Now consider the phenomenology of the $VV$ and $YY$ pair production. 
As mentioned above, there will be a DM candidate or LLP in the 331 models, depending on $\beta$. Table~\ref{tbl:DM} shows the possible DM candidate or LLP in different 331 models. We will discuss collider signatures of $VV$ and $YY$ pair productions. 

Note that the $V$ and $Y$ bosons are almost degenerate and $\big|M_{Y^{\pm Q_Y}}-M_{V^{\pm Q_V}}\big|\ll m_W$; see Eq. \eqref{deltaM_VY}. If the $Y$ boson is the lightest particle, then the $V$ boson can only decay into a $Y$ boson and an off-shell $W$ boson which will further decay into a pair of the SM fermions. It is indeed a three-body decay, i.e. $V\to Y W^* \to Y q\bar{q}^\prime$ or  $V\to Y W^* \to Y \ell^+\nu$. The SM quarks and charged leptons in $V$ boson decay tend to fail the experimental cuts simply because they are not energetic. The same conclusion also applies to the case of the $V$ boson being the lightest particle.  If other particle is the DM candidate, then the $V$ and $Y$ bosons decay in the way as shown in Fig.~\eqref{vbr}.

\renewcommand{\arraystretch}{1.0}
\begin{table}
\centering
\caption{Collider signatures of $VV$ ($YY$) pair productions for $\beta=\pm 1/\sqrt{3}$ and DM candidate, where the star symbol in the superscript denotes the particle being off-shell.}
\label{tbl:DM}
\begin{tabular}{|c|c|c|l|}
\hline
 & DM candidate & Production channel & Collider signature \\
\hline
\multirow{4}{1.6cm}{$\beta=-\frac{1}{\sqrt{3}}$} & \multirow{2}{0.4cm}{$Y$} & $VV$  & $W^{*+}W^{*-} + {E\!\!\!\!/}_T$\\
\cline{3-4}
 &  & $YY+j(\gamma)$  & $j/\gamma+{E\!\!\!\!/}_T$\\
\cline{2-4}
 & \multirow{2}{0.4cm}{$H^{Q_Y}$} & $VV$ & $WW + {E\!\!\!\!/}_T$\\
 \cline{3-4}
 & & $YY$  & $ZZ/hh/Zh +{E\!\!\!\!/}_T$\\
\hline
\multirow{6}{1.6cm}{$\beta=\frac{1}{\sqrt{3}}$} & \multirow{2}{0.4cm}{$V$} & $VV+j(\gamma)$ & $j/\gamma+{E\!\!\!\!/}_T$\\
\cline{3-4}
 & & $YY$ & $W^{*+}W^{*-} + {E\!\!\!\!/}_T$\\
\cline{2-4} 
 & \multirow{2}{0.4cm}{$H^{Q_V}$} & $VV$  & $ZZ/hh/Zh +{E\!\!\!\!/}_T$\\
\cline{3-4}
 & & $YY$  & $WW + {E\!\!\!\!/}_T$\\
\cline{2-4} 
 & \multirow{2}{0.4cm}{$E_\ell$} & $VV$  &  $j/\gamma+{E\!\!\!\!/}_T$\\
\cline{3-4}
 & & $YY$  & $\ell^+ \ell^- + {E\!\!\!\!/}_T$\\
\cline{2-4}
\hline
\end{tabular}
\end{table}

For $\beta=-1/\sqrt{3}$, either the $Y$ boson or the $H^{Q_Y}$ scalar can be a DM candidate. First, consider the $Y$ boson as the DM candidate. The $VV$ pair signal event are produced via  
\beq
pp \to VV \to YYW^{*+}W^{*-} \to YY jjjj,~YY jj \ell^\pm \nu,~YY\ell^+\ell^-\nu\bar{\nu},
\eeq
where $j$ denotes the QCD jet, $\ell^\pm$ the charged leptons, and the star symbol in the superscript denotes the particle being off-shell. As the DM candidate $Y$ boson and neutrinos are not detectable, we end up with a collider signature of 
\beq
jjjj+\met, ~jj \ell^+ +\met,~\ell^+\ell^-+\met.  
\eeq
The $YY$ pair can also be produced through the electroweak interaction, but it is not detectable. In order to trigger the $YY$ pair event, we require additional jet or photon radiation from the initial quark, yielding the so-called mono-jet or mono-photon event as following
\beq
pp \to YY +\gamma(j) \to \gamma (j)+ \met.
\eeq

Second, consider the case that the $H^{Q_Y}$ scalar is the DM candidate.  The $V$ boson can decay into a $H^{Q_Y}$ scalar and an on-shell $W$ boson as the mass difference of $V$ and $H^{Q_Y}$ can be quite large. Hence, the signal event of $VV$ pairs are produced via 
\bea
pp \to VV &\to& H^{Q_Y}H^{Q_Y}W^{+}W^{-} \nn\\
&\to& H^{Q_Y}H^{Q_Y}+jjjj,~H^{Q_Y}H^{Q_Y}+ jj \ell^\pm \nu,~H^{Q_Y}H^{Q_Y}+\ell^+\ell^-\nu\bar{\nu},
\eea
yielding collider signature of 
\beq
jjjj+\met, ~jj \ell^+ +\met,~\ell^+\ell^-+\met. 
\eeq
The event topology is similar to the previous case, but now the jets and charged leptons are much energetic as they arise from on-shell $W$ bosons. For the $YY$ pair production, the signal processes are 
\bea
pp \to YY &\to& H^{Q_Y}H^{Q_Y}ZZ, ~H^{Q_Y}H^{Q_Y}hh, ~H^{Q_Y}H^{Q_Y}Zh, 
\eea
with the subsequent decays of $Z\to jj/\ell^+\ell^-/\nu\bar{\nu}$ and $h\to b\bar{b}$. That yields very rich collider phenomenology.

The collider signature of $\beta=+1/\sqrt{3}$ is similar to that of $\beta=-1/\sqrt{3}$. However, there is one more DM candidate, new charge neutral leptons ($E_\ell$). The $Y$ boson decays into a $E_\ell$ lepton and a $\ell^+$ lepton, therefore, the $YY$ pair production yields a collider signature of 
\beq
pp \to YY \to E_\ell \bar{E}_\ell \ell^+ \ell^- \to \ell^+ \ell^- +\met.
\eeq
On the other hand, the $V$ boson decays into a $E_\ell$ lepton and a neutrino, both of which are invisible in the detector. In such a case one has to rely on the mono-jet or mono-photon from the initial state radiation to trigger the event. It yields a collider signature of 
\beq
pp \to VV + j (\gamma) \to E_\ell \nu\bar{E}_\ell \nu + j(\gamma) \to j(\gamma) + \met.
\eeq
Table \ref{tbl:DM} summarizes the possible DM candidates and the corresponding interesting collider signature of $VV$ and $YY$ pair productions in the $\beta=\pm 1/\sqrt{3}$ models.

\renewcommand{\arraystretch}{1.0}
\begin{table}\centering
\caption{Collider signature of $VV$ ($YY$) pair production for $\beta=\pm \sqrt{3}$ and lightest particle, where $R_h$ denotes $R$-hadron and can be found in Table~\ref{tbl:Rhadron}.}
\label{tbl:LLP}
\begin{tabular}{|c|c|c|c|}
\hline
 & Lightest particle & Production channel & Collider signature \\
\hline
\multirow{4}{1.6cm}{$\beta=\pm\sqrt{3}$} & \multirow{2}{1cm}{$D(S)$} & $VV$ & $R_hR_h+2j$\\
\cline{3-4}
 &  & $YY$ & $R_h R_h+2j$\\
\cline{2-4}
 & \multirow{2}{0.4cm}{$T$} & $VV$ & $R_hR_h+t\bar{t}$\\
 \cline{3-4}
 & & $YY$ & $R_hR_h+b\bar{b}$\\
\hline
\end{tabular}

\end{table}

In the $\beta=\pm\sqrt{3}$ models, there is no DM candidate but long-live charged particles (LLPs). It leads to quite interesting phenomenology at colliders. For example,  consider one of the new heavy quarks ($D, S, T$) as the LLP. If it has a mean lifetime longer than the typical hadronization time scale, the LLP might form QCD bound states with partons (quarks and/or gluons) in analogy with the ordinary hadrons. Such an exotic phenomenon is usually referred as $R$-hadron as it is often used in searches of $R$-parity violating or splitting supersymmetry models.
Searches for a long-lived gluino, stop or sbottom have been carried out by both the ATLAS and CMS collaborations~\cite{Beringer:1900zz}. 
Table~\ref{tbl:Rhadron} lists many kinds of $R$-hadrons for different LLPs in the $\beta=\pm \sqrt{3}$ models. 
Needless to say, the normal case would be that only one of them will be long-lived enough to form $R$-hadrons.

Depending on its lifetime and hadronization models, an $R$-hadron generates very diverse signature that are experimentally accessible, e.g. by the secondary vertex, or energy loss, charge exchanges, etc. In Table~\ref{tbl:LLP}, we list possible collider signatures of $VV$ ($YY$) pair production consisting of $R$-hadrons.

\section{Phenomenology of Heavy quarks}\label{sec:sec6}
\subsection{Production of Heavy quarks}

\begin{figure}[h!]
\centering
\includegraphics[width=0.7\textwidth]{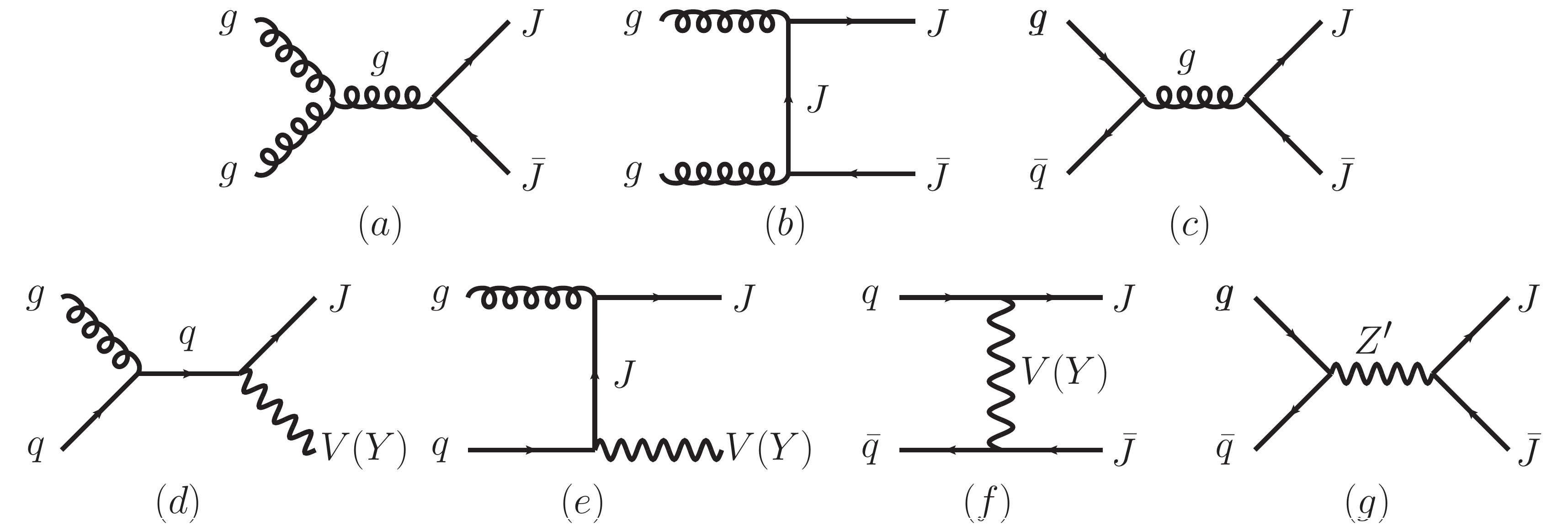}
\caption{\it Representative Feynman diagrams for heavy quark $J=D, S, T$ production: (a, b, c) pair production through the QCD interaction (a, b, c); (d) and (e) single production associated with another heavy gauge boson;  (f, g) pair production through the electroweak interation.}\label{hap}
\end{figure}

At the LHC, heavy quarks can be produced in pair or singly in the 331 models. Figure~\ref{hap} displays three kinds of processes of heavy quark productions. First, heavy quarks can be produced in pair through the QCD interaction which is the predominant production channel of heavy quarks; see Figs.~\ref{hap}(a, b, c). Second, heavy quark can also be produced in association with a heavy gauge boson; see Figs.~\ref{hap}(d, e). Third, heavy quarks can be produced in pair through electroweak interaction. Different from the QCD channel, the electroweak channel can produced two heavy quarks with different flavors; see Figs.~\ref{hap}(f, g). 

The $V$-$J$-$q$ and $Y$-$J$-$q$ couplings are both equal to $g/\sqrt{2}$ which can be neglected comparing to QCD coupling $g_s$. Hence, the single heavy quark production is less important, compared to the QCD induced heavy quark pair production. Similarly, the diagram (f) gives rise to the smallest production rate. However, the diagram (g) is different from others. The couplings of $Z^\prime$ to the heavy quarks could be large, therefore, the electroweak process through a $Z^\prime$ boson is not negligible. Hereafter we consider the heavy quark pair production through the QCD interaction and the $Z^\prime$ mediated electroweak process.

\begin{figure}
\centering
\includegraphics[width=0.3\textwidth]{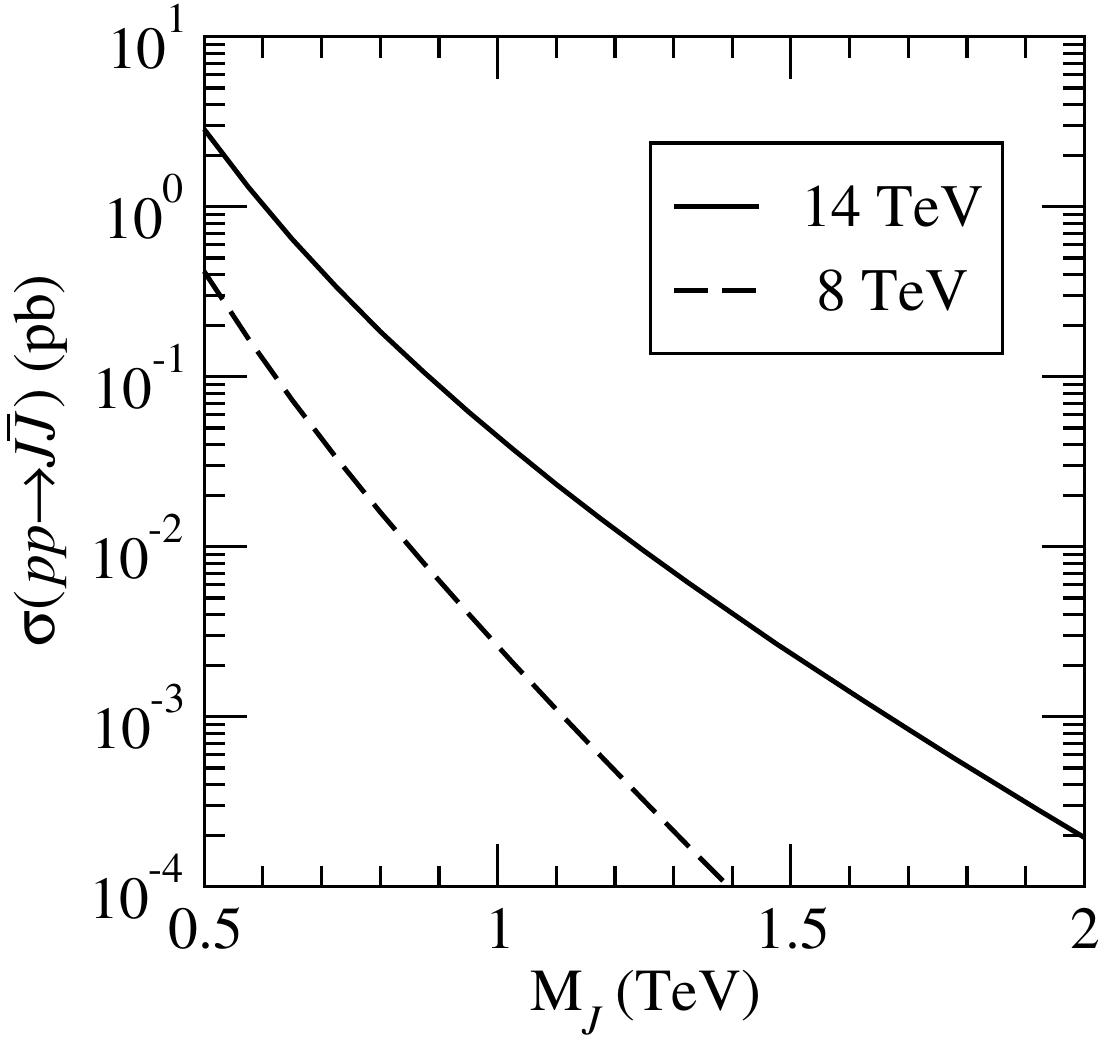}
\caption{\it The cross section of the $J\bar{J}$ ($J=D, S, T$)  pair production through the QCD interaction as a function of $M_J$ at the 8~TeV and the 14~TeV LHC.}
\label{QCD}
\end{figure}

\begin{figure}
\centering
\includegraphics[width=0.6\textwidth]{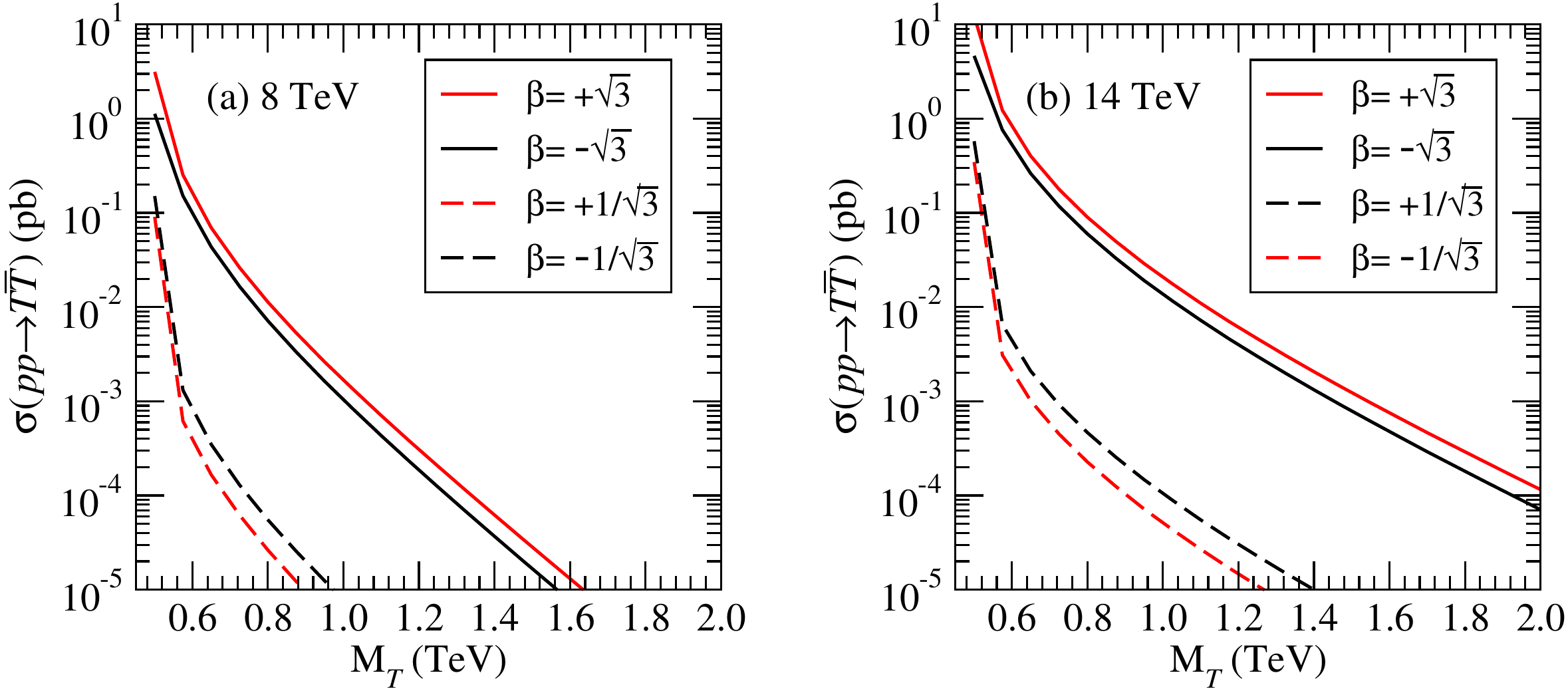}
\includegraphics[width=0.6\textwidth]{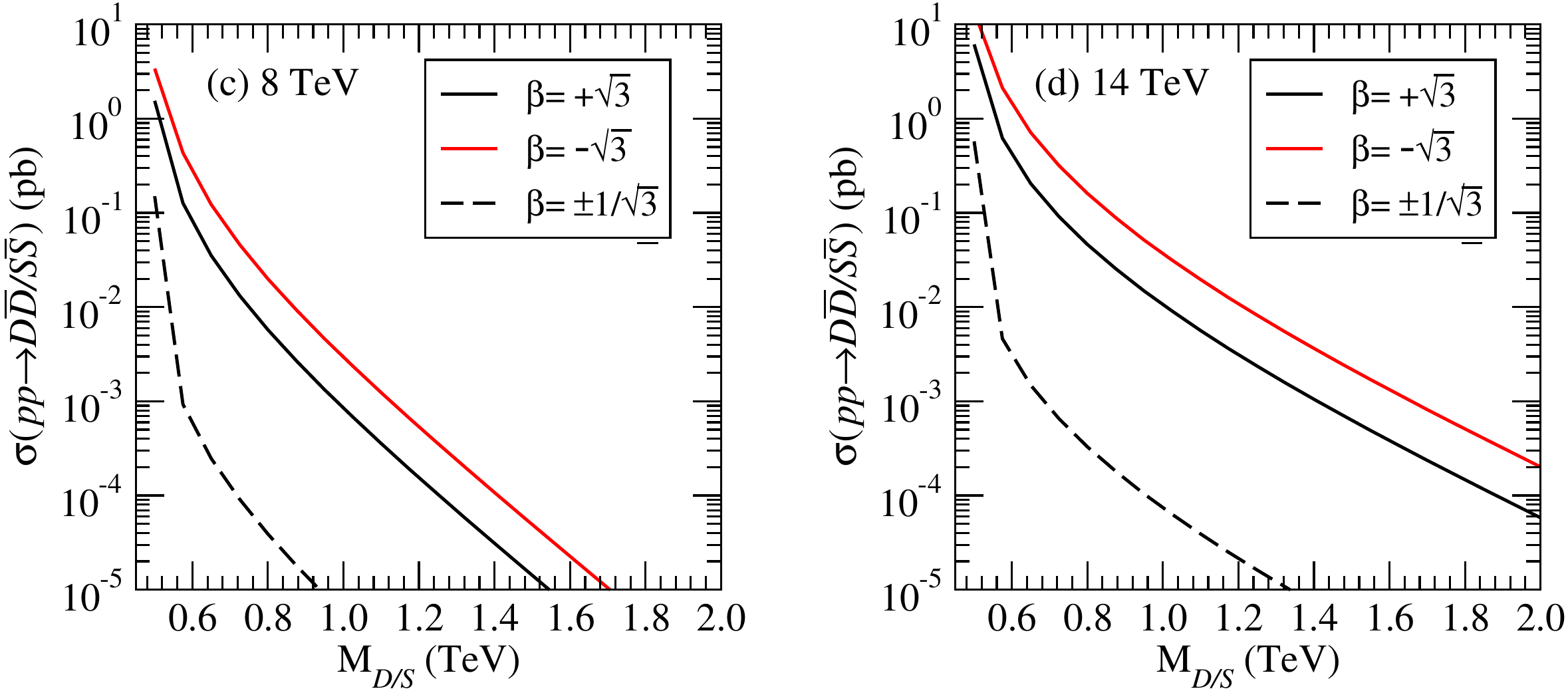}
\caption{\it The cross section of the $T\bar{T}$ pair production (a, b) and $D\bar{D}/S\bar{S}$ pair production (c, d) through the electroweak interaction in the 331 model at the 8~TeV LHC and the 14~TeV LHC, respectively. Here, we choose $M_{Z^\prime}=1\tev$ for which the $Z^\prime$ can decay only into the SM fermions.}
\label{EWppTT}
\end{figure}

There is no interference between the QCD channel and the $Z^\prime$ channel. We consider them separately below. Figure~\ref{QCD} displays the cross sections of heavy quark pair productions via the QCD interactions. Owing to the large coupling of the QCD interaction, the heavy quark pair can be produced copiously at the 8~TeV and 14~TeV. Figure~\ref{EWppTT} shows the production cross sections of $T\bar{T}$ pairs (a, b) and $D\bar{D}/S\bar{S}$ pairs (c, d) through a $Z^\prime$ mediation in the 331 models, respectively, where we fix $M_{Z^\prime}=1\tev$. The production cross section is negligible for $\beta=\pm 1/\sqrt{3}$, but it can be comparable with the QCD channel for $\beta=\pm\sqrt{3}$, thanks to the large electroweak couplings of $Z^\prime$ to both the SM quarks and heavy quarks.

\subsection{Decay of heavy quarks}

The decay modes of heavy quarks can be classified into two categories according to final state particles. One mode, named as gauge-boson mode, involves a new gauge boson and a SM quark. The other mode involves a 331 charged Higgs boson and a SM quark, named as charged Higgs mode. We take $T$ quark as an example to illustrate its decay branching ratio.

\begin{figure}\centering
\includegraphics[width=0.3\textwidth]{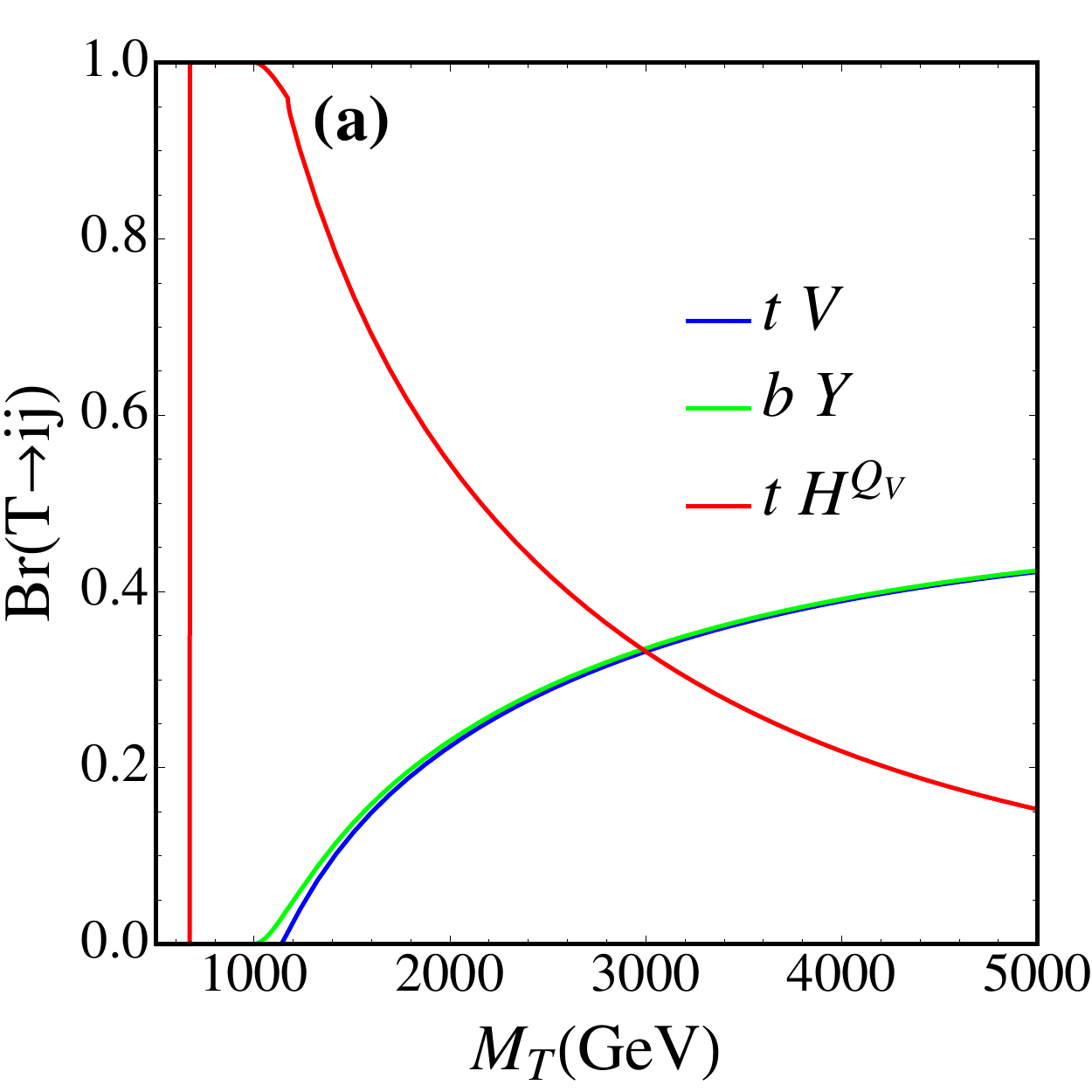}
\includegraphics[width=0.3\textwidth]{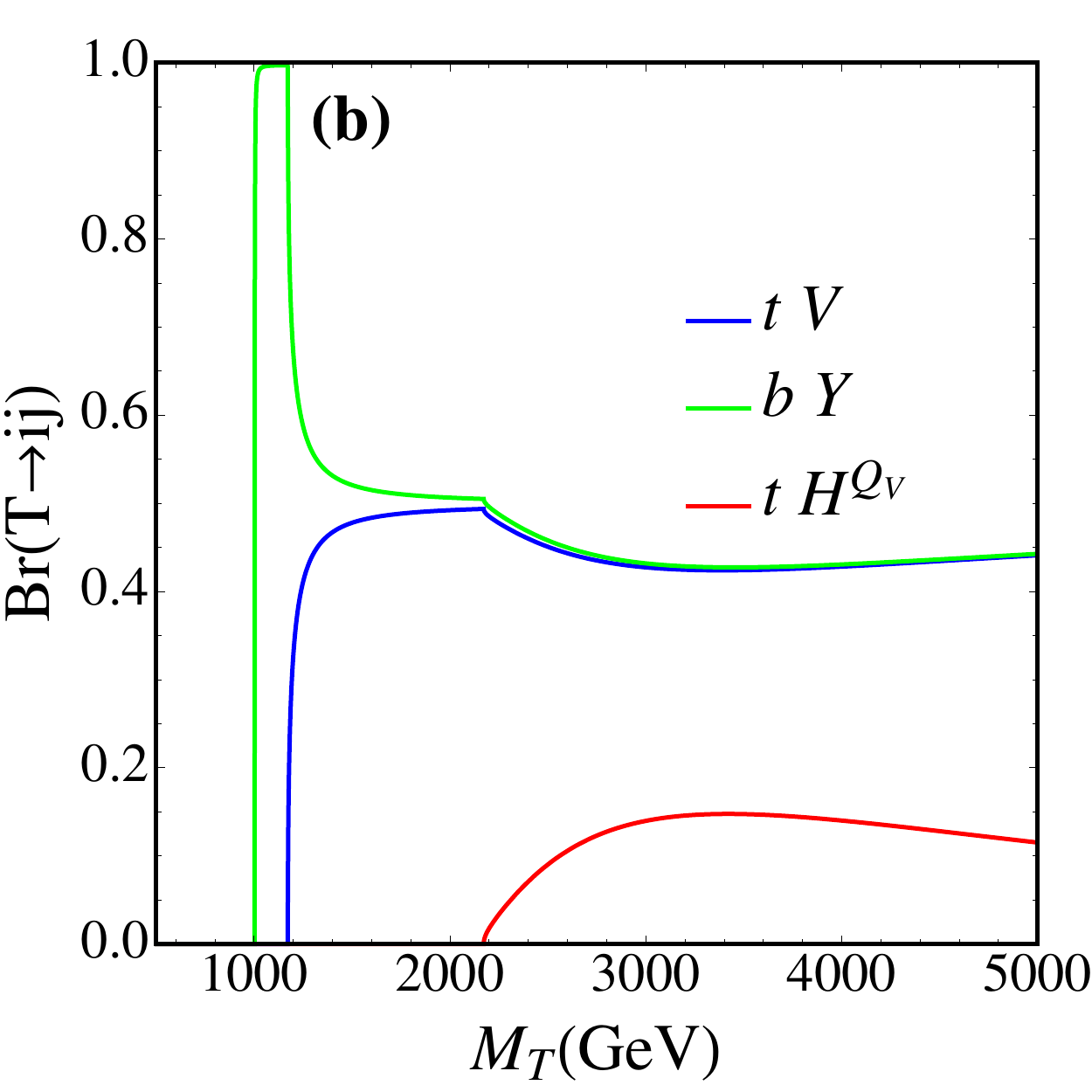}
\includegraphics[width=0.3\textwidth]{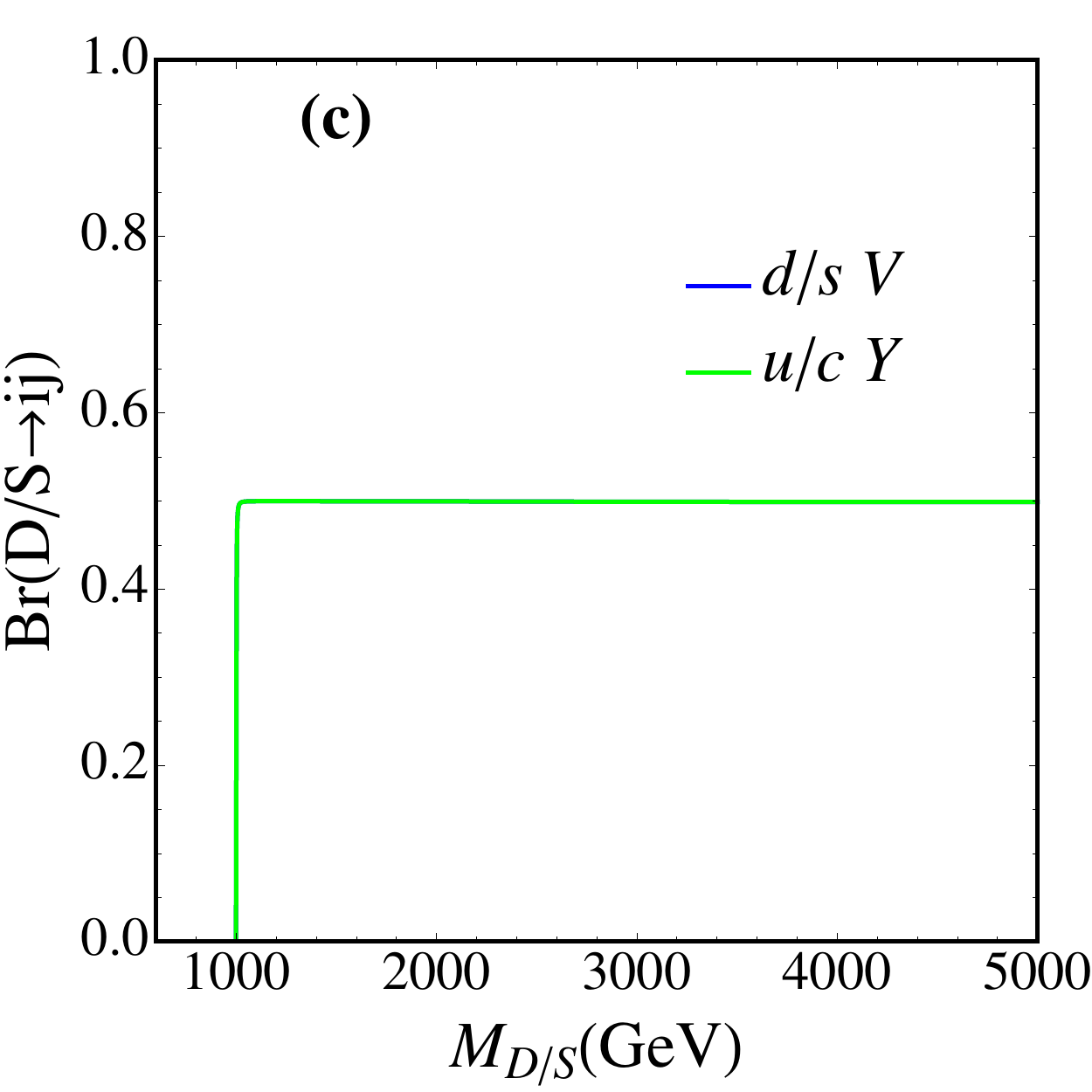}
\caption{\it Decay branching ratios of $T$ with respect to $M_T$ for (a)  $M_{V,Y}=1000\gev$ and $M_{H^{Q_V}}=500\gev$ and (b) $M_{V,Y}=1000\gev$ and $M_{H^{Q_V}}=2000\gev$; decay branching ratios of $D/S$ with respect to $M_{D/S}$ for $M_V=M_Y=1000\gev$.
}\label{Tbr}
\end{figure}

The partial decay width of $T$ quark in the gauge-boson mode is 
\bea
\Gamma(T\to q_i U_i)&=&\frac{g^2}{64 \pi  M_T^3 M_{U_i}^2}\left[\left(M_T^2-m_{q_i}^2\right)^2+m_{q_i}^2 M_{U_i}^2+M_T^2 M_{U_i}^2-2 M_{U_i}^4\right]\\ \nn
&\times & \lambda^{1/2}\left(M_T^2, m_{q_i}^2, M_{U_i}^2\right)
\eea
where $q_1=t$, $q_2=b$, $U_1=V$ and $U_2=Y$, respectively. We learn from Eq.~$\eqref{deltaM_VY}$ that  $|M_Y-M_V|\ll m_W$ and one can ignore the mass difference of $Y$ and $V$  bosons. As a result, $\Gamma(T\to bY)>\Gamma(T\to tV)$ for a light $T$-quark owing to the phase space suppression from the top-quark. For a heavy $T$ quark, $\Gamma(T\to bY)\simeq \Gamma(T\to tV)$.

In the charge-Higgs mode, the partial decay width of $T$ quark is 
\bea
\Gamma(T\to q_i H^{Q_i})&=&\frac{1}{16 \pi  M_T^3}
\left[\left(M_T^2-M_{H^{Q_i}}^2+m_{q_i}^2\right) \left(\frac{c_{i 3}^2 M_T^2}{v_3^2}+\frac{s_{i 3}^2 m_{q_i}^2 }{v_i^2}\right)+\frac{4 M_T^2 m_{q_i}^2 s_{i 3}c_{i 3}}{v_i v_3}\right]\nn \\
&\times&\lambda^{1/2}\left(M_T^2, m_{q_i}^2, M_{H^{Q_i}}^2\right),
\eea
where $q_1=t$, $q_2=b$, $Q_1=Q_V$ and $Q_2=Q_Y$. In the limit $v_3\gg v_1$, $c_{i 3}\to 0$ and $s_{i 3}\to 1$, yielding the partial width proportional to the SM quark mass.  Therefore, in all mass region of $T$, $\Gamma(T\to tH^{Q_V})\gg \Gamma(T\to bH^{Q_Y})$ and we can neglect the $bH^{Q_Y}$ mode.

The decay modes compete with each other, and their effects highly rely on the choice of $M_{V}$, $M_{Y}$ and $H^{Q_V}$. 
For a heavy $T$ quark, $\Gamma(T\to q_i U_i)$ is enhanced by $M_T^2/M_{U_i}^2$, which can be understood from the Goldstone equivalence theorem, however, such an enhancement is absent for a light $T$ quark such that 
$\Gamma(T\to q_i U_i)$ and $\Gamma(T\to tH^{Q_V})$ are comparable. Figure~\ref{Tbr} displays the branching ratio of $T$ quark versus $M_T$ for two benchmark scenarios: (a) $M_{V,Y}=1000~{\rm GeV}$, $M_{H^{Q_V}}=500~{\rm GeV}$, and (b) $M_{V,Y}=1000\gev$, $M_{H^{Q_V}}=2000\gev$. In the first benchmark scenario, as the gauge-boson modes are forbidden or highly suppressed by the phase space, a light $T$ quark mainly decays into a pair of $t$ and $H^{Q_V}$; see Fig.~\ref{Tbr}(a). In the second scenario, the charge-Higgs mode is forbidden in the light $T$ quark decay, therefore, the gauge-boson mode dominates.

Consider the $D$ and $S$ quarks. The charge-Higgs mode is negligible, therefore, the $D$ quark decays into either $uY$ or $dV$ pairs at half of the time, so does the $S$ quark. See Fig.~\ref{Tbr}.

\subsection{Searching for Heavy quarks at the LHC}

Heavy quark pair production have a very rich collider phenomenology. As a colored object, the heavy quark cannot be the DM candidate. There are a DM candidate or LLP in the decay products of heavy quarks. Table~\ref{tbl:JDM} shows the possible DM candidate, production channel and the corresponding collider signature in the $\beta=\pm 1/\sqrt{3}$ models. 

\subsubsection{$D\bar{D}$ ($S\bar{S}$) pair production}

Consider the $D\bar{D}$ pair production. The result is also valid to the $S\bar{S}$ pair production. The $D$ quark decays into $uY$ or $dV$ pairs at half of the time. First, consider the case that the $Y$ boson is the DM candidate in the $\beta=-1/\sqrt{3}$ model. The $D$ quark exhibit two decay chains as follows:
\bea
D&\to& uY,\nn\\
D&\to & dV \to  dY W^*\to dY jj, ~dY\ell^\pm \nu.
\eea 
Owing the degeneracy of $M_{V}$ and $M_{Y}$, the charged leptons or jets in the decay chain of $D\to dV$ are often too soft to pass experimental cuts and contribute to missing transverse momentum in analogy with the DM candidate $Y$ and neutrinos. Hence, both decay chains of $D$ quark can be treated as the $D$ quark decays into a quark and an invisible particle. Hereafter, we will not distinguish the two chains and treat the $D$ quark completely decays into a quark and a DM candidate (an invisible particle). Consequently, the $D\bar{D}$ pair exhibit a collider signature of
\beq
pp \to D \bar{D} \to j j +\met.
\eeq
The same result also holds for the case of the $V$ boson being the DM candidate in the $\beta=+1/\sqrt{3}$ model. 

Second, consider $H^{Q_Y}$ as the DM candidate. In the $\beta=-1/\sqrt{3}$ model, the cascade decay chains of the $D$ quark are given as follows: 
\bea
D & \to & uY \to u +Z H^{Q_Y},~u+ hH^{Q_Y},\nn\\
D & \to & dV \to d +W H^{Q_Y}.
\eea
Therefore, the $D\bar D$ pair have rich collider signatures as follows:
\bea
pp \to  D\bar{D} &\to& u \bar{u}YY, ~u\bar{d} YV, ~d\bar{d}VV\nn\\
&\to & u\bar{u} + ZZ H^{Q_Y}H^{Q_Y},~u\bar{u} + hh H^{Q_Y}H^{Q_Y},~u\bar{u} + Zh H^{Q_Y}H^{Q_Y},\nn\\
&& u\bar{d} + ZW H^{Q_Y}H^{Q_Y}, u\bar{d} + hW H^{Q_Y}H^{Q_Y}+h.c.\nn\\
&& d\bar{d} + WW H^{Q_Y}H^{Q_Y}.
\eea
The $\beta=+1/\sqrt{3}$ model with $H^{Q_V}$ being the DM candidate also has exactly the same signatures.

Third, consider the $E_\ell$ lepton as the DM candidate in the $\beta=+1/\sqrt{3}$ model. The decay chains of the $D$ quark are
\bea
D &\to& u Y \to u \ell^+ E_\ell, \nn\\
D &\to& d V \to d \nu E_\ell.
\eea
That yields the following collider signatures of the $D\bar{D}$ pair
\bea
pp \to D\bar{D} &\to& u\bar{u} Y^{Q_Y}Y^{-Q_Y}, ~u \bar{d} Y^{Q_Y} V^{-Q_V}, ~\bar{u} d Y^{-Q_Y} V^{Q_V}, ~d\bar{d} V^{Q_V} V^{-Q_V}\nn\\
&\to & u\bar{u}\ell^+\ell^- E_\ell \bar{E}_\ell, ~u\bar{d}\ell^+\nu E_\ell \bar{E}_\ell, ~\bar{u}d \ell^-\bar{\nu} E_\ell \bar{E}_\ell, ~d\bar{d}\nu\bar{\nu} E_\ell \bar{E}_\ell,\nn\\
&\to& jj \ell^+\ell^- +\met, ~jj \ell^++\met, ~jj \ell^-+\met, ~jj +\met.
\eea

Last, consider the $\beta=\pm\sqrt{3}$ models which have no DM candidate but LLP. Likewise, we focus on the $R$-hadron phenomenon. Table~\ref{tbl:JLLP} display the LLP candiate in the model and the corresponding signature of heavy quark pair production. 

If the $D$ quark is the LLP, the $D\bar{D}$ pair contribute to two $R$-hadrons in the detector. On the other hand, the $S\bar{S}$ pair have a rather long decay chain as following: 
\bea
pp \to S\bar{S} &\to& c \bar{c} ~Y^{Q_Y} Y^{-Q_Y} \to c\bar{c}  u \bar{u} D \bar{D},
\eea
yielding a signature of $jjjj + \met$. If the $T$ quark is the LLP, the $D$ quark decay chains are
\bea
D \to u Y \to u b T, ~~ D\to d V \to d t T.
\eea
Therefore, the $D\bar{D}$ pair events are produced via 
\bea
pp \to D\bar{D} &\to& u \bar{u} b\bar{b} T\bar{T},~u\bar{d} b \bar{t} T\bar{T}, ~d\bar{u}\bar{b}tT\bar{T},~d\bar{d} t\bar{t} T\bar{T}\nn\\
&\to& jj + R_h R^\prime_h + t\bar{t}/b\bar{b}/t\bar{b}/\bar{t}b.
\eea

\subsubsection{$T\bar{T}$ pair production}

The $T\bar{T}$ pair production is different from the $D\bar{D}$ and $S\bar{S}$ pair production. The $T$ quark decay into $bY$ and $tV$. First, consider the case that the $Y$ boson is the DM candidate in the $\beta=-1/\sqrt{3}$ model. The $T\bar{T}$ pair events are produced via 
\bea
pp \to T\bar{T} &\to& b\bar{b} Y Y \to b\bar{b} +\met. 
\eea 
In the $\beta=+1/\sqrt{3}$ model with the $V$ boson being the DM candidate, the $T\bar{T}$ pair production has the following signature
\beq
pp \to T\bar{T} \to t \bar{t} VV \to t\bar{t} +\met. 
\eeq

\renewcommand{\arraystretch}{1.0}
\begin{table}
\caption{DM candidate in the $\beta(\pm 1/\sqrt{3})$ models and collider signature of $DD$ ($SS$, $TT$) pair production.}
\label{tbl:JDM}
\centering
\begin{tabular}{|c|c|c|l|}
\hline
 & DM candidate & Production & Collider signature \\
\hline
\multirow{4}{1.6cm}{$\beta=-\frac{1}{\sqrt{3}}$} & \multirow{2}{0.4cm}{$Y$} & $DD$,~$SS$ & $2j + {E\!\!\!\!/}_T$\\
\cline{3-4}
 &  & $TT$ & $b\bar{b}+{E\!\!\!\!/}_T$\\
\cline{2-4}
 & \multirow{3}{0.4cm}{$H^{Q_Y}$} & $DD$,~$SS$ & $2j+WW/2Z/2h/WZ/Wh/Zh + {E\!\!\!\!/}_T$\\
 \cline{3-4}
 & & \multirow{2}{*}{$TT$} & $t\bar{t}+WW +{E\!\!\!\!/}_T,~b\bar{b}+2Z/2h/Zh+{E\!\!\!\!/}_T$,\\
 & & & $t\bar{b}/\bar{t}b+WZ/Wh+{E\!\!\!\!/}_T$\\
\hline
\multirow{7}{1.6cm}{$\beta=+\frac{1}{\sqrt{3}}$} & \multirow{2}{0.4cm}{$V$} & $DD$,~$SS$ & $2j+{E\!\!\!\!/}_T$\\
\cline{3-4}
 & & $TT$ & $t\bar{t}+{E\!\!\!\!/}_T$\\
\cline{2-4} 
 & \multirow{4}{0.4cm}{$H^{Q_V}$} & $DD$,~$SS$ & $2j+WW/2Z/2h/WZ/Wh/Zh+{E\!\!\!\!/}_T$\\
\cline{3-4}
 & & \multirow{3}{*}{$TT$} & $t\bar{t}+{E\!\!\!\!/}_T,~b\bar{b}+WW + {E\!\!\!\!/}_T$,\\
 & & & $t\bar{t}+2Z/2h/Zh/Z/h+{E\!\!\!\!/}_T$,\\
 & & & $t\bar{b}/\bar{t}b+WZ/Wh/W+{E\!\!\!\!/}_T$\\
 \cline{2-4}
 & \multirow{2}{0.4cm}{$E_\ell$} & $DD$,~$SS$ & $2j+{E\!\!\!\!/}_T,~2j+\ell^+ \ell^- + {E\!\!\!\!/}_T$, $2j+\ell^\pm +\met$ \\
\cline{3-4}
 & & $TT$ & $t\bar{t}+{E\!\!\!\!/}_T,~b\bar{b}+\ell^+\ell^- +{E\!\!\!\!/}_T$,~$t\bar{b}\ell^-(\bar{t}b\ell^+) +\met$\\
\cline{2-4}
\hline
\end{tabular}
\end{table}

Second, consider $H^{Q_Y}$ as the DM candidate. In the $\beta=-1/\sqrt{3}$ model, the cascade decay chains of the $T$ quark are  
\bea
T & \to & t V, ~~V\to WH^{Q_Y}, \nn\\
T & \to & b Y, ~~Y \to Z(h) H^{Q_Y},\nn\\
T &\to & t H^{Q_V}, ~~H^{Q_V} \to W H^{Q_Y}.
\eea
That yields the following signature of the $T\bar T$ pair production:
\bea
pp \to  T\bar{T} &\to& t \bar{t}VV, ~b\bar{b} YY, ~t\bar{b}VY, ~t\bar{t} VH^{-Q_V}\nn\\
&\to & t\bar{t} + WW H^{Q_Y}H^{Q_Y},~b\bar{b} + hh H^{Q_Y}H^{Q_Y},~b\bar{b} + ZZ H^{Q_Y}H^{Q_Y},~b\bar{b} + Zh H^{Q_Y}H^{Q_Y},\nn\\
&& t\bar{b} + WZ H^{Q_Y} H^{Q_Y},~ t\bar{b} + Wh H^{Q_Y} H^{Q_Y}\nn\\
&\to& t\bar{t} + WW +\met, ~b\bar{b} + hh/ZZ/Zh + \met, ~t\bar{b}/\bar{t}b+ WZ/Wh +\met.
\eea
Note that the $\beta=+1/\sqrt{3}$ model with $H^{Q_V}$ being the DM candidate exhibits similar signatures. In such a case the decay chains of $T$ quark are
\bea
&& T \to t V, ~V \to hH^{Q_V}, ZH^{Q_V}\nn\\
&& T \to bY, ~Y \to WH^{Q_V}, \nn\\
&& T \to t H^{Q_V},
\eea
which give rise to the signature of $T\bar T$ pair production as follows:
\bea
pp \to T\bar{T} &\to& t \bar{t} H^{Q_V} H^{Q_V},~b\bar{b}YY,~t\bar{t}VV, t\bar{t} V H^{Q_V}, t\bar{b} VY+h.c.\nn\\
&\to& t\bar{t}+H^{Q_V} H^{Q_V}, b\bar{b}WW+H^{Q_V} H^{Q_V},~t\bar{b} WZ/Wh/Z/h+H^{Q_V} H^{Q_V}, \nn\\
&& t\bar{t}ZZ/Zh/hh/Z/h + H^{Q_V} H^{Q_V}\nn\\
&\to& t\bar{t} +\met, ~b\bar{b} + \met, ~t\bar{b}/b\bar{t} + WZ/Wh/Z/h + \met, t\bar{t}+ZZ/Zh/hh/Z/h +\met.\nn
\eea

Third, consider the $E_\ell$ lepton as the DM candidate in the $\beta=+1/\sqrt{3}$ model. The decay chains of the $T$ quark are
\bea
T &\to& t V \to t \nu E_\ell, \nn\\
T &\to& b Y \to b \ell^+ E_\ell.
\eea
That yields the following collider signatures of the $T\bar{T}$ pair
\bea
pp \to T\bar{T} &\to& t\bar{t} E_\ell \bar{E}_\ell\nu\bar{\nu}, ~b\bar{b} \ell^+ \ell^- E_\ell\bar{E}_\ell, ~t\bar{b}\ell^-\bar{\nu}E_\ell\bar{E}_\ell, ~~\bar{t}b\ell^+\nu E_\ell\bar{E}_\ell,\nn\\
&\to& t\bar{t} +\met, b\bar{b} \ell^+\ell^- +\met, t\bar{b}+\ell^- +\met, \bar{t}b+\ell^++\met.
\eea

Last but not the least, consider the $\beta=\pm\sqrt{3}$ models which only have the lightest charged particle. Likewise, we focus on the $R$-hadron phenomenon. Table~\ref{tbl:JLLP} display the LLP candiate in the model and the corresponding signature of heavy quark pair production. If the $T$ quark is the LLP, the $T\bar{T}$ pair contribute to two $R$-hadrons in the detector. If the $D$ quark is the LLP,  the $T\bar{T}$ pair have a rather long decay chain as following: 
\bea
pp \to T\bar{T} &\to & t \bar{t} ~V^{Q_V} V^{-Q_V}, ~b\bar{b} Y^{Q_Y} Y^{-Q_Y}\nn\\
&\to & t\bar{t} d \bar{d} + D\bar{D},~b\bar{b} u \bar{u}+ D\bar{D} \nn\\
&\to& t\bar{t}+jj + R_h R_h, ~b\bar{b}+jj+R_h R_h.
\eea

\renewcommand{\arraystretch}{1.0}
\begin{table}
\caption{The LLP candidate in the $\beta=\pm \sqrt{3}$ models and collider signature of $DD$, $SS$, $TT$ pair productions, where $R_h$ labels $R$-hadron shown in Table~\ref{tbl:Rhadron}.}
\label{tbl:JLLP}
\centering
\begin{tabular}{|c|c|c|c|}
\hline
 & Lightest particle & Production process & Collider signature \\
\hline
\multirow{5}{1.6cm}{$\beta=\pm\sqrt{3}$} & \multirow{3}{1cm}{$D(S)$} & $DD(SS)$ & $R_h R_h$\\
\cline{3-4}
 &  & $S(D)$-pair & $R_h R_h+4j$\\
\cline{3-4}
 &  & $T$-pair & $R_h R_h+2j+t\bar{t}/b\bar{b}$\\
\cline{2-4}
 & \multirow{2}{0.4cm}{$T$} & $D,~S$-pair & $R_h R_h+2j+t\bar{t}/b\bar{b}/t\bar{b}/\bar{t}b$\\
 \cline{3-4}
 & & $T$-pair & $R_h R_h$\\
\hline
\end{tabular}

\end{table}

\section{Conclusions}\label{sec:sec7}

In this work, we have studied the different versions of the so-called 331 model based on the gauge group $SU(3)_C\otimes SU(3)_L\otimes U(1)_X$. The spontaneous symmetry breaking is realized by three Higgs triplets with two steps. Assuming the energy scale of the first step is much higher than the electroweak scale, the 331 model restores to the SM at the low energy scale. 
Based on the different breaking pattern at the first step, the 331 model is characterized by the parameter $\beta$. 
The deviation of the SM-like couplings will be modified by a suppressing factor $1/v_3$. Specifically, we calculate all the tree-level and loop-induced effective couplings of the Higgs boson to the SM particles, which has a natural decoupling limit at $v_3\gg v_1,v_2$. The latest global fit data of the Higgs boson couplings is used to set constraints on the model parameters, namely, the symmetry breaking scale $v_1$, the scalar mixing $U_{11}$, $U_{21}$. As a result, $v_1$ tends to be at the vicinity of $150{~\rm GeV}$, $U_{11}$ and $U_{21}$ are close to 0.7. The allowed 95\% confidence level region depends much on $v_3$, but not quite on $\beta$. With the allowed region, we make a prediction on $R_{Z\gamma}$. 
The largest value of $R_{Z\gamma}$ can only achieve to 1.2 for $\beta=\pm\sqrt{3}$, while most parameter space yields $R_{Z\gamma}<1$.
We also predict the double Higgs production rate within the allowed single Higgs parameter spaces. The production rate can be enhanced by a factor of 3 maximally, but it is still undetectable at the 14~TeV LHC with an integrated luminosity of  $3000{~\rm fb^{-1}}$. On the other hand, a future 100~TeV $pp$-collider can cover the entire parameter space. 

Because of the residual ``$Z_2$" symmetry after spontaneously symmetry breaking, the lightest new physics particle can be a dark matter candidate or a long-lived charged particle, depending on $\beta$. The $Y$ and $H^{Q_Y}$ are suitable for the DM candidate for $\beta=-1/\sqrt{3}$ while $V$, $H^{Q_V}$ and $E_\ell$ could be the DM candidate for $\beta=+1/\sqrt{3}$. In the $\beta=\pm\sqrt{3}$ models, there is a lightest charged particle which is absolutely stable or long-lived. If the long-live charged particle has a mean lifetime longer than the typical hadronization time scale, they might form QCD bound states with partons (quarks and/or gluons) in analogy with the ordinary hadrons. It is usually named as $R$-hadron.  We list out all the possible $R$-hadrons in the 331 models.

Due to the extension of the gauge symmetry, there are additional gauge bosons $Z^\prime$, $V$, $Y$ and quarks $D$, $S$, $T$. We have studied their productions and decays at the 8~TeV and 14~TeV LHC. In the case of $Z^\prime$, the production rate and decay mode both depend much on $\beta$. The rate is much larger than SSM's for $\beta=\pm \sqrt{3}$, while much smaller than SSM's for $\beta=\pm 1/\sqrt{3}$. The $Z^\prime$ predominantly decays into $D\bar{D}/S\bar{S}$ pairs for $\beta=+1/\sqrt{3}$ and $-\sqrt{3}$ and into $T\bar{T}$ for $\beta=-1/\sqrt{3}$ and $\sqrt{3}$. 
Since the couplings of $Z^\prime$ to SM fermions are large for $\beta=\pm\sqrt{3}$, the decay width of $Z^\prime$ can be as large as 10\%$M_{Z^\prime}$, which could lead to significant interference effects. We study the Forward Backward Asymmetry ($A_{FB}$) of the lepton pair in the $Z^\prime$ production through the Drell-Yan channel. 

We also study the production of $VV$ and $YY$ pair at the 8~TeV and 14~TeV LHC. 
We note that the dominant decay mode of $V$ and $Y$ bosons is fermion pair if kinematically allowed. The $W/Z/h$ plus scalar modes are also important, especially when the fermion mode is forbidden. 
Based on the production and decay of $V(Y)$, we summarize the different collider signatures in various 331 models. 

Finally, we examine heavy quark ($D$, $S$ and $T$) productions. The heavy quark can be produced in pair either through the QCD interaction or mediated by a heavy $Z^\prime$ boson. The two processes are comparable in the $\beta=\pm\sqrt{3}$ models, owing to the large coupling of $Z^\prime$ to the SM quarks. The dominant decay mode of $D$ quark is a $dV$ or $uY$ while the $S$ quark mainly decays into $sV$ or $cY$ pair. The $T$ quark can decay into $tV$, $bY$ and $tH^{Q_V}$ pairs. We explicitly list all the possible collider signatures of heavy quark pair productions in the 331 models. A detailed collider simulation including the SM backgrounds and realistic detector effects is in progress and will be presented elsewhere.

\section*{Acknowledgement}
We would like to thank Yan-Dong Liu and Bin-Yan for helpful discussions. This work was supported in part by the National Science Foundation of China under Grand No. 11275009.

\appendix

\section{ Feynman rules interaction vertices}\label{app_a}

For the convenience of further phenomenological exploration, we
list the Feynman rules of the interaction vertices in unitary gauge
among the new scalar sector, the new gauge bosons, the new fermions
and the SM particles. All particles are the mass eigenstates.
In the Feynman rules, all particles are assumed to be outgoing, and
we adopt the convention Feynman rule $= i \mathcal{L}$. Furthermore,
we define 
\beq
a_0=\sqrt{1-\left(1+\beta^2\right)s_W^2}, \qquad a_{\pm}=\sqrt{3}\beta\pm1,\qquad b_{\pm}=\sqrt{3}\beta\pm2.
\eeq

\subsection{Gauge Boson-Scalar Couplings}

a.~Three-point vertices in Tables  \ref{1gauge-2scalars},
 \ref{2gauges-1scalar}.
 
\renewcommand{\arraystretch}{1.5}
\large
\begin{longtable}{c|c}
\hline
particles & vertices \\
\hline
$A H^{+Q_V} H^{-Q_V}$ & $-\frac{i  e \left(p_1-p_2\right)^{\mu }a_-}{2}  $ \\
\hline
$A H^+ H^-$ & $i e \left(p_1-p_2\right)^{\mu } $ \\
\hline
$V^{+Q_V} H^{-Q_V} h$ & $-\frac{e \left(p_1-p_2\right)^{\mu } \left(s_{13}  U_{11}- U_{31} c_{13}\right)}{2 s_W}$ \\
\hline
$V^{+Q_V} H^{-Q_V} H_0$ & $-\frac{i e \left(p_1-p_2\right)^{\mu } \left(  O_{31} c_{13}+s_{13}   O_{11}\right)}{2 s_W}$ \\
\hline
$V^{+Q_V} H^{-Q_V} H_2$ & $-\frac{e \left(p_1-p_2\right)^{\mu } \left(s_{13}   U_{12}-  U_{32} c_{13}\right)}{2 s_W}$ \\
\hline
 $V^{+Q_V} H^{-Q_V} H_3$ & $-\frac{e \left(p_1-p_2\right)^{\mu } \left(s_{13}   U_{13}-  U_{33} c_{13}\right)}{2 s_W}$ \\
\hline
$V^{+Q_V} H^- H^{-Q_Y}$ & $-\frac{i e \left(p_1-p_2\right)^{\mu } c_{12}  s_{23}}{2 s_W}$ \\
\hline
 $W^+ H^- h$ & $-\frac{e \left(p_1-p_2\right)^{\mu } \left(s_{12}   U_{11}-  U_{21} c_{12}\right)}{2 s_W}$ \\
\hline
$W^+ H^- H_0$ & $-\frac{i e \left(p_1-p_2\right)^{\mu } \left(  O_{21} c_{12}+s_{12}   O_{11}\right)}{2 s_W}$ \\
\hline
 $W^+ H^- H_2$ & $-\frac{e \left(p_1-p_2\right)^{\mu } \left(s_{12}   U_{12}-  U_{22} c_{12}\right)}{2 s_W}$ \\
\hline
$W^+ H^- H_3$ & $-\frac{e \left(p_1-p_2\right)^{\mu } \left(s_{12}   U_{13}-  U_{23} c_{12}\right)}{2 s_W}$ \\
\hline
 $W^+ H^{-Q_V} H^{-Q_Y}$ & $\frac{i e \left(p_1-p_2\right)^{\mu } c_{23} c_{13}}{\sqrt{2} s_W}$ \\
\hline
$Y^{+Q_Y} H^{-Q_Y} h$ & $-\frac{e \left(p_1-p_2\right)^{\mu } \left(s_{23}   U_{21}-  U_{31} c_{23}\right)}{2 s_W}$ \\
\hline
 $Y^{+Q_Y} H^{-Q_Y} H_0$ & $-\frac{i e \left(p_1-p_2\right)^{\mu } \left(  O_{31} c_{23}+s_{23}   O_{21}\right)}{2 s_W}$ \\
\hline
$Y^{+Q_Y} H^{-Q_Y} H_2$ & $-\frac{e \left(p_1-p_2\right)^{\mu } \left(s_{23}   U_{22}-  U_{32} c_{23}\right)}{2 s_W}$ \\
\hline
 $Y^{+Q_Y} H^{-Q_Y} H_3$ & $-\frac{e \left(p_1-p_2\right)^{\mu } \left(s_{23}   U_{23}-  U_{33} c_{23}\right)}{2 s_W}$ \\
\hline
$Y^{+Q_Y} H^{-Q_V} H^+$ & $-\frac{i e \left(p_1-p_2\right)^{\mu } s_{12} s_{13}}{\sqrt{2} s_W}$ \\
\hline
$Z h H_0$ & $-\frac{e \left(p_1-p_2\right)^{\mu }  \left(  O_{11}   U_{11}-  O_{21}   U_{21}\right)}{2 c_W s_W}$ \\
\hline
 $Z H_2 H_0$ & $\frac{e \left(p_1-p_2\right)^{\mu }  \left(  O_{11}   U_{12}-  O_{21}   U_{22}\right)}{2 c_W s_W}$ \\
\hline
$Z H_3 H_0$ & $\frac{e \left(p_1-p_2\right)^{\mu }  \left(  O_{11}   U_{13}-  O_{21}   U_{23}\right)}{2 c_W s_W}$ \\
\hline
 $Z H^{+Q_V} H^{-Q_V}$ & $\frac{i e \left(p_1-p_2\right)^{\mu } \left[c_{13}^2 \left(1+a_- s_W^2\right)+a_- s_{13}^2 s_W^2\right]}{2 c_W s_W}$ \\
\hline
$Z H^+ H^-$ & $\frac{i e \left(p_1-p_2\right)^{\mu } \left(c_W^2-s_W^2\right) }{2 c_W s_W}$ \\
\hline
 $Z H^{+Q_Y} H^{-Q_Y}$ & $\frac{i e \left(p_1-p_2\right)^{\mu } \left[c_{23}^2 \left(a_+ s_W^2-1\right)+a_+ s_{23}^2 s_W^2\right]}{2 c_W s_W}$ \\
\hline
$Z' h H_0$ & $\frac{e \left(p_1-p_2\right)^{\mu } \left[   \left(a_+ s_W^2-1\right)O_{11}   U_{11}-  \left(a_- s_W^2+1\right)O_{21}   U_{21} -2 \left(s_W^2-1\right)   O_{31}   U_{31}\right]}{2\sqrt{3} c_W s_W a_0}$ \\
\hline
 $Z' H_2 H_0$ & $\frac{e \left(p_1-p_2\right)^{\mu } \left[   \left(a_+ s_W^2-1\right)O_{11}   U_{12}-  \left(a_- s_W^2+1\right)O_{21}   U_{22} -2 \left(s_W^2-1\right)   O_{31}   U_{32}\right]}{2\sqrt{3} c_W s_W a_0}$ \\
\hline
$Z' H_3 H_0$ & $\frac{e \left(p_1-p_2\right)^{\mu } \left[   \left(a_+ s_W^2-1\right)O_{11}   U_{13}-  \left(a_- s_W^2+1\right)O_{21}   U_{23} -2 \left(s_W^2-1\right)   O_{31}   U_{33}\right]}{2\sqrt{3} c_W s_W a_0}$\\
\hline
$Z' H^{+Q_V} H^{-Q_V}$ & $-\frac{i e \left(p_1-p_2\right)^{\mu } \left[ c_{13}^2 \left(\left(a_+a_- +2\right) s_W^2-1\right)+s_{13}^2 \left(\left(a_+^2-a_-^2\right) s_W^2-2 \right)\right]}{2\sqrt{3} c_W s_W a_0}$ \\
\hline
$Z' H^+ H^-$ & $\frac{i e \left(p_1-p_2\right)^{\mu } \left[c_{12}^2 \left(1+a_- s_W^2\right)+s_{12}^2 \left(a_+ s_W^2-1\right)\right]}{2\sqrt{3} c_W s_W a_0}$ \\
\hline
 $Z' H^{+Q_Y} H^{-Q_Y}$ &$-\frac{i e \left(p_1-p_2\right)^{\mu } \left[ c_{23}^2 \left(\left(a_+a_- +2\right) s_W^2-1\right)+s_{23}^2 \left(\left(a_+^2-a_-^2\right) s_W^2-2 \right)\right]}{2\sqrt{3} c_W s_W a_0}$ \\
\hline
\caption{Three-point couplings of one gauge boson to two scalars ($VSS$).
All particles are the mass eigenstates. The momenta are assigned according to $V_\mu S_1(p_1)S_2(p_2)$.
\label{1gauge-2scalars}
}
\end{longtable}

\begin{longtable}{c|c}
\hline
particles & vertices \\
\hline
$V^{+Q_V} V^{-Q_V} h$ & $\frac{i e^2 g_{\mu \nu}\left(v_1   U_{11}+v_3   U_{31}\right)}{2 s_W^2}$ \\
\hline
$V^{+Q_V} V^{-Q_V} H_2$ & $\frac{i e^2 g_{\mu \nu}\left(v_1   U_{12}+v_3   U_{32}\right)}{2 s_W^2}$ \\
\hline
 $V^{+Q_V} V^{-Q_V} H_3$ & $\frac{i e^2 g_{\mu \nu}\left(v_1   U_{13}+v_3   U_{33}\right)}{2 s_W^2}$ \\
\hline
$A W^+ H^-$ & $\frac{e^2 g_{\mu \nu}\left(v_2 c_{12}-v_1 s_{12}\right)}{2 s_W}$ \\
\hline
 $W^+ V^{-Q_V} H^{-Q_Y}$ & $\frac{e^2 g_{\mu \nu}\left(v_3 c_{23}+v_2 s_{23}\right)}{2 \sqrt{2} s_w^2}$ \\
\hline
$W^+ W^- h$ & $\frac{i e^2 g_{\mu \nu}\left(v_1   U_{11}+v_2   U_{21}\right)}{2 s_W^2}$ \\
\hline
 $W^+ W^- H_2$ & $\frac{i e^2 g_{\mu \nu}\left(v_1   U_{12}+v_2   U_{22}\right)}{2 s_W^2}$ \\
\hline
$W^+ W^- H_3$ & $\frac{i e^2 g_{\mu \nu}\left(v_1   U_{13}+v_2   U_{23}\right)}{2 s_W^2}$ \\
\hline
 $A Y^{+Q_Y} H^{-Q_Y}$ & $\frac{a_+ e^2 g_{\mu \nu}\left(v_2 s_{23}-v_3 c_{23}\right)}{4 s_W}$ \\
\hline
$Y^{+Q_Y} V^{-Q_V} H^+$ & $\frac{e^2 g_{\mu \nu}\left(v_2 c_{12}+v_1 s_{12}\right)}{2 \sqrt{2} s_W^2}$ \\
\hline
 $W^+ Y^{+Q_Y} H^{-Q_V}$ & $-\frac{e^2 g_{\mu \nu}\left(v_3 c_{13}+v_1 s_{13}\right)}{2 \sqrt{2} s_w^2}$ \\
\hline
$Y^{+Q_Y} Y^{-Q_Y} h$ & $\frac{i e^2 g_{\mu \nu}\left(v_2   U_{21}+v_3   U_{31}\right)}{2 s_W^2}$ \\
\hline
 $Y^{+Q_Y} Y^{-Q_Y} H_2$ & $\frac{i e^2 g_{\mu \nu}\left(v_2   U_{22}+v_3   U_{32}\right)}{2 s_W^2}$ \\
\hline
$Y^{+Q_Y} Y^{-Q_Y} H_3$ & $\frac{i e^2 g_{\mu \nu}\left(v_2   U_{23}+v_3   U_{33}\right)}{2 s_W^2}$ \\
\hline
 $Z V^{+Q_V} H^{-Q_V}$ & $\frac{e^2 g_{\mu \nu}\left[c_W^2 \left(v_3 c_{13}+v_1 s_{13}\right)+s_W^2 \left(\left(a_- +1\right)  v_3 c_{13}-b_- v_1 s_{13}\right)\right]}{4 c_W s_W^2}$ \\
\hline
 $Z Y^{+Q_Y} H^{-Q_Y}$ & $-\frac{e^2 g_{\mu \nu}\left[c_W^2 \left(v_3 c_{23}+v_2 s_{23}\right)+s_W^2 \left(b_+ v_2 s_{23}-\left(a_- +1\right)  v_3 c_{23}\right)\right]}{4 c_W s_W^2}$ \\
\hline
$Z Z h$ & $\frac{i e^2 g_{\mu \nu} \left(v_1   U_{11}+v_2   U_{21}\right)}{2 c_W^2 s_W^2}$ \\
\hline
 $Z Z H_2$ & $\frac{i e^2 g_{\mu \nu} \left(v_1   U_{12}+v_2   U_{22}\right)}{2 c_W^2 s_W^2}$ \\
\hline
$Z Z H_3$ & $\frac{i e^2 g_{\mu \nu} \left(v_1   U_{13}+v_2   U_{23}\right)}{2 c_W^2 s_W^2}$ \\
\hline
 $Z' V^{+Q_V} H^{-Q_V}$ & $\frac{e^2 g_{\mu \nu}\left[ -v_3 c_{13} \left(a_+a_- s_W^2+1\right)+v_1 s_{13} \left(a_-^2 s_W^2-1\right)\right]}{4\sqrt{3} c_W s_W^2 a_0}$ \\
\hline
$Z' W^+ H^-$ & $\frac{e^2 g_{\mu \nu}\left[v_2 c_{12} \left(1+a_- s_W^2\right)+v_1 s_{12} \left(1-a_+ s_W^2\right)\right]}{2\sqrt{3} c_W s_W^2 a_0}$ \\
\hline
$Z' Y^{+Q_Y} H^{-Q_Y}$ & $\frac{e^2 g_{\mu \nu}\left[ -v_3 c_{23} \left(a_+a_- s_W^2+1\right)+v_2 s_{23} \left(a_-^2 s_W^2-1\right)\right]}{4\sqrt{3} c_W s_W^2 a_0}$  \\
\hline
$Z Z' h$ & $-\frac{i e^2 g_{\mu \nu} \left[v_1   U_{11} \left(a_+ s_W^2-1\right)+v_2   U_{21} \left(1+a_- s_W^2\right)\right]}{2\sqrt{3} c_W^2 s_W^2 a_0}$ \\
\hline
 $Z Z' H_2$ & $-\frac{i e^2 g_{\mu \nu} \left[v_1   U_{12} \left(a_+ s_W^2-1\right)+v_2   U_{22} \left(1+a_- s_W^2\right)\right]}{2\sqrt{3} c_W^2 s_W^2 a_0}$ \\
\hline
$Z Z' H_3$ & $-\frac{i e^2 g_{\mu \nu} \left[v_1   U_{13} \left(a_+ s_W^2-1\right)+v_2   U_{23} \left(1+a_- s_W^2\right)\right]}{2\sqrt{3} c_W^2 s_W^2 a_0}$ \\
\hline
$Z' Z' h$ & $\frac{i e^2 g_{\mu \nu}\left[v_1   U_{11} \left(a_+s_W^2-1\right)^2+v_2   U_{21} \left(a_-s_W^2+1\right)^2+4 v_3 \left(s_W^2-1\right)^2   U_{31}\right]}{6 c_W^2 s_W^2 a_0^2}$ \\
\hline
$Z' Z' H_2$ & $\frac{i e^2 g_{\mu \nu}\left[v_1   U_{12} \left(a_+s_W^2-1\right)^2+v_2   U_{22} \left(a_-s_W^2+1\right)^2+4 v_3 \left(s_W^2-1\right)^2   U_{32}\right]}{6 c_W^2 s_W^2 a_0^2}$ \\
\hline
 $Z' Z' H_3$ & $\frac{i e^2 g_{\mu \nu}\left[v_1   U_{13} \left(a_+s_W^2-1\right)^2+v_2   U_{23} \left(a_-s_W^2+1\right)^2+4 v_3 \left(s_W^2-1\right)^2   U_{33}\right]}{6 c_W^2 s_W^2 a_0^2}$ \\
\hline
\caption{Three-point couplings of two gauge bosons to one scalar ($VVS$).
\label{2gauges-1scalar}
}

\end{longtable}

b.~Four-point vertices in Table \ref{2gauges-2scalars}.
 
\begin{longtable}{c|c}
\hline
particles & vertices \\
\hline
$A A H^{+Q_Y} H^{-Q_Y}$ & $\frac{i a_+^2 e^2 \eta _{\mu \nu }}{2}  $ \\
\hline
$A A H^+ H^-$ & $2 i e^2 g_{\mu \nu}$ \\
\hline
$A A H^{+Q_V} H^{-Q_V}$ & $\frac{i a_-^2 e^2 \eta _{\mu \nu }}{2}  $ \\
\hline
$A V^{+Q_V} H^- H^{-Q_Y}$ & $\frac{i \sqrt{3}a_+e^2 g_{\mu \nu}s_{23} c_{12}}{2 \sqrt{2} s_W}$ \\
\hline
$A V^{+Q_V} H_0 H^{-Q_V}$ & $\frac{i a_- e^2 g_{\mu \nu}\left( c_{13}O_{31}+s_{13} O_{11}\right)}{4 s_W}$ \\
\hline
$A V^{+Q_V} h H^{-Q_V}$ & $\frac{a_-e^2 g_{\mu \nu}\left(s_{13} U_{11}- c_{13}U_{31}\right)}{4 s_W}$ \\
\hline
$A V^{+Q_V} H_2 H^{-Q_V}$ & $\frac{a_-e^2 g_{\mu \nu}\left(s_{13} U_{12}- c_{13}U_{32}\right)}{4 s_W}$ \\
\hline
$A V^{+Q_V} H_3 H^{-Q_V}$ & $\frac{a_- e^2 g_{\mu \nu}\left(s_{13} U_{13}- c_{13}U_{33}\right)}{4 s_W}$ \\
\hline
$V^{+Q_V} V^{-Q_V} H^{+Q_Y} H^{-Q_Y}$ & $\frac{i e^2 g_{\mu \nu}s_{23}^2}{2 s_W^2}$ \\
\hline
$V^{+Q_V} V^{-Q_V} H^+ H^-$ & $\frac{i e^2 g_{\mu \nu}c_{12}^2}{2 s_W^2}$ \\
\hline
$V^{+Q_V} V^{-Q_V} H^{+Q_V} H^{-Q_V}$ & $\frac{i e^2 g_{\mu \nu}}{2 s_W^2}$ \\
\hline
$V^{+Q_V} V^{-Q_V} H_0 H_0$ & $\frac{i e^2 g_{\mu \nu}\left[\left(O_{11}\right)^2+\left(O_{31}\right)^2\right]}{2 s_W^2}$ \\
\hline
$V^{+Q_V} V^{-Q_V} h h$ & $\frac{i e^2 g_{\mu \nu}\left[\left(U_{11}\right)^2+\left(U_{31}\right)^2\right]}{2 s_W^2}$ \\
\hline
$V^{+Q_V} V^{-Q_V} h H_2$ & $\frac{i e^2 g_{\mu \nu}\left[U_{11} U_{12}+U_{31} U_{32}\right]}{2 s_W^2}$ \\
\hline
$V^{+Q_V} V^{-Q_V} H_2 H_2$ & $\frac{i e^2 g_{\mu \nu}\left[\left(U_{12}\right)^2+\left(U_{32}\right)^2\right]}{2 s_W^2}$ \\
\hline
$V^{+Q_V} V^{-Q_V} h H_3$ & $\frac{i e^2 g_{\mu \nu}\left[U_{11} U_{13}+U_{31} U_{33}\right]}{2 s_W^2}$ \\
\hline
$V^{+Q_V} V^{-Q_V} H_2 H_3$ & $\frac{i e^2 g_{\mu \nu}\left[U_{12} U_{13}+U_{32} U_{33}\right]}{2 s_W^2}$ \\
\hline
$V^{+Q_V} V^{-Q_V} H_3 H_3$ & $\frac{i e^2 g_{\mu \nu}\left[\left(U_{13}\right)^2+\left(U_{33}\right)^2\right]}{2 s_W^2}$ \\
\hline
$A W^+ H^{-Q_Y} H^{-Q_V}$ & $\frac{i (a_- +1)  e^2 g_{\mu \nu}c_{23} c_{13}}{\sqrt{2}s_W}$ \\
\hline
$A W^+ H_0 H^-$ & $-\frac{i e^2 g_{\mu \nu}\left( c_{12}O_{21}+s_{12} O_{11}\right)}{2 s_W}$ \\
\hline
$A W^+ h H^-$ & $\frac{e^2 g_{\mu \nu}\left( c_{12}U_{21}-s_{12} U_{11}\right)}{2 s_W}$ \\
\hline
$A W^+ H_2 H^-$ & $\frac{e^2 g_{\mu \nu}\left( c_{12}U_{22}-s_{12} U_{12}\right)}{2 s_W}$ \\
\hline
$A W^+ H_3 H^-$ & $\frac{e^2 g_{\mu \nu}\left( c_{12}U_{23}-s_{12} U_{13}\right)}{2 s_W}$ \\
\hline
$W^+ V^{-Q_V} H^{+Q_V} H^-$ & $\frac{i e^2 g_{\mu \nu}s_{12} s_{13}}{2 s_W^2}$ \\
\hline
$W^+ V^{-Q_V} H_0 H^{-Q_Y}$ & $-\frac{i e^2 g_{\mu \nu}\left(s_{23} O_{21}- c_{23}O_{31}\right)}{2 \sqrt{2} s_W^2}$ \\
\hline
$W^+ V^{-Q_V} h H^{-Q_Y}$ & $\frac{e^2 g_{\mu \nu}\left( c_{23}U_{31}+s_{23} U_{21}\right)}{2 \sqrt{2} s_W^2}$ \\
\hline
$W^+ V^{-Q_V} H_2 H^{-Q_Y}$ & $\frac{e^2 g_{\mu \nu}\left( c_{23}U_{32}+s_{23} U_{22}\right)}{2 \sqrt{2} s_W^2}$ \\
\hline
$W^+ V^{-Q_V} H_3 H^{-Q_Y}$ & $\frac{e^2 g_{\mu \nu}\left( c_{23}U_{33}+s_{23} U_{23}\right)}{2 \sqrt{2} s_W^2}$ \\
\hline
$W^+ W^- H^{+Q_Y} H^{-Q_Y}$ & $\frac{i e^2 g_{\mu \nu}c_{23}^2}{2 s_W^2}$ \\
\hline
$W^+ W^- H^+ H^-$ & $\frac{i e^2 g_{\mu \nu}}{2 s_W^2}$ \\
\hline
$W^+ W^- H^{+Q_V} H^{-Q_V}$ & $\frac{i e^2 g_{\mu \nu}c_{13}^2}{2 s_W^2}$ \\
\hline
$W^+ W^- H_0 H_0$ & $\frac{i e^2 g_{\mu \nu}\left[\left(O_{11}\right)^2+\left(O_{21}\right)^2\right]}{2 s_W^2}$ \\
\hline
$W^+ W^- h h$ & $\frac{i e^2 g_{\mu \nu}\left[\left(U_{11}\right)^2+\left(U_{21}\right)^2\right]}{2 s_W^2}$ \\
\hline
$W^+ W^- h H_2$ & $\frac{i e^2 g_{\mu \nu}\left[U_{11} U_{12}+U_{21} U_{22}\right]}{2 s_W^2}$ \\
\hline
$W^+ W^- H_2 H_2$ & $\frac{i e^2 g_{\mu \nu}\left[\left(U_{12}\right)^2+\left(U_{22}\right)^2\right]}{2 s_W^2}$ \\
\hline
$W^+ W^- h H_3$ & $\frac{i e^2 g_{\mu \nu}\left[U_{11} U_{13}+U_{21} U_{23}\right]}{2 s_W^2}$ \\
\hline
$W^+ W^- H_2 H_3$ & $\frac{i e^2 g_{\mu \nu}\left[U_{12} U_{13}+U_{22} U_{23}\right]}{2 s_W^2}$ \\
\hline
$W^+ W^- H_3 H_3$ & $\frac{i e^2 g_{\mu \nu}\left[\left(U_{13}\right)^2+\left(U_{23}\right)^2\right]}{2 s_W^2}$ \\
\hline
$A Y^{+Q_Y} H^+ H^{-Q_V}$ & $-\frac{i \left(b_--1\right) e^2 g_{\mu \nu}s_{12} s_{13}}{2 \sqrt{2} s_W}$ \\
\hline
$A Y^{+Q_Y} H_0 H^{-Q_Y}$ & $\frac{i a_+ e^2 g_{\mu \nu}\left( c_{23}O_{31}+s_{23} O_{21}\right)}{4 s_W}$ \\
\hline
$A Y^{+Q_Y} h H^{-Q_Y}$ & $\frac{a_+ e^2 g_{\mu \nu}\left(s_{23} U_{21}- c_{23}U_{31}\right)}{4 s_W}$ \\
\hline
$A Y^{+Q_Y} H_2 H^{-Q_Y}$ & $\frac{a_+ e^2 g_{\mu \nu}\left(s_{23} U_{22}- c_{23}U_{32}\right)}{4 s_W}$ \\
\hline
$A Y^{+Q_Y} H_3 H^{-Q_Y}$ & $\frac{a_+ e^2 g_{\mu \nu}\left(s_{23} U_{23}- c_{23}U_{33}\right)}{4 s_W}$ \\
\hline
$W^+ Y^{+Q_Y} H^- H^{-Q_Y}$ & $-\frac{i e^2 g_{\mu \nu}s_{23} c_{12}}{2 s_W^2}$ \\
\hline
$W^+ Y^{+Q_Y} H_0 H^{-Q_V}$ & $-\frac{i e^2 g_{\mu \nu}\left(s_{13} O_{11}- c_{13}O_{31}\right)}{2 \sqrt{2} s_W^2}$ \\
\hline
$W^+ Y^{+Q_Y} h H^{-Q_V}$ & $-\frac{e^2 g_{\mu \nu}\left( c_{13}U_{31}+s_{13} U_{11}\right)}{2 \sqrt{2} s_W^2}$ \\
\hline
$W^+ Y^{+Q_Y} H_2 H^{-Q_V}$ & $-\frac{e^2 g_{\mu \nu}\left( c_{13}U_{32}+s_{13} U_{12}\right)}{2 \sqrt{2} s_W^2}$ \\
\hline
$W^+ Y^{+Q_Y} H_3 H^{-Q_V}$ & $-\frac{e^2 g_{\mu \nu}\left( c_{13}U_{33}+s_{13} U_{13}\right)}{2 \sqrt{2} s_W^2}$ \\
\hline
$V^{+Q_V} Y^{-Q_Y} H^{-Q_Y} H^{-Q_V}$ & $\frac{i e^2 g_{\mu \nu}c_{23} c_{13}}{2 s_W^2}$ \\
\hline
$V^{+Q_V} Y^{-Q_Y} H_0 H^-$ & $-\frac{i e^2 g_{\mu \nu}\left(s_{12} O_{11}- c_{12}O_{21}\right)}{2 \sqrt{2} s_W^2}$ \\
\hline
$V^{+Q_V} Y^{-Q_Y} h H^-$ & $-\frac{e^2 g_{\mu \nu}\left( c_{12}U_{21}+s_{12} U_{11}\right)}{2 \sqrt{2} s_W^2}$ \\
\hline
$V^{+Q_V} Y^{-Q_Y} H_2 H^-$ & $-\frac{e^2 g_{\mu \nu}\left(c_{12}U_{22} +s_{12} U_{12}\right)}{2 \sqrt{2} s_W^2}$ \\
\hline
$V^{+Q_V} Y^{-Q_Y} H_3 H^-$ & $-\frac{e^2 g_{\mu \nu}\left( c_{12}U_{23}+s_{12} U_{13}\right)}{2 \sqrt{2} s_W^2}$ \\
\hline
$Y^{+Q_Y} Y^{-Q_Y} H^{+Q_Y} H^{-Q_Y}$ & $\frac{i e^2 g_{\mu \nu}}{2 s_W^2}$ \\
\hline
$Y^{+Q_Y} Y^{-Q_Y} H^+ H^-$ & $\frac{i e^2 g_{\mu \nu}s_{12}^2}{2 s_W^2}$ \\
\hline
$Y^{+Q_Y} Y^{-Q_Y} H^{+Q_V} H^{-Q_V}$ & $\frac{i e^2 g_{\mu \nu}s_{13}^2}{2 s_W^2}$ \\
\hline
$Y^{+Q_Y} Y^{-Q_Y} H_0 H_0$ & $\frac{i e^2 g_{\mu \nu}\left[\left(O_{21}\right)^2+\left(O_{31}\right)^2\right]}{2 s_W^2}$ \\
\hline
$Y^{+Q_Y} Y^{-Q_Y} h h$ & $\frac{i e^2 g_{\mu \nu}\left[\left(U_{21}\right)^2+\left(U_{31}\right)^2\right]}{2 s_W^2}$ \\
\hline
$Y^{+Q_Y} Y^{-Q_Y} h H_2$ & $\frac{i e^2 g_{\mu \nu}\left[U_{21} U_{22}+U_{31} U_{32}\right]}{2 s_W^2}$ \\
\hline
$Y^{+Q_Y} Y^{-Q_Y} H_2 H_2$ & $\frac{i e^2 g_{\mu \nu}\left[\left(U_{22}\right)^2+\left(U_{32}\right)^2\right]}{2 s_W^2}$ \\
\hline
$Y^{+Q_Y} Y^{-Q_Y} h H_3$ & $\frac{i e^2 g_{\mu \nu}\left(U_{21} U_{23}+U_{31} U_{33}\right)}{2 s_W^2}$ \\
\hline
$Y^{+Q_Y} Y^{-Q_Y} H_2 H_3$ & $\frac{i e^2 g_{\mu \nu}\left(U_{22} U_{23}+U_{32} U_{33}\right)}{2 s_W^2}$ \\
\hline
$Y^{+Q_Y} Y^{-Q_Y} H_3 H_3$ & $\frac{i e^2 g_{\mu \nu}\left[\left(U_{23}\right)^2+\left(U_{33}\right)^2\right]}{2 s_W^2}$ \\
\hline
$A Z H^{+Q_Y} H^{-Q_Y}$ & $-\frac{i e^2 g_{\mu \nu}\left(a_+^2s_W^2-a_+c_{23}^2\right)}{2 c_W s_W}$ \\
\hline
$A Z H^+ H^-$ & $\frac{i e^2 g_{\mu \nu}\left(c_W^2-s_W^2\right) }{c_W s_W}$ \\
\hline
$A Z H^{+Q_V} H^{-Q_V}$ & $-\frac{i e^2 g_{\mu \nu}\left(a_-^2s_W^2+a_-c_{13}^2\right)}{2 c_W s_W}$ \\
\hline
$V^{+Q_V} Z H^- H^{-Q_Y}$ & $\frac{i e^2 g_{\mu \nu}s_{23} c_{12} \left(c_W^2-b_+ s_W^2\right)}{2 \sqrt{2} c_W s_W^2}$ \\
\hline
$V^{+Q_V} Z H_0 H^{-Q_V}$ & $\frac{i e^2 g_{\mu \nu}\left[c_W^2 \left(s_{13} O_{11}- c_{13}O_{31}\right)-s_W^2 \left((a_- +1)   c_{13}O_{31}+b_- s_{13} O_{11}\right)\right]}{4 c_W s_W^2}$ \\
\hline
$V^{+Q_V} Z h H^{-Q_V}$ & $\frac{e^2 g_{\mu \nu}\left[c_W^2 \left( c_{13}U_{31}+s_{13} U_{11}\right)+s_W^2 \left((a_- +1)   c_{13}U_{31}-b_-s_{13} U_{11}\right)\right]}{4 c_W s_W^2}$ \\
\hline
$V^{+Q_V} Z H_2 H^{-Q_V}$ & $\frac{e^2 g_{\mu \nu}\left[c_W^2 \left( c_{13}U_{32}+s_{13} U_{12}\right)+s_W^2 \left((a_- +1)   c_{13}U_{32}-b_-s_{13} U_{12}\right)\right]}{4 c_W s_W^2}$ \\
\hline
$V^{+Q_V} Z H_3 H^{-Q_V}$ & $\frac{e^2 g_{\mu \nu}\left[c_W^2 \left( c_{13}U_{33}+s_{13} U_{13}\right)+s_W^2 \left( (a_- +1)   c_{13}U_{33}-b_-s_{13} U_{13}\right)\right]}{4 c_W s_W^2}$ \\
\hline
$W^+ Z H^{-Q_Y} H^{-Q_V}$ & $-\frac{i (a_- +1)  e^2 g_{\mu \nu}c_{23} c_{13}}{\sqrt{2}c_W}$ \\
\hline
$W^+ Z H_0 H^-$ & $\frac{i e^2 g_{\mu \nu}\left(O_{21} c_{12}+s_{12} O_{11}\right)}{2 c_W}$ \\
\hline
$W^+ Z h H^-$ & $\frac{e^2 g_{\mu \nu}\left(s_{12} U_{11}-U_{21} c_{12}\right)}{2 c_W}$ \\
\hline
$W^+ Z H_2 H^-$ & $\frac{e^2 g_{\mu \nu}\left(s_{12} U_{12}-U_{22} c_{12}\right)}{2 c_W}$ \\
\hline
$W^+ Z H_3 H^-$ & $\frac{e^2 g_{\mu \nu}\left(s_{12} U_{13}-U_{23} c_{12}\right)}{2 c_W}$ \\
\hline
$Y^{+Q_Y} Z H^+ H^{-Q_V}$ & $\frac{i e^2 g_{\mu \nu}s_{12} s_{13} \left(c_W^2+b_- s_W^2\right)}{2 \sqrt{2} c_W s_W^2}$ \\
\hline
$Y^{+Q_Y} Z H_0 H^{-Q_Y}$ & $-\frac{i e^2 g_{\mu \nu}\left[c_W^2 \left(s_{23} O_{21}- c_{23}O_{31}\right)+s_W^2 \left((a_- +1)   c_{23}O_{31}+b_+ s_{23} O_{21}\right)\right]}{4 c_W s_W^2}$ \\
\hline
$Y^{+Q_Y} Z h H^{-Q_Y}$ & $-\frac{e^2 g_{\mu \nu}\left[c_W^2 \left( c_{23}U_{31}+s_{23} U_{21}\right)+s_W^2 \left(b_+ s_{23} U_{21}-(a_- +1)   c_{23}U_{31}\right)\right]}{4 c_W s_W^2}$ \\
\hline
$Y^{+Q_Y} Z H_2 H^{-Q_Y}$ & $-\frac{e^2 g_{\mu \nu}\left[c_W^2 \left( c_{23}U_{32}+s_{23} U_{22}\right)+s_W^2 \left(b_+ s_{23} U_{22}-(a_- +1)   c_{23}U_{32}\right)\right]}{4 c_W s_W^2}$ \\
\hline
$Y^{+Q_Y} Z H_3 H^{-Q_Y}$ & $-\frac{e^2 g_{\mu \nu}\left[c_W^2 \left( c_{23}U_{33}+s_{23} U_{23}\right)+s_W^2 \left(b_+ s_{23} U_{23}-(a_- +1)   c_{23}U_{33}\right)\right]}{4 c_W s_W^2}$ \\
\hline
$Z Z H^{+Q_Y} H^{-Q_Y}$ & $\frac{i e^2 g_{\mu \nu}\left(a_+^2s_W^4-2c_{23}^2a_+s_W^2+c_{23}^2\right)}{2 c_W^2 s_W^2}$ \\
\hline
$Z Z H^+ H^-$ & $\frac{i e^2 g_{\mu \nu}\left(c_W^2-s_W^2\right)^2 }{2 c_W^2 s_W^2}$ \\
\hline
$Z Z H^{+Q_V} H^{-Q_V}$ & $\frac{i e^2 g_{\mu \nu}\left(a_-^2s_W^4+2c_{13}^2a_-s_W^2+c_{13}^2\right)}{2 c_W^2 s_W^2}$ \\
\hline
$Z Z H_0 H_0$ & $\frac{i e^2 g_{\mu \nu} \left[\left(O_{11}\right)^2+\left(O_{21}\right)^2\right]}{2 c_W^2 s_W^2}$ \\
\hline
$Z Z h h$ & $\frac{i e^2 g_{\mu \nu} \left[\left(U_{11}\right)^2+\left(U_{21}\right)^2\right]}{2 c_W^2 s_W^2}$ \\
\hline
$Z Z h H_2$ & $\frac{i e^2 g_{\mu \nu} \left[U_{11} U_{12}+U_{21} U_{22}\right]}{2 c_W^2 s_W^2}$ \\
\hline
$Z Z H_2 H_2$ & $\frac{i e^2 g_{\mu \nu} \left[\left(U_{12}\right)^2+\left(U_{22}\right)^2\right]}{2 c_W^2 s_W^2}$ \\
\hline
$Z Z h H_3$ & $\frac{i e^2 g_{\mu \nu} \left(U_{11} U_{13}+U_{21} U_{23}\right)}{2 c_W^2 s_W^2}$ \\
\hline
$Z Z H_2 H_3$ & $\frac{i e^2 g_{\mu \nu} \left(U_{12} U_{13}+U_{22} U_{23}\right)}{2 c_W^2 s_W^2}$ \\
\hline
$Z Z H_3 H_3$ & $\frac{i e^2 g_{\mu \nu} \left[\left(U_{13}\right)^2+\left(U_{23}\right)^2\right]}{2 c_W^2 s_W^2}$ \\
\hline
$A Z' H^{+Q_Y} H^{-Q_Y}$ & $\frac{i e^2 g_{\mu \nu} \left[a_+\left(\left(a_+a_- +2\right)s_W^2-1\right)+s_{23}^2\left(a_+^2s_W^2-a_+\right)\right]}{2\sqrt{3} c_W s_W a_0}$ \\
\hline
$A Z' H^+ H^-$ & $\frac{i e^2 g_{\mu \nu}\left(1+a_- s_W^2+2s_{12}^2 c_W^2\right)}{\sqrt{3} c_W s_W a_0}$ \\
\hline
$A Z' H^{+Q_V} H^{-Q_V}$ & $\frac{i e^2 g_{\mu \nu}\left[a_- \left(1-\left(a_+ a_- +2\right)s_W^2\right)+s_{13}^2\left(a_- + a_-^2s_W^2\right)\right]}{2\sqrt{3} c_W s_W a_0}$ \\
\hline
$V^{+Q_V} Z' H^- H^{-Q_Y}$ & $\frac{i e^2 g_{\mu \nu}s_{23} c_{12} \left(a_+^2 s_W^2-1\right)}{2\sqrt{6} c_W s_W^2 a_0}$ \\
\hline
$V^{+Q_V} Z' H_0 H^{-Q_V}$ & $\frac{i e^2 g_{\mu \nu}\left[ c_{13} O_{31} \left(a_+a_- s_W^2+1\right)+s_{13} O_{11} \left(a_-^2 s_W^2-1\right)\right]}{4\sqrt{3} c_W s_W^2 a_0}$ \\
\hline
$V^{+Q_V} Z' h H^{-Q_V}$ & $\frac{e^2 g_{\mu \nu}\left[ c_{13} U_{31}\left(1-a_+a_- s_W^2\right)+s_{13} U_{11} \left(a_-^2 s_W^2-1\right)\right]}{4\sqrt{3} c_W s_W^2 a_0}$ \\
\hline
$V^{+Q_V} Z' H_2 H^{-Q_V}$ & $\frac{e^2 g_{\mu \nu}\left[ c_{13} U_{32}\left(1-a_+a_- s_W^2\right)+s_{13} U_{12} \left(a_-^2 s_W^2-1\right)\right]}{4\sqrt{3} c_W s_W^2 a_0}$ \\
\hline
$V^{+Q_V} Z' H_3 H^{-Q_V}$ & $\frac{e^2 g_{\mu \nu}\left[ c_{13} U_{33}\left(1-a_+a_- s_W^2\right)+s_{13} U_{13} \left(a_-^2 s_W^2-1\right)\right]}{4\sqrt{3} c_W s_W^2 a_0}$ \\
\hline
$W^+ Z' H^{-Q_Y} H^{-Q_V}$ & $\frac{i e^2 g_{\mu \nu}c_{23} c_{13} \left(\left(a_+a_- +2\right) s_W^2-1\right)}{\sqrt{6} c_W s_W^2 a_0}$ \\
\hline
$W^+ Z' H_0 H^-$ & $-\frac{i e^2 g_{\mu \nu}\left[ c_{12}O_{21} \left(1+a_- s_W^2\right)+s_{12} O_{11} \left(a_+ s_W^2-1\right)\right]}{2\sqrt{3} c_W s_W^2 a_0}$ \\
\hline
$W^+ Z' h H^-$ & $\frac{e^2 g_{\mu \nu}\left[ c_{12}U_{21} \left(1+a_- s_W^2\right)+s_{12} U_{11} \left(1-a_+ s_W^2\right)\right]}{2\sqrt{3} c_W s_W^2 a_0}$ \\
\hline
$W^+ Z' H_2 H^-$ & $\frac{e^2 g_{\mu \nu}\left[ c_{12}U_{22} \left(1+a_- s_W^2\right)+s_{12} U_{12} \left(1-a_+ s_W^2\right)\right]}{2\sqrt{3} c_W s_W^2 a_0}$ \\
\hline
$W^+ Z' H_3 H^-$ & $\frac{e^2 g_{\mu \nu}\left[ c_{12}U_{23} \left(1+a_- s_W^2\right)+s_{12} U_{13} \left(1-a_+ s_W^2\right)\right]}{2\sqrt{3} c_W s_W^2 a_0}$ \\
\hline
$Y^{+Q_Y} Z' H^+ H^{-Q_V}$ & $-\frac{i e^2 g_{\mu \nu}s_{12} s_{13} \left(a_-^2 s_W^2-1\right)}{2\sqrt{6} c_W s_W^2 a_0}$ \\
\hline
$Y^{+Q_Y} Z' H_0 H^{-Q_Y}$ & $\frac{i e^2 g_{\mu \nu}\left[  c_{23} O_{31}\left(a_+a_- s_W^2+1\right)+s_{23} O_{21} \left(a_+^2 s_W^2-1\right)\right]}{4\sqrt{3} c_W s_W^2 a_0}$ \\
\hline
$Y^{+Q_Y} Z' h H^{-Q_Y}$ & $-\frac{e^2 g_{\mu \nu}\left[  c_{23} U_{31}\left(a_+a_-s_W^2+1\right)-s_{23} U_{21} \left(a_+^2 s_W^2-1\right)\right]}{4\sqrt{3} c_W s_W^2 a_0}$ \\
\hline
$Y^{+Q_Y} Z' H_2 H^{-Q_Y}$ &  $-\frac{e^2 g_{\mu \nu}\left[  c_{23} U_{32}\left(a_+a_-s_W^2+1\right)-s_{23} U_{22} \left(a_+^2 s_W^2-1\right)\right]}{4\sqrt{3} c_W s_W^2 a_0}$ \\
\hline
$Y^{+Q_Y} Z' H_3 H^{-Q_Y}$ &  $-\frac{e^2 g_{\mu \nu}\left[  c_{23} U_{33}\left(a_+a_-s_W^2+1\right)-s_{23} U_{23} \left(a_+^2 s_W^2-1\right)\right]}{4\sqrt{3} c_W s_W^2 a_0}$ \\
\hline
$Z Z' H^{+Q_Y} H^{-Q_Y}$ & $-\frac{i e^2 g_{\mu \nu}\left[s_{23}^2 s_W^2 \left(a_+\left(a_-b_+ +4\right) s_W^2-2 a_+\right)-c_{23}^2 \left(\left(a_+a_- +2\right) s_W^2-1\right) \left(1-a_+ s_W^2\right)\right]}{2\sqrt{3} c_W^2 s_W^2 a_0}$ \\
\hline
$Z Z' H^+ H^-$ & $\frac{i e^2 g_{\mu \nu}\left(c_W^2-s_W^2\right) \left[c_{12}^2 \left(1+a_- s_W^2\right)+s_{12}^2 \left(a_+ s_W^2-1\right)\right]}{2\sqrt{3} c_W^2 s_W^2 a_0}$ \\
\hline
$Z Z' H^{+Q_V} H^{-Q_V}$ & $-\frac{i e^2 g_{\mu \nu}\left[c_{13}^2 \left(\left(a_+a_- +2\right) s_W^2-1\right) \left(1+a_- s_W^2\right)+s_{13}^2 s_W^2 \left(-2 a_- +a_-\left(a_+b_- +4\right)s_W^2\right)\right]}{6 c_W^2 s_W^2 a_0}$ \\
\hline
$Z Z' H_0 H_0$ & $-\frac{i e^2 g_{\mu \nu} \left[\left(O_{11}\right)^2 \left(a_+ s_W^2-1\right)+\left(O_{21}\right)^2 \left(1+a_- s_W^2\right)\right]}{2\sqrt{3} c_W^2 s_W^2 a_0}$ \\
\hline
$Z Z' h h$ & $-\frac{i e^2 g_{\mu \nu} \left[\left(U_{11}\right)^2 \left(a_+ s_W^2-1\right)+\left(U_{21}\right)^2 \left(1+a_- s_W^2\right)\right]}{2\sqrt{3} c_W^2 s_W^2 a_0}$ \\
\hline
$Z Z' h H_2$ & $-\frac{i e^2 g_{\mu \nu} \left[U_{11} U_{12} \left(a_+ s_W^2-1\right)+U_{21} U_{22} \left(1+a_- s_W^2\right)\right]}{2\sqrt{3} c_W^2 s_W^2 a_0}$ \\
\hline
$Z Z' H_2 H_2$ & $-\frac{i e^2 g_{\mu \nu} \left[\left(U_{12}\right)^2 \left(a_+ s_W^2-1\right)+\left(U_{22}\right)^2 \left(1+a_- s_W^2\right)\right]}{2\sqrt{3} c_W^2 s_W^2 a_0}$ \\
\hline
$Z Z' h H_3$ & $-\frac{i e^2 g_{\mu \nu} \left[U_{11} U_{13} \left(a_+ s_W^2-1\right)+U_{21} U_{23} \left(1+a_- s_W^2\right)\right]}{2\sqrt{3} c_W^2 s_W^2 a_0}$ \\
\hline
$Z Z' H_2 H_3$ & $-\frac{i e^2 g_{\mu \nu} \left[U_{12} U_{13} \left(a_+ s_W^2-1\right)+U_{22} U_{23} \left(1+a_- s_W^2\right)\right]}{2\sqrt{3} c_W^2 s_W^2 a_0}$ \\
\hline
$Z Z' H_3 H_3$ & $-\frac{i e^2 g_{\mu \nu} \left[\left(U_{13}\right)^2 \left(a_+ s_W^2-1\right)+\left(U_{23}\right)^2 \left(1+a_- s_W^2\right)\right]}{2\sqrt{3} c_W^2 s_W^2 a_0}$ \\
\hline
$Z' Z' H^{+Q_Y} H^{-Q_Y}$ & $\frac{i e^2 g_{\mu \nu}\left[c_{23}^2 \left(\left(a_+a_- +2\right) s_W^2-1\right)^2+s_{23}^2 \left(\left(a_-b_+  +4\right)s_W^2-2\right)^2\right]}{6 c_W^2 s_W^2 a_0^2}$ \\
\hline
$Z' Z' H^+ H^-$ & $\frac{i e^2 g_{\mu \nu}\left[c_{12}^2 \left(a_- s_W^2\right)^2+s_{12}^2 \left(a_+ s_W^2\right)^2\right]}{6 c_W^2 s_W^2 a_0^2}$ \\
\hline
$Z' Z' H^{+Q_V} H^{-Q_V}$ & $\frac{i e^2 g_{\mu \nu}\left[c_{13}^2 \left(\left(a_+a_- +2\right) s_W^2-1\right)^2+s_{13}^2 \left(\left(a_+b_- +4\right)s_W^2-2\right)^2\right]}{6 c_W^2 s_W^2 a_0^2}$ \\
\hline
$Z' Z' H_0 H_0$ & $\frac{i e^2 g_{\mu \nu}\left[\left(O_{11}\right)^2 \left(a_+ s_W^2-1\right)^2+\left(O_{21}\right)^2 \left(a_- s_W^2+1\right)^2+\left(4 c_W^4 \left(O_{31}\right)^2\right)\right]}{6 c_W^2 s_W^2 a_0^2}$ \\
\hline
$Z' Z' h h$ & $\frac{i e^2 g_{\mu \nu}\left[\left(U_{11}\right)^2 \left(a_+ s_W^2-1\right)^2+\left(U_{21}\right)^2 \left(a_- s_W^2+1\right)^2+\left(4 c_W^4 \left(U_{31}\right)^2\right)\right]}{6 c_W^2 s_W^2 a_0^2}$ \\
\hline
$Z' Z' h H_2$ &  $\frac{i e^2 g_{\mu \nu}\left[U_{11} U_{12} \left(a_+ s_W^2-1\right)^2+U_{21} U_{22} \left(a_- s_W^2+1\right)^2+\left(4 c_W^4 U_{31} U_{32}\right)\right]}{6 c_W^2 s_W^2 a_0^2}$\\
\hline
$Z' Z' H_2 H_2$ & $\frac{i e^2 g_{\mu \nu}\left[\left(U_{12}\right)^2 \left(a_+ s_W^2-1\right)^2+\left(U_{22}\right)^2 \left(a_- s_W^2+1\right)^2+\left(4 c_W^4 \left(U_{32}\right)^2\right)\right]}{6 c_W^2 s_W^2 a_0^2}$  \\
\hline
$Z' Z' h H_3$ & $\frac{i e^2 g_{\mu \nu}\left[U_{11} U_{13} \left(a_+ s_W^2-1\right)^2+U_{21} U_{23} \left(a_- s_W^2+1\right)^2+\left(4 c_W^4 U_{31} U_{33}\right)\right]}{6 c_W^2 s_W^2 a_0^2}$ \\
\hline
$Z' Z' H_2 H_3$ &$\frac{i e^2 g_{\mu \nu}\left[U_{12} U_{13} \left(a_+ s_W^2-1\right)^2+U_{22} U_{23} \left(a_- s_W^2+1\right)^2+\left(4 c_W^4 U_{32} U_{33}\right)\right]}{6 c_W^2 s_W^2a_0^2}$  \\
\hline
$Z' Z' H_3 H_3$ & $\frac{i e^2 g_{\mu \nu}\left[\left(U_{13}\right)^2 \left(a_+ s_W^2-1\right)^2+\left(U_{23}\right)^2 \left(a_- s_W^2+1\right)^2+\left(4 c_W^4 \left(U_{33}\right)^2\right)\right]}{6 c_W^2 s_W^2 a_0^2}$ \\
\hline
\caption{Four-point couplings of two gauge bosons to two scalars.
\label{2gauges-2scalars}
}
\end{longtable}

\subsection{Gauge Boson Self-couplings}
The gauge boson self-couplings are given here, with all momenta out-going. The three-point and four point couplings take the following form
\begin{eqnarray}
V_1^\mu(k_1) V_2^\nu(k_2) V_3^\rho(k_3)&&g_{V_1V_2V_3}\left[\left(k_1-k_2\right)^{\rho } \eta _{\mu ,\nu }+\left(k_2-k_3\right)^{\mu } \eta _{\nu ,\rho }+\left(k_3-k_1\right)^{\nu } \eta _{\mu ,\rho }\right], \nn\\
V_1^\mu V_2^\nu V_3^\rho V_4^\sigma &&g_{V_1V_2V_3V_4}^1(\eta _{\mu ,\sigma } \eta _{\nu ,\rho }+\eta _{\mu ,\rho } \eta _{\nu ,\sigma }-2 \eta _{\mu ,\nu } \eta _{\rho ,\sigma }),  \nonumber \\
V_1^\mu V_2^\nu V_3^\rho V_4^\sigma &&g_{V_1V_2V_3V_4}^2[\left(3-\left(a_+ +2\right)s_W^2\right)\eta _{\mu ,\sigma } \eta _{\nu ,\rho } +2 \left(a_- +1\right)  s_W^2 \eta _{\mu ,\rho } \eta _{\nu ,\sigma } \nn \\
&&-\left(3+\left(a_- -2\right)s_W^2\right)\eta _{\mu ,\nu } \eta _{\rho ,\sigma}],  \nonumber \\
V_1^\mu V_2^\nu V_3^\rho V_4^\sigma &&g_{V_1V_2V_3V_4}^3[\left(\sqrt{3} \beta -3\right) \eta _{\mu ,\sigma } \eta _{\nu ,\rho }-2 \sqrt{3} \beta  \eta _{\mu ,\rho } \eta _{\nu ,\sigma } \nn \\
&&+\left(\sqrt{3} \beta +3\right) \eta _{\mu ,\nu } \eta _{\rho ,\sigma}].
\end{eqnarray}
The coefficients $g_{V_1V_2V_3}$, $g_{V_1V_2V_3V_4}^1$, $g_{V_1V_2V_3V_4}^2$ and $g_{V_1V_2V_3V_4}^3$ are given in Table~\ref{gauges}.

\begin{longtable}{c|c||c|c}
\hline
particles & $g_{V_1V_2V_3}$ & particles & $g_{V_1V_2V_3}$ \\
\hline
$AV^{+Q_V}V^{-Q_V}$ & $\frac{1}{2} i a_- e$ & $A Y^{+Q_Y} Y^{-Q_Y}$ & $\frac{1}{2} i a_+ e$ \\
\hline
$AWW$ & $-ie$ & $ZV^{+Q_V}V^{-Q_V}$ & $-\frac{i e \left(1+a_- s_W^2\right)}{2 c_W s_W}$ \\
\hline
$ZY^{+Q_Y}Y^{-Q_Y}$ & $\frac{i e \left(1-a_+ s_W^2\right)}{2 c_W s_W}$ & $ZWW$ & $-\frac{i e c_W}{s_W}$ \\
\hline
$Z'V^{+Q_V}V^{-Q_V}$ & $-\frac{i \sqrt{3} e a_0}{2 c_W s_W}$ & $Z'Y^{+Q_Y}Y^{-Q_Y}$ & $-\frac{i \sqrt{3} e a_0}{2 c_W s_W}$ \\
\hline
$V^{-Q_V} W^+Y^{+Q_Y}$ & $\frac{i e}{\sqrt{2} s_W}$ & $V^{+Q_V}W^- Y^{-Q_Y}$ & $-\frac{i e}{\sqrt{2} s_W}$ \\
\hline \hline
particles & $g_{V_1V_2V_3V_4}^1$ & particles & $g_{V_1V_2V_3V_4}^1$ \\
\hline
$AAV^{+Q_V}V^{-Q_V}$ & $\frac{i e^2 a_-^2}{4}$ & $V^{+Q_V}V^{+Q_V}V^{-Q_V} V^{-Q_V}$ & $-\frac{i e^2}{s_W^2}$ \\
\hline
$AAWW$ & $i e^2$ & $W^+WW W^-$ & $-\frac{i e^2}{s_W^2}$ \\
\hline
$AAY^{+Q_Y}Y^{-Q_Y}$ & $\frac{i e^2 a_+^2}{4}$ & $Y^{+Q_Y}Y^{+Q_Y}Y^{-Q_Y} Y^{-Q_Y}$ & $-\frac{i e^2}{s_W^2}$ \\
\hline
$V^{+Q_V}V^{-Q_V} ZZ$ & $\frac{i e^2 \left(1+a_- s_W^2\right)^2}{4 c_W^2 s_W^2}$ & $WW ZZ$ & $\frac{i e^2 c_W^2}{s_W^2}$ \\
\hline
$Y^{+Q_Y}Y^{-Q_Y} ZZ$ & $\frac{i e^2 \left(-1+a_+ s_W^2\right)^2}{4 c_W^2 s_W^2}$ & $V^{+Q_V}V^{-Q_V} ZZ'$ & $\frac{i e^2 \sqrt{3}a_0 \left(1+a_- s_W^2\right)}{4 c_W^2 s_W^2}$ \\
\hline
$Y^{+Q_Y}Y^{-Q_Y} ZZ'$ & $-\frac{i e^2 \sqrt{3}a_0 \left(1-a_+ s_W^2\right)}{4 c_W^2 s_W^2}$ & $V^{+Q_V}V^{-Q_V} Z'Z'$ & $\frac{3 i e^2 a_0^2}{4 c_W^2 s_W^2}$ \\
\hline
$Y^{+Q_Y}Y^{-Q_Y} Z'Z'$ & $\frac{3 i e^2 a_0^2}{4 c_W^2 s_W^2}$ & $AZ'Y^{-Q_Y} Y^{+Q_Y}$ & $-\frac{i \sqrt{3}a_+ e^2 a_0}{4 c_W s_W}$ \\
\hline
$W^+Y^{-Q_Y} Y^{+Q_Y}W^-$ & $-\frac{i e^2}{2 s_W^2}$ & $AZW^- W^+$ & $\frac{i e^2 c_W}{s_W}$ \\
\hline
$AZV^{-Q_V} V^{+Q_V}$ & $-\frac{i e^2 \left(a_- + a_-^2 s_W^2\right)}{4 c_W s_W}$ & $AZY^{-Q_Y} Y^{+Q_Y}$ & $\frac{i e^2 \left(a_+ - a_+^2s_W^2\right)}{4 c_W s_W}$ \\
\hline
$AZ'V^{-Q_V} V^{+Q_V}$ & $-\frac{i \sqrt{3}a_- e^2 a_0}{4 c_W s_W}$ & $V^{+Q_V}W^+V^{-Q_V} W^-$ & $-\frac{i e^2}{2 s_W^2}$ \\
\hline
$V^{+Q_V}Y^{+Q_Y}V^{-Q_V} Y^{-Q_Y}$ & $-\frac{i e^2}{2 s_W^2}$ & $V^{-Q_V} Y^{+Q_Y}W^+Z'$ & $-\frac{i \sqrt{\frac{3}{2}} e^2 a_0}{2 c_W s_W^2}$\\
\hline
\hline
particles & $g_{V_1V_2V_3V_4}^2$ & particles & $g_{V_1V_2V_3V_4}^3$ \\
\hline
$V^{-Q_V} W^+Y^{+Q_Y}Z$ & $\frac{i e^2}{2 \sqrt{2} c_W s_W^2}$ &  $AV^{-Q_V} W^+Y^{+Q_Y}$ & $\frac{i e^2}{2 \sqrt{2} s_W}$\\
\hline
\caption{Gauge boson self-couplings.}
\label{gauges}
\end{longtable}

\subsection{Gauge Boson-Fermion Couplings}
The couplings between gauge bosons and fermions are given in
Table~\ref{gauge-fermion}. The vertices are $g_L\gamma^\mu P_L+g_R\gamma^\mu P_R$,
where $P_R=(1+\gamma^5)/2$ and $P_L=(1-\gamma^5)/2$ are the projection operators.

\begin{longtable}{c|c|c}
\hline
particles & $g_L$ & $g_R$ \\
\hline
$\bar e E_e Y^{+Q_Y}, \bar \mu E_\mu Y^{+Q_Y}, \bar \tau E_\tau Y^{+Q_Y}$ & $-\frac{i e }{\sqrt{2} s_W}$ & 0 \\
\hline
$\bar D u Y^{+Q_Y}, \bar S c Y^{+Q_Y}, \bar t T V^{+Q_V}$ & $-\frac{i e }{\sqrt{2} s_W}$ & 0 \\
\hline
$\bar D d V^{+Q_V}, \bar S s V^{+Q_V}, \bar T b Y^{-Q_Y}$ & $-\frac{i e V_{ud}}{\sqrt{2} s_W}, -\frac{i e V_{cs}}{\sqrt{2} s_W}, -\frac{i e V_{tb}}{\sqrt{2} s_W}$ & 0 \\
\hline
$\bar v_e E_e V^{+Q_V}, \bar v_\mu E_\mu V^{+Q_V}, \bar v_\tau E_\tau V^{+Q_V}$ & $-\frac{i e }{\sqrt{2} s_W}$ & 0 \\
\hline
$\bar e e Z', \bar \mu \mu Z', \bar \tau \tau Z'$ & $-\frac{i e \left(a_+ s_W^2-1\right)}{2\sqrt{3} c_W s_W a_0}$ & $-\frac{i \beta  e s_W}{c_W a_0}$ \\
\hline
$\bar v_e v_e Z', \bar v_\mu v_\mu Z', \bar v_\tau v_\tau Z'$ & $-\frac{i e \left(a_+ s_W^2-1\right)}{2\sqrt{3} c_W s_W a_0}$ & $0$ \\
\hline
$\bar u u Z', \bar c c Z'$ & $\frac{i e \left[\left(\beta +\sqrt{3}\right) s_W^2-\sqrt{3}\right]}{6 c_W s_W a_0}$ & $\frac{2 i \beta  e s_W}{3 c_W a_0}$ \\
\hline
$\bar t t Z'$ & $\frac{i e \left[\left(\beta -\sqrt{3}\right) s_W^2+\sqrt{3}\right]}{6 c_W s_W a_0}$ & $\frac{2 i \beta  e s_W}{3 c_W a_0}$ \\
\hline
$\bar d d Z', \bar s s Z'$ & $\frac{i e \left[\left(\beta +\sqrt{3}\right) s_W^2-\sqrt{3}\right]}{6 c_W s_W a_0}$ & $-\frac{i \beta  e s_W}{3 c_W a_0}$ \\
\hline
$\bar b b Z'$ & $\frac{i e \left[\left(\beta -\sqrt{3}\right) s_W^2+\sqrt{3}\right]}{6 c_W s_W a_0}$ & $-\frac{i \beta  e s_W}{3 c_W a_0}$ \\
\hline
$\bar D D A, \bar S S A$ & $-\frac{ i \left(3 \sqrt{3} \beta -1\right) e}{6}$ & $-\frac{ i \left(3 \sqrt{3} \beta -1\right) e}{6}$ \\
\hline
$\bar T T A$ & $\frac{ i \left(3 \sqrt{3} \beta +1\right) e}{6}$ & $\frac{ i \left(3 \sqrt{3} \beta +1\right) e}{6}$ \\
\hline
$\bar D D Z, \bar S S Z$ & $\frac{i \left(3 \sqrt{3} \beta -1\right) e s_W}{6 c_W}$ & $\frac{i \left(3 \sqrt{3} \beta -1\right) e s_W}{6 c_W}$ \\
\hline
$\bar T T Z$ & $-\frac{i \left(3 \sqrt{3} \beta +1\right) e s_W}{6 c_W}$ & $-\frac{i \left(3 \sqrt{3} \beta +1\right) e s_W}{6 c_W}$ \\
\hline
$\bar D D Z', \bar S S Z'$ & $-\frac{i e \left(\left(3 \sqrt{3} \beta ^2-\beta +2 \sqrt{3}\right) s_W^2-2 \sqrt{3}\right)}{6 c_W s_W a_0}$ & $-\frac{i \beta  \left(3 \sqrt{3} \beta -1\right) e s_W}{6 c_W a_0}$ \\
\hline
$\bar T T Z'$ & $\frac{i e \left(\left(3 \sqrt{3} \beta ^2+\beta +2 \sqrt{3}\right) s_W^2-2 \sqrt{3}\right)}{6 c_W s_W a_0}$ & $\frac{i \beta  \left(3 \sqrt{3} \beta +1\right) e s_W}{6 c_W a_0}$ \\
\hline
$\bar E_e E_e A, \bar E_\mu E_\mu A, \bar E_\tau E_\tau A$ & $\frac{i a_- e}{2} $ &$\frac{i a_- e}{2} $ \\
\hline
$\bar E_e E_e Z, \bar E_\mu E_\mu Z, \bar E_\tau E_\tau Z$ & $-\frac{i a_- e s_W}{2 c_W}$ & $-\frac{i a_- e s_W}{2 c_W}$ \\
\hline
$\bar E_e E_e Z', \bar E_\mu E_\mu Z', \bar E_\tau E_\tau Z'$ & $\frac{i e \left[\left(a_-^2+1\right) s_W^2-2\right]}{2\sqrt{3} c_W s_W a_0}$ & $\frac{i \sqrt{3} \beta ^2 e s_W}{2 c_W a_0}$ \\
\hline
\caption{Gauge boson-fermion couplings.}
\label{gauge-fermion}
\end{longtable}

\subsection{Scalar-Fermion Couplings}
The couplings between scalars and fermions are given in
Table~\ref{scalar-fermion}, where we only include the flavor-diagonal interactions.

\begin{longtable}{c|c}
\hline
particles & vertices \\
\hline
$H_0\bar{E}_\ell E_\ell$ & $\frac{O_{31}M_{E_\ell}}{v_3}$ \\
\hline
$h\bar{E}_\ell E_\ell$ & $-i\frac{U_{31}M_{E_\ell}}{v_3}$ \\
\hline
$H_2\bar{E}_\ell E_\ell$ & $-i\frac{U_{32}M_{E_\ell}}{v_3}$ \\
\hline
$H_3\bar{E}_\ell E_\ell$ & $-i\frac{U_{33}M_{E_\ell}}{v_3}$ \\
\hline
$H_0\bar{\ell}\ell$ & $\frac{O_{21} m_\ell }{v_2}$ \\
\hline
$h\bar{\ell}\ell$ & $-i \frac{U_{21} m_\ell}{v_2} $ \\
\hline
$H_2\bar{\ell}\ell$ & $-i \frac{U_{22} m_\ell}{v_2} $ \\
\hline
$H_3\bar{\ell}\ell$ & $-i \frac{U_{23} m_\ell}{v_2} $ \\
\hline
$H^+ \bar{\nu_\ell} \ell$ & $\frac{\sqrt{2}m_\ell c_{12}}{v_2}P_R$ \\
\hline
$H^{+Q_V}\bar{E}_\ell \nu_\ell$ & $-\frac{\sqrt{2}c_{13}M_{E_\ell}}{v_3}P_L$ \\
\hline
$H^{-Q_Y}\bar{E_\ell}\ell$ & $-\sqrt{2} \left(\frac{s_{23} m_\ell}{v_2}P_L+\frac{c_{23} M_{E_\ell}}{v_3}P_R\right)$ \\
\hline
$H_0\bar{u}u$ & $\frac{O_{21} m_u}{v_2}$ \\
\hline
$h\bar{u}u$ & $-\frac{i U_{21} m_u}{v_2}$ \\
\hline
$H_2\bar{u}u$ & $-\frac{i U_{22} m_u}{v_2}$ \\
\hline
$H_3\bar{u}u$ & $-\frac{i U_{23} m_u}{v_2}$ \\
\hline
$H_0\bar{d}d$ & $\frac{O_{11} m_u}{v_1}$ \\
\hline
$h\bar{d}d$ & $-\frac{i U_{11} m_u}{v_1}$ \\
\hline
$H_2\bar{d}d$ & $-\frac{i U_{12} m_u}{v_1}$ \\
\hline
$H_3\bar{d}d$ & $-\frac{i U_{13} m_u}{v_1}$ \\
\hline
$H^+ \bar{u}(\bar{c})d(s)$ & $-\sqrt{2} V_{\text{u(c)d(s)}} \left(\frac{c_{12} m_{u(c)}}{v_2}P_L+\frac{s_{12} m_{d(s)}}{v_1}P_R\right)$ \\
\hline
$H^{-Q_V}\bar{D}(\bar{S})d(s)$ & $-\sqrt{2} V_{\text{u(c)d(s)}} \left(\frac{c_{13} M_{D(S)}}{v_3}P_L+\frac{s_{13} m_{d(s)}}{v_1}P_R\right)$ \\
\hline
$H^{+Q_Y}\bar{u}(\bar{c})D(S)$ & $\sqrt{2} \left(\frac{s_{23} m_{u(c)}}{v_2}P_L+\frac{c_{23} M_{D(S)}}{v_3}P_R\right)$ \\
\hline
$H_0\bar{t}t$ & $\frac{O_{11} m_u}{v_1}$ \\
\hline
$h\bar{t}t$ & $-i\frac{U_{11} m_u}{v_1}$ \\
\hline
$H_2\bar{t}t$ & $-i\frac{U_{12} m_u}{v_1}$ \\
\hline
$H_3\bar{t}t$ & $-i\frac{U_{13} m_u}{v_1}$ \\
\hline
$H_0\bar{b}b$ & $\frac{O_{21} m_b}{v_2}$ \\
\hline
$h\bar{b}b$ & $-i\frac{U_{21} m_b}{v_2}$ \\
\hline
$H_2\bar{b}b$ & $-i\frac{U_{22} m_b}{v_2}$ \\
\hline
$H_3\bar{b}b$ & $-i\frac{U_{23} m_b}{v_2}$ \\
\hline
$H^+ \bar{t}b$ & $\sqrt{2} V_{\text{tb}} \left(\frac{s_{12} m_t}{v_1}P_L+\frac{c_{12} m_b}{v_2}P_R\right)$ \\
\hline
$H^{-Q_V}\bar{t}T$ & $\sqrt{2} \left(\frac{s_{13} m_t}{v_1}P_L +\frac{c_{13} M_T}{v_3}P_R\right)$ \\
\hline
$H^{+Q_Y}\bar{T}b$ & $\sqrt{2} V_{\text{tb}} \left(\frac{c_{23} M_T}{v_3}P_L+\frac{s_{23} m_b}{v_2}P_R\right)$ \\
\hline
\caption{Scalar-fermion couplings.}
\label{scalar-fermion}
\end{longtable}

\newpage
\section{The rotation matrix for the neutral scalars }\label{app_b}
\subsection{The functions $g_{i1}(v)$ in the CP-even scalar mixing matrix}
\bea
g_{11}(v)&=&-\frac{v_1 v_2^2 \left(-2 k v_2 v_1+3 \lambda _{13} v_1^2+\lambda _{23} v_2^2\right)}{2 \lambda _3 v^3}-\frac{2 \lambda _1 v_2 v_1^4+\lambda _{12} v_2^3 v_1^2}{k v^3}\\
                  &-&\frac{v_1 \left(-2 k v_2 v_1+\lambda _{13} v_1^2+\lambda _{23} v_2^2\right)^2}{8 \lambda _3^2 v^3}+\frac{\lambda _{13} v_1^2 v_2 \left(\lambda _{13} v_1^2+\lambda _{23} v_2^2\right)}{2 k \lambda _3 v^3},\nn\\
g_{21}(v)&=&-\frac{v_2 \left(-2 k v_2 v_1+\lambda _{13} v_1^2+\lambda _{23} v_2^2\right)^2}{8 \lambda _3^2 v^3}-\frac{v_2^3 \left(-2 k v_2 v_1+3 \lambda _{13} v_1^2+\lambda _{23} v_2^2\right)}{2 \lambda _3 v^3}\nn\\
                   &+&\frac{-2 k v^2 v_1+2 k v_1^3+\lambda _{23} v^2 v_2+\left(3 \lambda _{13}-\lambda _{23}\right) v_2 v_1^2}{2 \lambda _3 v}\nn\\
                   &+&\frac{v_1 v_2^2 \left(\left(\lambda _{13}^2-4 \lambda _1 \lambda _3\right) v_1^2+\left(\lambda _{13} \lambda _{23}-2 \lambda _3 \lambda _{12}\right) v_2^2\right)}{2 k \lambda _3 v^3}\nn\\
                   &+&\frac{v_1 \left(-\lambda _{13} \lambda _{23} v^2+2 \lambda _3 \lambda _{12} v_2^2+\left(4 \lambda _1 \lambda _3+\lambda _{13} \left(\lambda _{23}-\lambda _{13}\right)\right) v_1^2\right)}{2 k \lambda _3 v},\\
g_{31}(v)&=&\frac{k v_1 v_2}{\lambda_3 v }-\frac{\lambda_{13}v_1^2+\lambda_{23}v_2^2}{2\lambda_3  v}.
\eea

\subsection{The CP-odd scalar mixing matrix $O$}
$O$ is defined as
\be
O=
\left(
\begin{array}{ccc}
O_{11} & O_{12} & O_{13} \\
O_{21} & O_{22} & O_{23} \\
O_{31} & O_{32} & O_{33} \\
\end{array}
\right)
\ee
with
\begin{align}
& O_{11}=\frac{v_2 v_3}{\sqrt{v_1^2 v_2^2+v^2 v_3^2}}, && O_{12}=-\frac{v_1 v_3^2}{\sqrt{\left(v_1^2+v_3^2\right) \left(v_1^2 v_2^2+v^2 v_3^2\right)}}, && O_{13}=-\frac{v_1}{\sqrt{v_1^2+v_3^2}}, \nn\\
& O_{21}=\frac{v_1 v_3}{\sqrt{v_1^2 v_2^2+v^2 v_3^2}}, && O_{22}=v_2 \sqrt{\frac{v_1^2+v_3^2}{v_1^2 v_2^2+v^2 v_3^2}}, && O_{23}=0,\nn\\
& O_{31}=\frac{v_1 v_2}{\sqrt{v_1^2 v_2^2+v^2 v_3^2}}, && O_{32}=\frac{v_1^2 v_3}{\sqrt{\left(v_1^2+v_3^2\right) \left( v_1^2 v_2^2+v^2 v_3^2\right)}}, && O_{33}=\frac{v_3}{\sqrt{v_1^2+v_3^2}}.\nn
\end{align}

\newpage
\section{The loop functions in the loop-induced decay width of the Higgs boson}\label{app_c}
The function $A_{1/2}^h(\tau)$ is given by
\be
A_{1/2}^h(\tau)=2[\tau+(\tau-1)f(\tau)]\tau^{-2},
\ee
where the function $f(\tau)$ is defined as
\be
f(\tau)=
   \begin{cases}
   \arcsin^2{\sqrt{\tau}} &\mbox{$\tau\leq$ 1}\\
   -\frac{1}{4}\bigg[\log{\frac{1+\sqrt{1-\tau^{-1}}}{1-\sqrt{1-\tau^{-1}}}-i\pi}\bigg]^2 &\mbox{$\tau>$ 1}
   \end{cases}.
\ee
The function $A_{1}^h(\tau)$ is given by
\be
A_{1}^h(\tau)=-[2\tau^2+3\tau+3(2\tau-1)f(\tau)]\tau^{-2}.
\ee
The functions $A_{1/2}^h(\tau,\lambda)$ and $A_{1}^h(\tau,\lambda)$ are defined as
\bea
A_{1/2}^h(\tau,\lambda)&=&I_1(\tau,\lambda)-I_2(\tau,\lambda)\\
A_1^h(\tau,\lambda)&=&c_W\left\{4\left(3-\frac{s_W^2}{c_W^2}\right)I_2(\tau,\lambda)+\left[\left(1+\frac{2}{\tau}\right)\frac{s_W^2}{c_W^2}-\left(5+\frac{2}{\tau}\right)\right]I_1(\tau,\lambda)\right\}.\nn
\eea
The functions $I_1$ and $I_2$ are given by
\bea
I_1(\tau,\lambda)&=&\frac{\tau\lambda}{2(\tau-\lambda)}+\frac{\tau^2\lambda^2}{2(\tau-\lambda)^2}[f(\tau^{-1})-f(\lambda^{-1})]+\frac{\tau^2\lambda}{(\tau-\lambda)^2}[g(\tau^{-1})-g(\lambda^{-1})]\nn\\
I_2(\tau,\lambda)&=&-\frac{\tau\lambda}{2(\tau-\lambda)}[f(\tau^{-1})-f(\lambda^{-1})],
\eea
where the function $g(\tau)$ can be expressed as
\be
g(\tau)=
   \begin{cases}
   \sqrt{\tau^{-1}-1}\arcsin{\sqrt{\tau}} &\mbox{$\tau\geq$ 1}\\
   \frac{\sqrt{1-\tau^{-1}}}{2}\bigg[\log{\frac{1+\sqrt{1-\tau^{-1}}}{1-\sqrt{1-\tau^{-1}}}-i\pi}\bigg]^2 &\mbox{$\tau<$ 1}
   \end{cases}.
\ee

\newpage
\bibliographystyle{apsrev}
\bibliography{reference}

\end{document}